\documentclass[12pt]{JHEP3}

\usepackage{graphicx}
\usepackage{amsmath,epsfig}
\usepackage{amssymb,amsfonts}
\usepackage{xcolor}
\let\normalcolor\relax
\usepackage{epsfig}
\usepackage{subfigure}
\usepackage{subfig}
\usepackage{array}

\relax
\renewcommand{\theequation}{\arabic{section}.\arabic{equation}}
\def\be{\begin{equation}}
\def\ee{\end{equation}}

\newcommand{\ha}{{1 \over 2}}

\newcommand{\de}{\partial}
\newcommand{\bear}{\begin{eqnarray}}
\newcommand{\bea}{\begin{eqnarray}}
\newcommand{\eear}{\end{eqnarray}}
\newcommand{\eea}{\end{eqnarray}}
\def\hri#1#2{\href{http://arxiv.org/abs/#1}{[ArXiv:#1]#2}}
\def\hre#1#2{\href{http://arxiv.org/abs/#1/#2}{[ArXiv:#1/#2]}}

\newbox\pippobox

\def\Pt{{\tilde P}}
\def\II{\relax{\rm I\kern-.18em I}}

\def\cO{{\cal O}}
\def\cU{{\cal U}}

\def\l{\lambda}
\def\m{\mu}
\def\n{\nu}
\def\mn{{\mu\nu}}
\def\r{\rho}
\def\g{\gamma}
\def\s{\sigma}

\def\t{\theta}
\def\sp{\;\;\;,\;\;\;}

\def\p{\partial}

\def\f{\varphi}
\def\z{\zeta}
\def\a{\alpha}
\def\b{\beta}

\def\t{\tau}

\def\o{\omega}
\def\le{\left}
\def\ri{\right}

\def\Vg{\sqrt{-g}} 
\def\emn{\eta_\mn} 

\newcommand{\eql}[2]
{ \begin{equation} \label{#1}
 #2
\end{equation}}

\title{Exotic RG flows from Holography}

\author{Elias Kiritsis$^{1,2,3}$, Francesco Nitti$^1$ and Leandro Silva Pimenta$^1$\\
 ~\\
 $^1$
\href{http://www.apc.univ-paris7.fr}
{APC, Universit\'e Paris 7}, CNRS/IN2P3, CEA/IRFU, Obs. de Paris, Sorbonne Paris Cit\'e, B\^atiment Condorcet, F-75205, Paris Cedex 13, France (UMR du CNRS 7164).\\
 ~\\
 $^2$
 \href{http://hep.physics.uoc.gr/}
 {Crete Center for Theoretical Physics}, Institute for Theoretical and Computational Physics, Department of Physics, University of Crete
 71003 Heraklion, Greece\\
  ~\\
$^3$
Crete Center for Quantum Complexity and Nanotechnology,
	Department of Physics, University of Crete, 71003 Heraklion, Greece.}

\abstract{Holographic RG flows are studied in an Einstein-dilaton theory with a general potential. The superpotential formalism is utilized in order to characterize and classify all solutions that are associated with asymptotically AdS space-times. Such solutions correspond to holographic RG flows and are characterized by their holographic $\beta$-functions. Novel solutions are found that have exotic properties from a RG point-of view. Some have $\beta$-functions that are defined patch-wise and lead to flows where the $\beta$-function changes sign without the flow stopping.
Others describe flows that end in non-neighboring extrema in field space. Finally others describe  regular flows between two minima  of the potential and correspond holographically to flows driven by the VEV of an irrelevant operator in the UV CFT.\\
\vspace{1cm}

{\em Dedicated to John H. Schwarz on his 75th birthday.}
}

\preprint{CCTP-2016-17\\
CCQCN-2016-175}

\begin{document}

\maketitle 

\section{Introduction}
\label{sec:gen}

The Wilsonian view of renormalization in quantum field theory (QFT) has led to the idea of the renormalization group (RG) that has changed dramatically our view on quantum field theories, \cite{W}. The RG  determines a local map of the space of quantum field theories. This mapping is organized by scale-invariant QFTs associated with fixed points of the RG flow and cutoff QFTs that fill the rest of the space of QFTs.

The conventional definition of the RG transformation typically proceeds by infinitesimal steps of integrating-out UV degrees of freedom, using most of the time perturbation theory and truncation of the space of possible operators\footnote{There are approaches where one truncates the space of operators but keeps the RG transformation exact. Such approaches come under the name of the exact RG, \cite{wt,morris}.}, \cite{polchinski}. Most of the time this is a well justified approach and it is fair to say that it has served us well during several  decades of theoretical efforts on QFT.

The holographic correspondence \cite{Malda,GKP,Witten98}, has provided a map between QFT and gravity/string-theories in higher dimensions at least in the limit of large $N$. The notion of RG flow is geometrized in the dual gravitational theories as the holographic dimension serves an effective RG scale in the dual QFT. Therefore, bulk evolution in the holographic dimensions seems to correspond to the QFT RG flow, \cite{GPPZ,Freedman,ceresole,MS}.

A priori, there seems to be a puzzle as the QFT RG flow equation is a first order differential equation while bulk gravity equations are second order.
A careful treatment of the radial flow in holography, \cite{ceresole,dVV,Bourdier13,rg1,rg2,ps,papa1,papa2,gkn,hp,flr} indicates that the proper formalism (at least for Lorentz invariant QFTs) is the Hamilton-Jacobi formalism. It  provides the appropriate framework for casting the bulk radial evolution equations into a first order (Hamiltonian) form, that is natural in order to show the correspondence with boundary RG flows.
Once the requirement of IR regularity of the bulk gravitational solutions (that correspond to the large-N saddle points) is taken into account, the gravitational flows seem to match the boundary first order flows, \cite{A,dVV,Bourdier13,rg1,rg2,ps,papa1,papa2,gkn,hp,flr}.

The formalism is capable of producing not only the RG flows but also the renormalized Schwinger source functional and (by Legendre transform) the quantum effective action, at least in Lorentz invariant cases and in an expansion in boundary derivatives, \cite{dVV,rg1,rg2,papa1,papa2,glueball}.
Moreover, it allows for the construction of holographic RG flows with space-time dependent couplings and provides all the Callan-Symanzik type equations known from QFT, \cite{rg2}.
Although it is not yet clear how this method can be applied to black-holes and other solutions, in order to compute the effective action, direct methods allow the computation of the effective potential at finite temperature and density, \cite{rg1,Lindgren}.

The attempts to bridge the QFT and holographic/gravitational pictures of RG flow suggest that there may be more to RG flows that what we currently believe. First, the Schwinger source functional in quantum field theory is not a well understood object. It is usually defined and studied in weakly coupled theories where the coupling to composite operators is usually not present. This is because  at weak coupling,  their correlators can be constructed from the basic Schwinger functional using functional differentiation (and renormalization)\footnote{The effective action for composite operators can be defined though and was done first in \cite{tom}.}.
Moreover, sources are non-dynamical.
Therefore, the analogue of multi-trace (or polynomial) operators is present but algebraically redundant.

In contrast, in holography, multi-trace operators are absent as dynamical variables from the classical saddle point, and correspond to multi-particle states in the string theory. Moreover, although the boundary values of the fields are non-dynamical, the bulk fields, dual to sources, are dynamical.
Various ideas have been put forward on how to obtain the gravitational picture of the Schwinger functional from the QFT one, by integrating out rather than dropping the multi-trace contributions in quantum field theory, \cite{polchinski,flr,LV}. This procedure was formalized and explored in \cite{sslee,sslee2} where a detailed algorithm was described to integrate out
multi-trace operators at the cost of making single trace sources dynamical. This was combined and intertwined with the usual integrating-out procedure  of the RG to produce effectively couplings in one more dimension, satisfying second order flow equations. This was called the ``quantum RG group" in \cite{sslee2}. Although many elements are missing from this procedure, it matches in its general lines the holographic setup.

The notion of the quantum RG equation as an evolution equation for couplings and eventually the whole of the effective action raises several interesting questions concerning what is possible in RG flows.
In quantum field theory, with the conventional first order RG evolution, the space of QFTs is organized by the zeros of the $\beta$-functions.
In the case of a single coupling for example, the evolution is always monotonic and it always  ends  at the fixed points (zeroes of the $\beta$-function).

 The space of RG flows can be quite complicated and the most complex example of a multidimensional space with an explicitly known set of $\beta$ functions, and an associated C-function is known in two-dimensions, \cite{C}, where the fixed points of the flow include all G/H coset CFTs and more\footnote{The flow is not the usual flow between Lorentz-Invariant CFTs but a kind of Hamiltonian flow between non-relativistic Hamiltonians, \cite{hal}.}. The set of such theories are general factorizations of the G WZW theory in 2d, and they contain CFTs with irrational central charge and conformal weights, \cite{ICFT}.
Such flows are stratified in two dimensions by the C-function and the associated C-theorem, but analogous statements hold in 4 dimensions and the associated $a$-theorem, \cite{Cardy,a}.

Although in two dimensions the RG picture seems clear, in four dimensions many open ends remain.
Although the $a$-theorem is now considered proven, \cite{a}, its strong form remains an open problem. It is known to be valid in perturbation theory, \cite{Osborn,JO2} but its general validity is still unknown.
A related important question is whether scale invariance and Poincar\'e invariance imply conformal invariance in four-dimensional positive QFTs.
Although no counterexample is known, a general proof is still lacking despite concrete recent progress in \cite{rychkov,lpr,kom1} and \cite{Nakayama}.

A related question is the possibility of limit cycles in unitary 4d relativistic QFTs. This has been linked to  the $a$-theorem as well as to the absence of conformal invariance.
There are a few examples of RG limit cycles in the literature, \cite{Wilson,Bedaque,doll} but all of them violate one of the assumptions (unitarity or relativistic invariance).
Limit cycles were recently claimed to exist  in 4d QFTs  with many scalars, \cite{RemoveC} but it was subsequently realized that the limit cycle behavior was an artifact of the potential global symmetry of the fixed point, \cite{lpr,RemoveC}.
The fact that an $a$-theorem does not necessarily forbid limit cycles was argued in \cite{cycles} by presenting toy multi-branched examples of $\beta$-functions that respect the $a$-theorem but have limit cycles.

Using holography, the analogous landscape of fixed points and solutions/flows of QFT becomes  the string landscape (restricted to regions when the potential is negative and the solutions AdS-like).
The conventional folk picture is that potential maxima correspond to unstable UV fixed points while minima to (potentially) stable IR fixed points. Of course, the generic potential extremum is a saddle with unstable directions corresponding to relevant operators in the dual CFT and stable directions corresponding to irrelevant operators.

This holographic picture leads  us to believe that the space of AdS-like solutions is composed of such RG-like flows, and there is a one to one map between potential extrema and RG-fixed points of dual QFTs.
However, the map between second order gravity equations (or ``quantum RG" (QRG) equations) and first order RG equations is valid patch-wise in the space of solutions and this allows a priori for more exotic possibilities.

Some of these possibilities were glimpsed in earlier attempts to understand YM as a holographic theory. In this effort, runaway  mildly singular solutions  with non-monotonic bulk scalar field were found  \cite{gkn,thermo}, but were not considered further because they did not fit with the goals at that time, as  non-monotonicity is not something we expect for the YM coupling constant.

The purpose of this paper is to undertake a systematic classification of gravitational solutions and the associated dual RG flows and to investigate whether all such solutions have conventional dual QFT analogues.  We will not address the most general case but in this paper we will confine to the case of a single coupling of a scalar operator, or in the gravitational side to an Einstein theory with a single scalar field.

We focus in particular on the following aspects:
\begin{enumerate}
\item A general characterization of gravitational solutions in terms of a superpotential, from which one can directly read-off the properties of the corresponding RG-flow;
\item An exploration of ``exotic'' holographic RG-flows: by this we mean those gravity solutions that result in unexpected behavior of the running coupling  when translated in the boundary field theory language. These include for example non-monotonic flows and flows that connect non-adjacent fixed points. Their existence is intrinsically related to the second-order nature of the bulk equations, and it can only arise non-perturbatively (if it at all) on the field theory side.
\item  A detailed  analysis of the perturbations around asymptotically $AdS$ RG flows, which shows that these solutions are always stable under small perturbations, provided that 1) the UV does not violate the BF bound and 2) the IR satisfies an appropriate regularity criterion.
\end{enumerate}

\subsection{Summary and discussion of results}

We  use a two-derivative $d+1$-dimensional model described by the action:
\be \label{intro1}
S =  \int du d^dx\,\left (R - {1\over 2}(\de \phi)^2 - V(\phi) \right).
\ee
A given Poincar\'e-invariant  solution  is characterized by the metric scale factor $e^{A(u)}$, which measures the field theory energy scale  and by a scalar field profile $\phi(u)$, which is interpreted as the running coupling:
\be
ds^2 = du^2 + e^{2A(u)}\eta_{\mu\nu}dx^\mu dx^\nu, \qquad \phi = \phi(u).\label{intro1-b}
\ee
Although the bulk theory is defined in terms of the scalar potential $V(\phi)$, the key to connecting a bulk geometry with an RG flow in the dual field theory is an auxiliary  scalar function $W(\phi)$ known as the {\em superpotential,} and satisfying the non-linear equation:
\be \label{intro2}
{1\over 2}\left(d W\over d\phi\right)^2 -  {d\over 4(d-1)} W^2 = V
\ee
We consider only $V<0$ because this will only allow for $AdS$ fixed points of the holographic RG flow of $\phi(u)$.
Both the energy scale and the running of the coupling are then determined by the superpotential through the following equations:
\be \label{intro2-b}
{dA \over du} = -{1\over 2(d-1)}W(\phi(u)), \qquad {d\phi \over du} = W(\phi(u)),
\ee
where $u$ is the holographic coordinate. In the framework we are discussing here, the $\beta$-function is directly related to the superpotential:
\be \label{intro2-c}
\beta(\phi)\equiv {d\phi\over dA} = -2(d-1) {d \log W \over d\phi}
\ee

Moreover $C(u) \propto 1/W^{d-1}$ can be taken as  the $c$-function of the holographic RG Flow \cite{GPPZ,Freedman}.

The space of superpotentials coincides with the space of possible RG-flows up to a choice of initial conditions for the running coupling (the latter only changes the ``speed'' of the flow but not its qualitative features) and the choice of scale for the boundary metric. Therefore, characterizing the possible RG-flows corresponding to a given bulk theory is equivalent to classifying all solutions to the non-linear equation (\ref{intro2}), with a given potential $V(\phi)$.

The superpotential provides a  connection between holography and the RG group and this connection has been explored in the past.
The superpotential  plays a central role in the Hamilton-Jacobi approach \cite{Freedman,dVV,papa1,papa2}.

\subsubsection{General properties of holographic RG flows}
In this paper we give a (as complete as possible) classification of the solutions of RG flows. Most of the properties can be deduced from the  superpotential equation. Some of these  are well known and we review them for completeness. We also point out some new features that have not been considered in the past. In this section we summarize the main points of this classification, leaving the details to the main body of the paper.

\paragraph{Monotonicity.}  $W(\phi)$ is always monotonically increasing  along the holographic RG flow.
\be
{dW\over du}~~\geq 0
\label{cc}\ee
 In particular,  a minimum of $W$ always corresponds to a UV fixed point, and a maximum to an IR fixed point. This ties well with the interpretation of $1/W^{d-1}$ as a C-function for the holographic RG Flow, \cite{Freedman}.

Because we use only strictly negative potentials, $W(\phi)$ is bounded from below by a curve $B(\phi)$ defined by:
\eql{defB}{B(\phi):=\sqrt{-{4(d-1)\over d}V(\phi)},}
as a consequence of equation \eqref{intro2}.

\begin{figure}[t]
\centering
\includegraphics[width=0.75\textwidth]{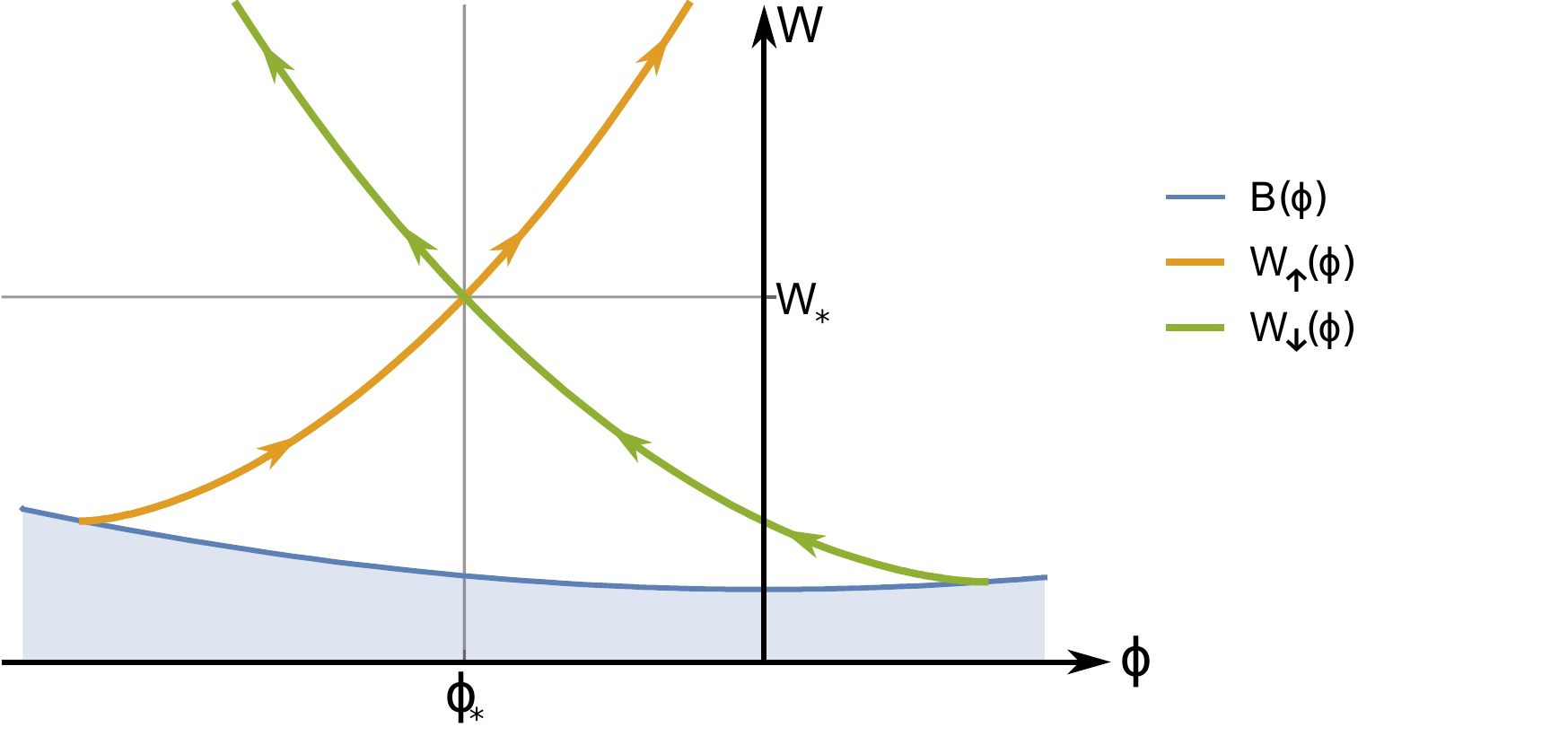}
\caption{Two branches. The two solutions $W_\uparrow(\phi)$ and $W_\downarrow(\phi)$ correspond to the two possible signs of $W'(\phi)$ in equation \protect\eqref{intro2}, for the initial condition $W(\phi_*)=W_*$. No solution can lie below the curve $B(\phi)$ defined in \protect\eqref{defB}, (here shown with $d=4$).
  The arrows indicate the direction of increasing holographic coordinate $u$.}
\label{fig:Crossing}
\end{figure}

\paragraph{Two branches} (Figure \ref{fig:Crossing}).
Equation \eqref{intro2} is equivalent to
\be
W'(\phi)\equiv {dW\over d\phi}=\pm\sqrt{{2d\over (d-1)}\le(W^2-B^2\ri)}, \label{+-}
\ee
which implies that through each generic point in the plane $(\phi, W)$ there are two intersecting solutions: a growing solution $W_\uparrow(\phi)$ and a decreasing solution $W_\downarrow(\phi)$. They correspond to an RG-flow in which $\phi$ increases or decreases with scale, respectively\footnote{Note that although $dW\over d\phi$ can have either sign, ${dW\over du}$ is always non-negative.}.

\paragraph{Fixed points.} As  is well known, extrema of $W(\phi)$ which coincide with extrema of $V(\phi)$ are fixed points of the holographic RG flow. One important difference between UV and IR fixed points is the following:
 \begin{enumerate}
\item {\bf UV:} Equation (\ref{intro2}) admits an infinite family of solutions describing  a relevant deformation away from a  UV fixed point, and characterized by an integration constant which sets the operator VEV in the dual field theory \cite{papa1}.
    However, only a finite subset of them can be extended to globally regular solutions/flows.

\item {\bf IR:} In contrast, the solutions reaching an IR fixed point are finite in number and therefore do not admit continuous deformations.

\end{enumerate}
As a consequence, any bulk theory with a finite number number of extrema of $V$ can only have a finite number of solutions reaching an IR fixed point. Moreover, if a solution reaches a given IR fixed point, no  other solution can have the same IR limit.

\begin{figure}[t]
\centering
\includegraphics[width=0.7\textwidth]{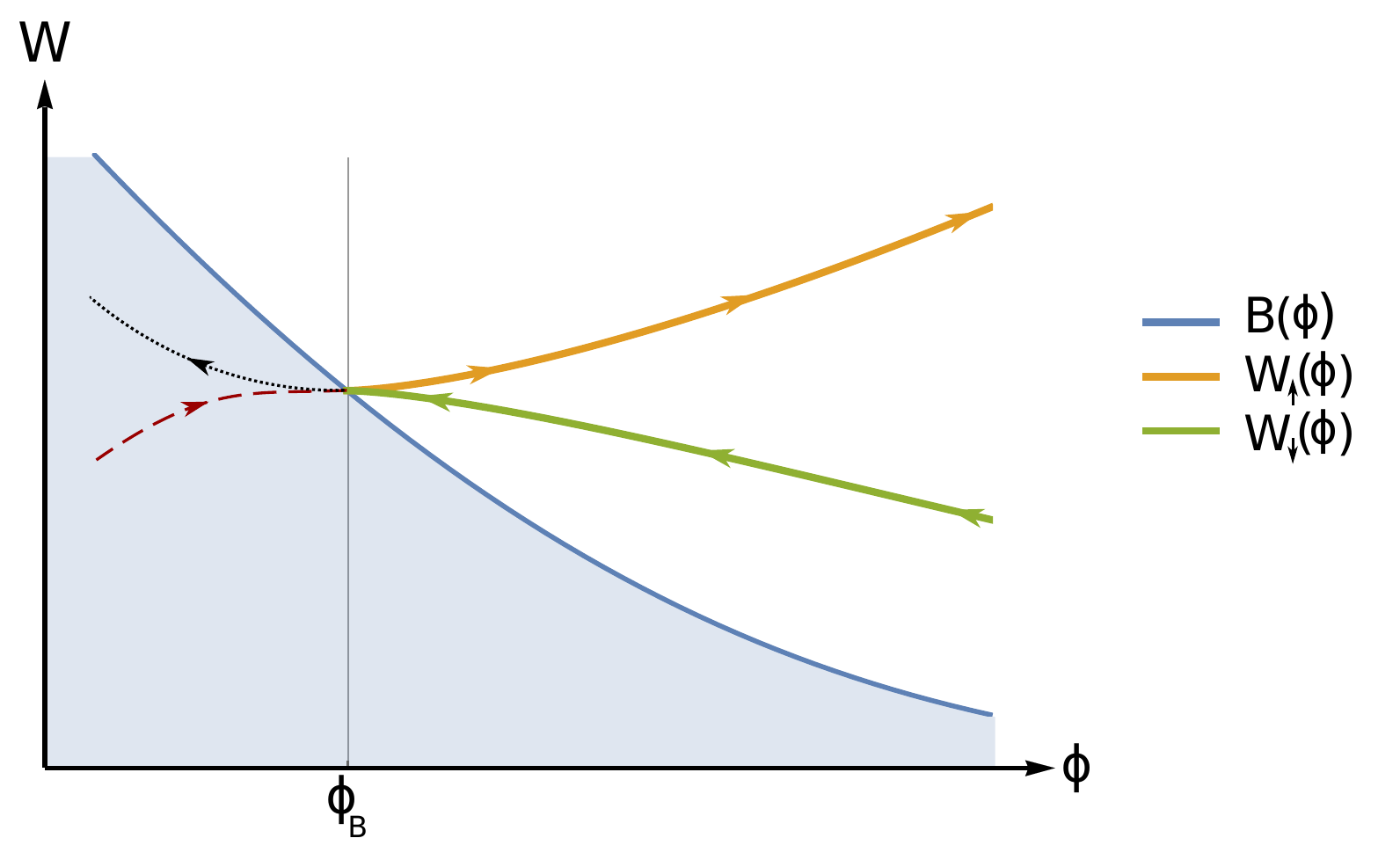}
\caption{Bounce. The two solutions $W_{\uparrow}(\phi)$ and $W_{\downarrow}(\phi)$
  reach the curve  $B(\phi)$ defined in \protect\eqref{defB} at the same point $\phi_B$ with vanishing derivative.
The solutions become complex, therefore unphysical, on the left of the bounce (dotted lines). The arrows  indicate the direction of the flow towards lower energies.}
\label{fig:bounce1}
\end{figure}

\paragraph{Bounces} (Figure \ref{fig:bounce1}).  Generically, one can also have extrema of $W(\phi)$ that do not coincide with extrema of $V(\phi)$. The two  branches of solutions to the superpotential equation (\ref{+-}) meet,  with vanishing derivative,  at a point $\phi_B$ on the critical curve $B(\phi)$. These are {\em not} standard fixed points. Rather, $\phi_B$  corresponds to a local maximum or minimum for the running coupling, in the interior of the bulk space-time:
at these points the superpotential (and therefore the $\beta$-function) becomes multi-branched as a function of the field  (the coupling). However,  we can smoothly connect  the two solutions corresponding to the two signs in equation (\ref{+-}), into a single regular solution. This results in a {\em bounce} in the corresponding RG-flow: at $\phi_B$  the running  passes through a maximum or a minimum and the flow inverts its direction. These non-monotonic solutions are a new type of holographic RG flows,  which we explore in detail in this paper.

\paragraph{Regularity.} Monotonic RG flow solutions in theories with an analytic scalar potential can be singular only when $\phi$ reaches infinity \cite{Bourdier13}. Here
 we extend this property to flows displaying a bounce: we show that at the bounce the geometry is  regular  (all curvature invariants are finite), and   that small perturbations around a bounce are  well-behaved (small non-homogeneities do not lead to large curvature corrections).

\begin{figure}[h]
\centering
\includegraphics[width=0.7\textwidth]{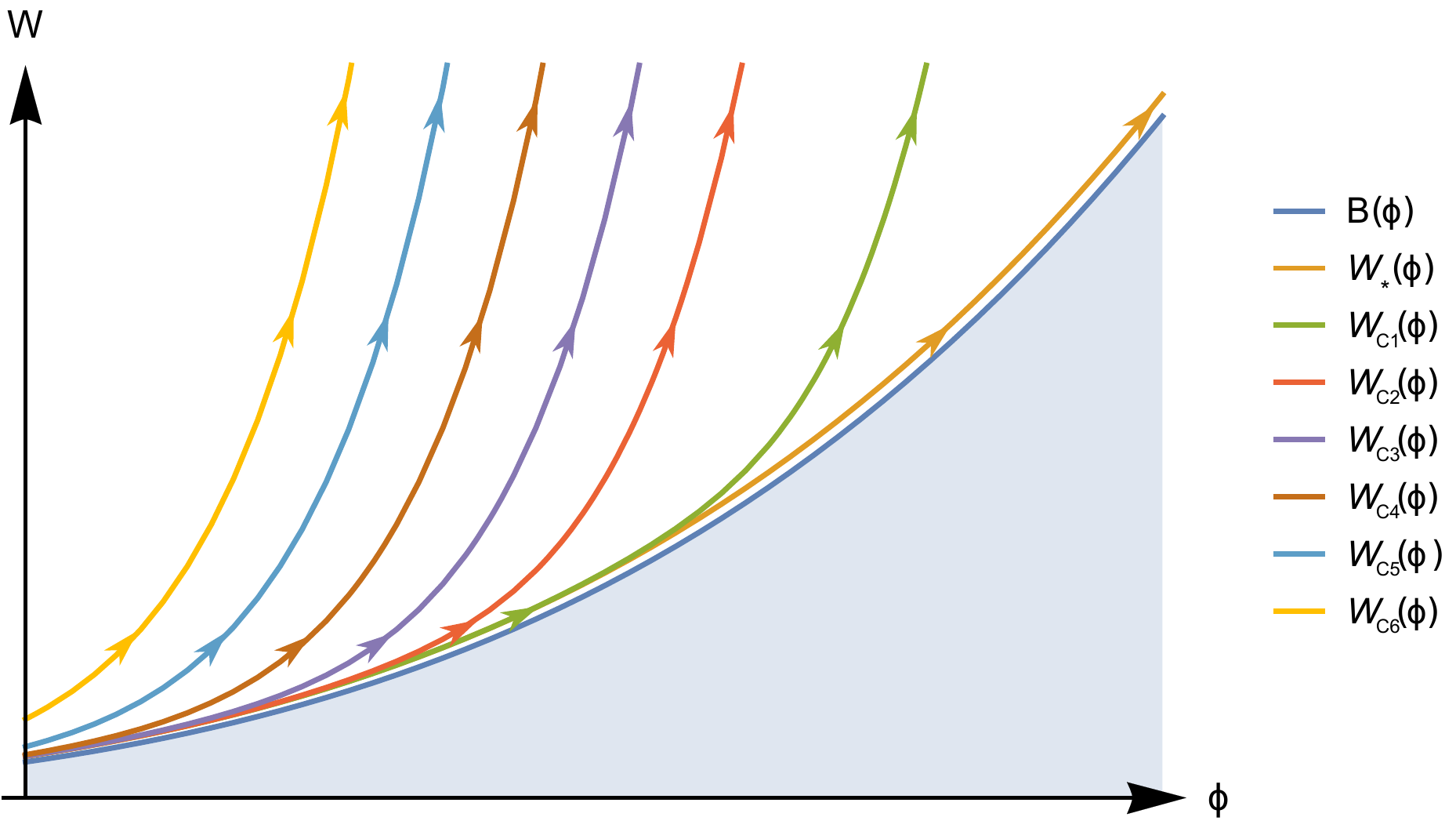}
\caption{Schematic view of the solutions of the superpotential equation that reach $\phi \to +\infty$. The curve $W_*$ represents the special solution. The curves labeled $C_1$ through $C_6$ are different curves belonging to the continuous family, and they grow exponentially faster than $W_*$.}
\label{fig:sing2}d
\end{figure}

\paragraph{Flows to infinity} (Figure \ref{fig:sing2}). Solutions for $W(\phi)$  can exist where the UV or the IR are reached for $\phi\to \pm \infty$. These cases have been less investigated in the literature, but they are relevant for example in phenomenological models of QCD \cite{IHQCDreview} and condensed matter, \cite{cgkkm}. For the branch with $W'>0$,   $\phi \to -\infty$ corresponds to a ``runaway'' UV fixed point, where the geometry is $AdS$-like if the potential asymptotes  to a finite (negative) constant.  In this case  there is still a one-parameter family of superpotentials that leaves the UV. This case includes the UV regime of the IHQCD model, \cite{gkn}.

On the other hand,  $\phi \to +\infty$ corresponds to the IR and typically the solution is singular in this region\footnote{For the  branch with $W'<0$ the results are the same, except that  the UV and IR are interchanged with  $\phi\to +\infty$ corresponding to the UV, and $\phi\to -\infty$ to the IR.}.

There are two kinds of singular solutions that reach infinity in the IR:
\begin{itemize}
\item A continuous family, parametrized by a constant $C$,  with exponential asymptotic behavior:
\be \label{intro4}
W_C \sim C \exp \left[{d\over 2 (d-1)} \phi \right] , \qquad \phi \to +\infty
\ee
This asymptotic behavior is independent of the potential $V(\phi)$ of the theory.

\item
A special isolated solution with asymptotics:
\be \label{intro5}
W_*(\phi) \sim \sqrt{-V(\phi)}, \qquad  \phi \to +\infty
\ee
but no arbitrary parameters.
\end{itemize}
Both kinds of  solutions (\ref{intro4}-\ref{intro5}) exist  only if the scalar potential grows slower than $\exp \left[\sqrt{2d\over (d-1)} \phi\right]$ as $\phi \to +\infty$.
If this is not the case, then all solutions bounce before reaching infinity.

\paragraph{IR Regularity.} Solutions such that $\phi \to +\infty$  in the IR are generically singular,  but this does not mean that they must necessarily be discarded: it has been argued that  certain types of singularities are acceptable in holography\footnote{This is because they can be resolved. Indeed, for example dropping higher dimensions and KK states leads to singular solutions in lower dimensions although one has started from a regular higher-dimensional solution.}. One   ``good-singularity'' criterion was formulated by  Gubser \cite{Gubcrit}, and states that one must be able to hide the singularity behind an infinitesimally small black-hole horizon. A stronger criterion may be called ``computability'' of the singularity:  the fluctuation problem around any solution must be well posed without need for IR boundary conditions other than normalizability\footnote{If this is not satisfied,  it does not mean that the solution is unacceptable, but rather that it is not predictive without embedding in a more complete framework where the singularity is resolved.}.

It turns out  that {\em only the special solution (\ref{intro5})} can satisfy either criteria. For Gubser's criterion, this had been shown in \cite{thermo}, whereas here we discuss the computability criterion in more generality.

In what follows we will adopt the  loose term ``IR-regular'' to indicate  both the strictly regular solutions and those with an acceptable singularity.

\paragraph{Vacuum Selection.} It is very important for the vacuum selection problem in holography that, for generic $V(\phi)$ with a finite number of extrema, {\em only a finite number of IR-regular solutions can exist:} these are  the solutions that reach  IR fixed points at the minima of the potential, plus eventually the special solutions that extend to infinity. Therefore, a holographic model corresponds to a field theory which has several (but a finite number of) ground states, each corresponding to one of the finite number of IR-regular flows. When two such flows leave the same UV, they are distinguished by a different VEV for the relevant operator driving the flow. Different VEVs correspond to different superpotentials, therefore to different holographic $\beta$-functions.

In this paper we show that there is a simple criterion that determines which, among the various regular RG flows, is the true vacuum of the theory: by computing the free energy at Euclidean signature, we show that the ground state is always the flow that has the largest VEV of the dual operator. For monotonic flows (i.e. flows which do not contain bounces) this is  the flow whose IR central charge is the smallest.
In particular, in case there exists a confining solution, such that in the IR $W\to+\infty$, this is the ground state  of the theory.

\paragraph{Stability.} Finally, we review the problem of small perturbations around an RG-flow solution in a generic Einstein-dilaton theory. This problem has been widely studied in the past (see e.g. \cite{gkn,csaki1,Kofman04,Kiritsis06}).  Using the same techniques, here  we show in complete generality that RG flow  solutions are  stable under small perturbations provided: 1) the UV satisfies the BF bound; 2)  the flow is regular in the IR or has a singularity that respects the computability bound\footnote{Strictly speaking, we show this is true for the theory in the standard quantization in the UV. However this can be generalized to different quantizations provided the contribution from the corresponding multi-trace deformation to the Hamiltonian is   bounded below.}.

\subsubsection{Exotic holographic RG flows}

\begin{figure}[h!]
\centering
\includegraphics[width=0.5\textwidth]{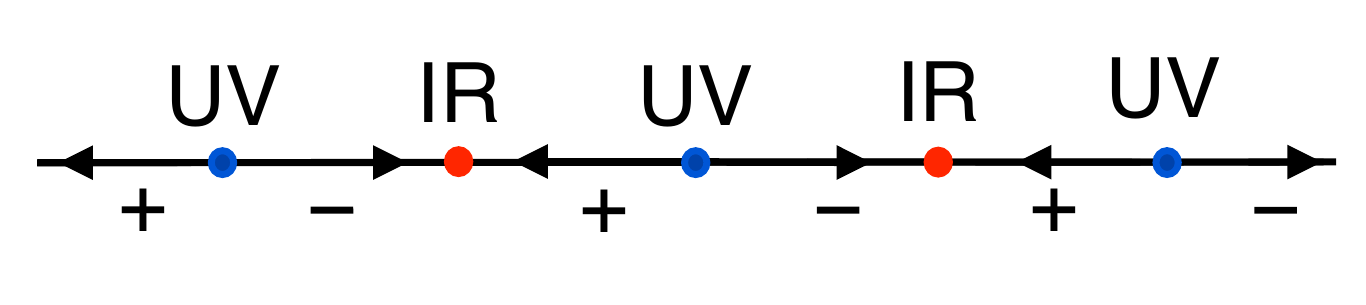}
\caption{Standard RG flow for a single coupling. The signs of the $\b$-function on the intervals between fixed points are indicated. The arrows correspond to the direction of the RG flow as the energy is lowered.}
\label{fig:RG_intro}
\end{figure}

The general picture outlined in the previous subsection is made concrete in several  examples, which we construct explicitly. In this paper we focus on IR-regular\footnote{in the generalized sense explained in the previous subsection}, {\em exotic} holographic RG-flows, i.e situations that naively, are not expected to occur  in standard (perturbative) field theory with one coupling, in which a flow starting at a UV fixed point reaches the nearest available IR fixed point (see figure \ref{fig:RG_intro}). Below we summarize these exotic solutions.

\begin{figure}[t]
\centering
\subfigure[]{
\includegraphics[width=0.4\textwidth]{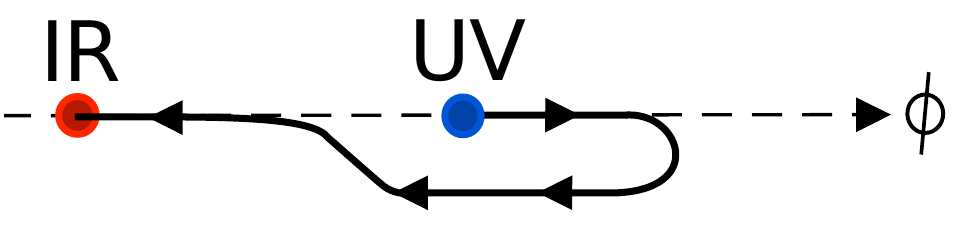}
}
\subfigure[]{
\includegraphics[width=0.5\textwidth]{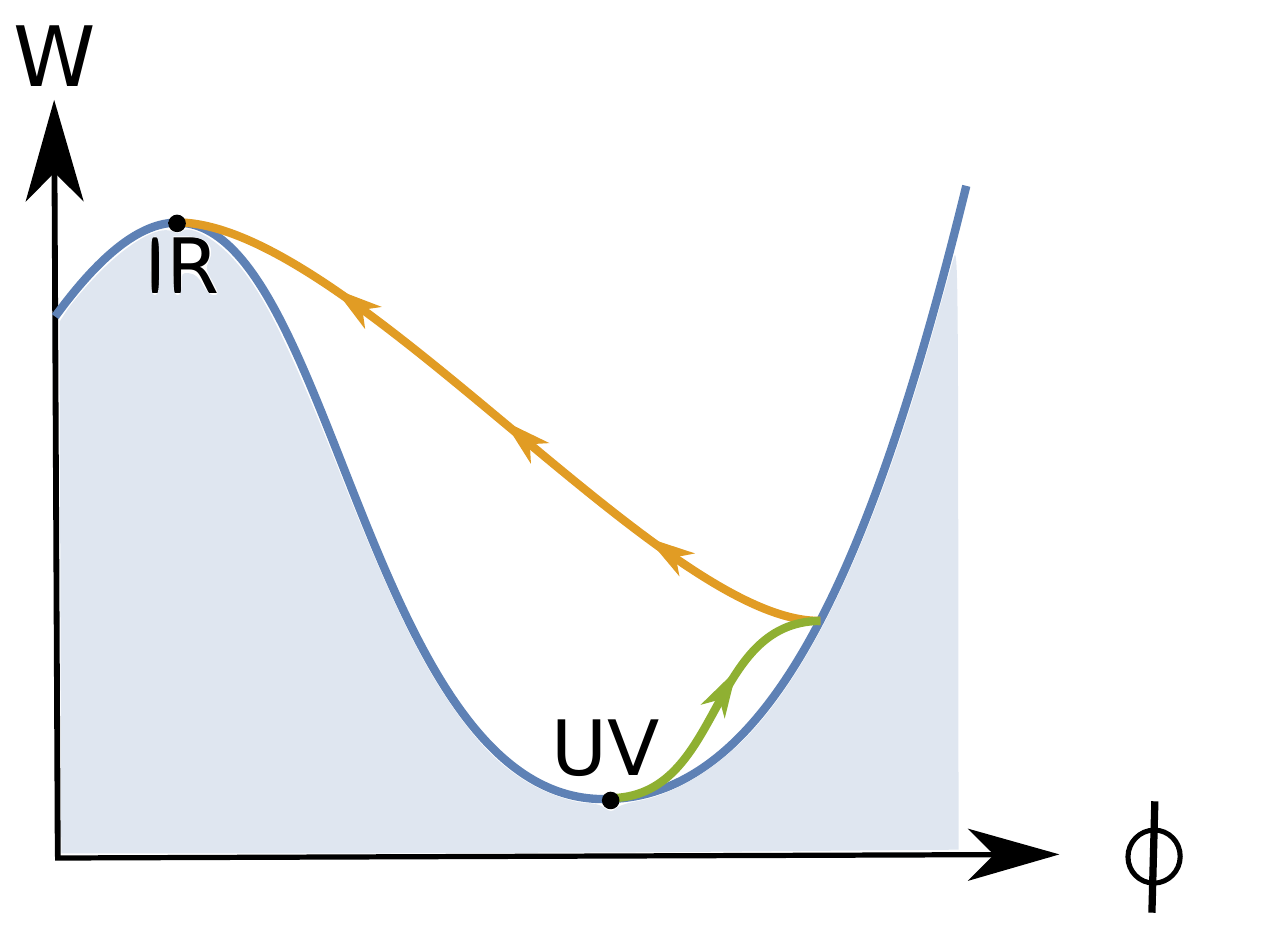}
}
\caption{Bouncing solution.  (a):  bouncing behavior along the $\phi$ axis for comparison with the standard RG flow of figure \protect\ref{fig:RG_intro}. (b):  a sketch of the corresponding superpotential with  two branches. The blue line is the critical curve $B(\phi)$ defined in \protect\eqref{defB}.  The arrows on (a) and (b) indicate the direction of increasing holographic coordinate $u$ or decreasing energy scale.}
\label{fig:sketch_intro}
\end{figure}

\paragraph{Bouncing RG flows} (Figure \ref{fig:sketch_intro}). These are RG flows that display one or more bounces before reaching the IR as shown in figure \ref{fig:sketch_intro}. We show that such a situation is rather easy to encounter in holography, but it has not been encountered so far in controllable QFTs. At the bounce, the  $\beta$-function vanishes. Normally this would imply that the flow stops at either infinite energy (UV fixed point) or zero energy (IR fixed point). On the other hand, in these solutions the bounce occurs at a finite scale factor (i.e. finite QFT  energy): the flow  continues beyond the bounce in the reversed direction,  as shown in figure \ref{fig:sketch_intro} (a). The corresponding superpotential has two or more branches, as sketched in figure \ref{fig:sketch_intro} (b), but the full solution is a regular geometry interpolating between two fixed points in the UV and IR.

The bouncing behavior could actually occur in QFTs where the $\beta$-function vanishes in the appropriate way, for example if there exist  a coupling value $g_B$ close to which:
\be  \label{intro8}
\beta_+ \simeq \sqrt{g_B-g}, \qquad g < g_B
\ee
In this case the coupling reaches  $g_B$ at finite energy, as one can see by integrating  the RG equation.
Note that the behavior in (\ref{intro8}) is necessarily non-perturbative in $g_B-g$. It is therefore not visible in perturbation theory.

 One can continue the flow past this point if furthermore one has a modified version of the theory where the $\beta$-function has the equal and opposite behavior, namely
\be \label{intro9}
\beta_- \simeq  -  \sqrt{g_B-g}, \qquad g < g_B
\ee
In this case the flow has two branches, and reaches $g_B$ as a local maximum. In fact, the behavior in equations (\ref{intro8}-\ref{intro9}) corresponds exactly to the holographic $\beta$-function obtained by the bouncing superpotential via the relation (\ref{intro2-c}). Such  a behavior has been suggested to occur in  the context of field theory, in association with {\em limit cycles} of the RG group, in which the coupling takes on a periodic behavior: near the turning points, the $\beta$-function looks locally like the one in equation (\ref{intro8}), see e.g.  \cite{cycles}. However, cycles seem so far not to occur in  full,  consistent field theories: the examples of limit cycles that we are aware of  are either non-unitary \cite{doll} or they are toy models based on truncations of full field theories to a subspace of the full Fock space  \cite{Wilson}.

 In other cases, RG cycles can be removed  by a  redefinition of the coupling \cite{RemoveC}.
Notice that, compared with the case discussed in \cite{cycles},  in our examples {\it the flow  is not cyclic}, and the beta-function has the form (\ref{intro8}-\ref{intro9}) only locally,  close to the turning points. We show in appendix \ref{app:cycles} that cycles can only arise if the potential $V(\phi)$ is multi-valued in a very special way. This is not however the case in this paper, where the bulk potential is assumed to be regular and single valued. In a sense, solutions that bounce a few times are intermediate between standard RG flows without bounces and limit cycles with an infinite number of bounces.

We should emphasize though that this behavior is easy to produce if we project the multitrace RG onto single trace RG along the lines of \cite{sslee}. A simple example of this projection is worked out in appendix \ref{single} where we show how non-monotonic RG flows can be generated.

\paragraph{Cascading solutions.} As is well known, extrema of $V(\phi)$ which violate the Breitenlohner-Freedman (BF) bound \cite{BF82},
\be\label{intro10}
{V'' \over V} \geqslant - {d^2\over 4d(d-1)}
\ee
do not correspond to unitary CFTs, since the corresponding operator dimension becomes complex. On the gravity side, this is signaled by the fact that fluctuations around an extremum of $V$  violating the bound (\ref{intro10}) display tachyonic instabilities (i.e. exponentially growing modes in time), as we review in appendix \ref{BF}.

In this paper we show that there are instead solutions that display an infinite ``cascade'' emanating from a BF-violating  extremum in the UV,  via an infinite number of bounces.  These are part  of the space of solutions to the bulk equations and they are  reminiscent of the  Klebanov-Strassler cascading ${\cal N}=1$ solutions in type IIB string theory \cite{KS}.

However, we also show that such solutions are not acceptable because, as one is approaching the top of the cascade, the fluctuations display the same kind of instability as the solutions sitting at the BF-violating extremum. Therefore, they are a curiosity, but have to be excluded from the consistent solutions of unitary holographic theories.

\begin{figure}[t]
\centering
\subfigure[]{
\includegraphics[width=0.35\textwidth]{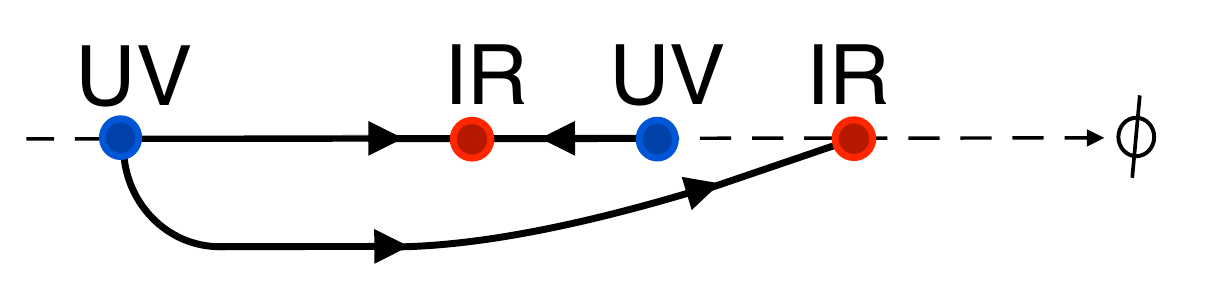}}
\subfigure[]{\includegraphics[width=0.62\textwidth]{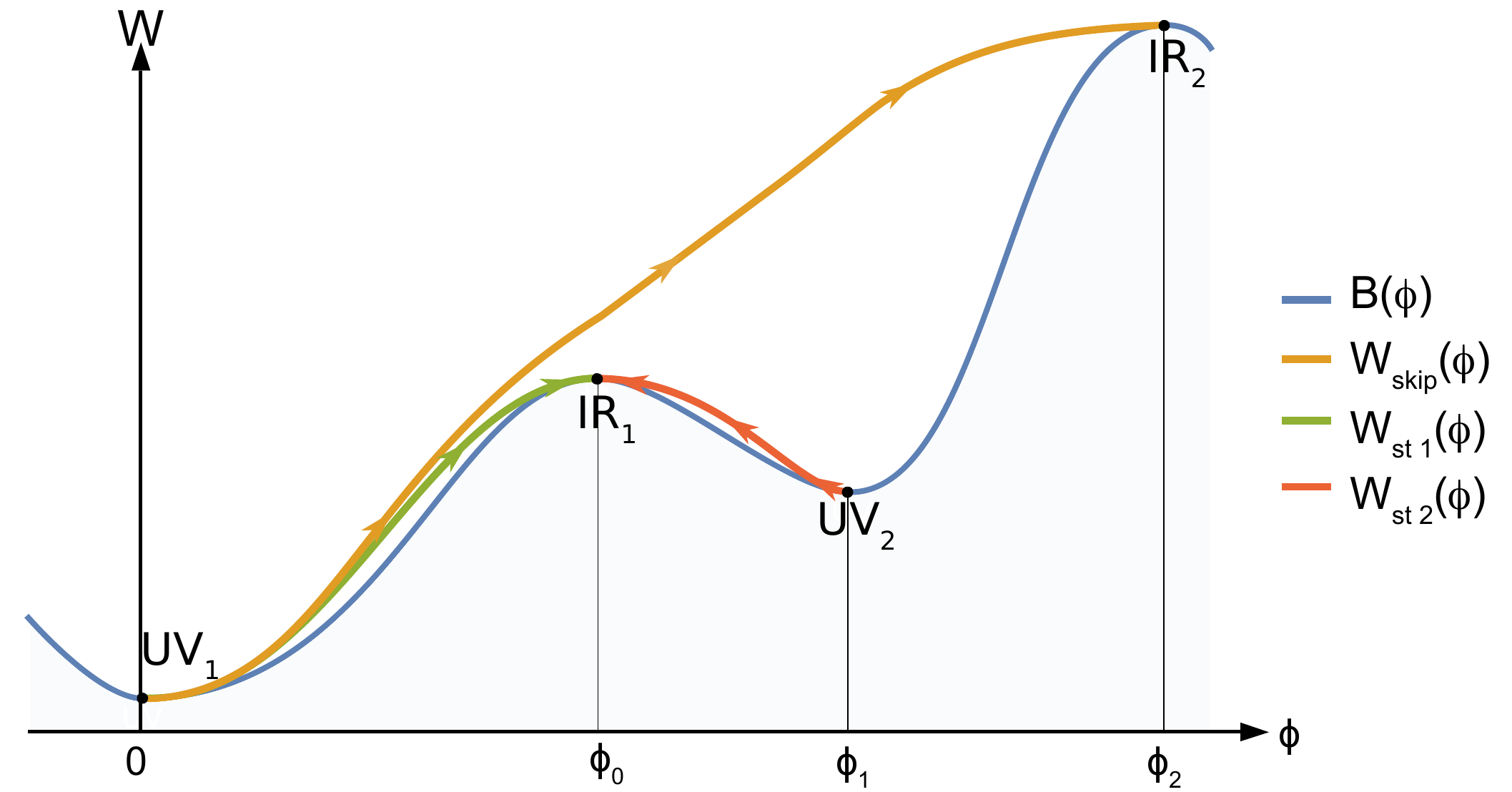}
}
\caption{A sketch of a holographic RG flow skipping fixed points. (a) Two different flows leave the UV fixed point on the left,  corresponding to the same deformation but different operator VEVs.  (b)  The solution  $W_{skip}$ corresponding to the yellow curve skips the intermediate IR fixed point and reaches a different IR fixed point farther away. The solutions $W_{st 1}$ and  $W_{st 2}$ correspond instead to standard holographic RG flows, as they connect neighboring fixed points.
}
\label{fig:SolSkip}
\end{figure}

\paragraph{Skipping RG-flows} (Figure \ref{fig:SolSkip}).  We provide examples of holographic RG flows which start at a UV fixed point, at a local maximum of $V(\phi)$, but do not stop at the nearest available IR fixed point: rather, they go through (without stopping) an even number of extrema of $V(\phi)$ which separate the starting point from the final IR endpoint.

This is unusual in  field theory, where  flows stop at the first IR point available, i.e. the one which is closest (in coupling space) to the starting UV fixed point, as in figure \ref{fig:RG_intro}. Here instead there may be  two IR-regular solutions of the superpotential equation (\ref{intro2}) which start at the same UV and reach two different extrema of $V(\phi)$ in the IR. These two solutions  differ by the VEV of the operator dual to $\phi$. In this sense these flows are non-perturbatively related: they describe two vacua/saddle points  of the field theory with the same UV (and same relevant deformation) but different VEVs and different IR fixed points.

In such cases we may have two distinct flows leaving a UV fixed point and reaching two different IR fixed points. These two flows differ only in their VEV's. In such a case one can compare their free energies to see which one is dominant.
We find that, in the absence of bounces,  the flow that ends furthest in scalar field space, i.e.  that has the lowest $a$-value in the IR,  is the dominant one.
The other flow is non-perturbatively unstable (by bubble nucleation).

In the single-field  case we examine in this paper this is more or less the full story. However, when field space is  multidimensional,  there is the possibility of quantum phase transitions if parameters in the potential change due to other scalar field VEVs.

\begin{figure}[t]
\centering
\includegraphics[width=0.5\textwidth]{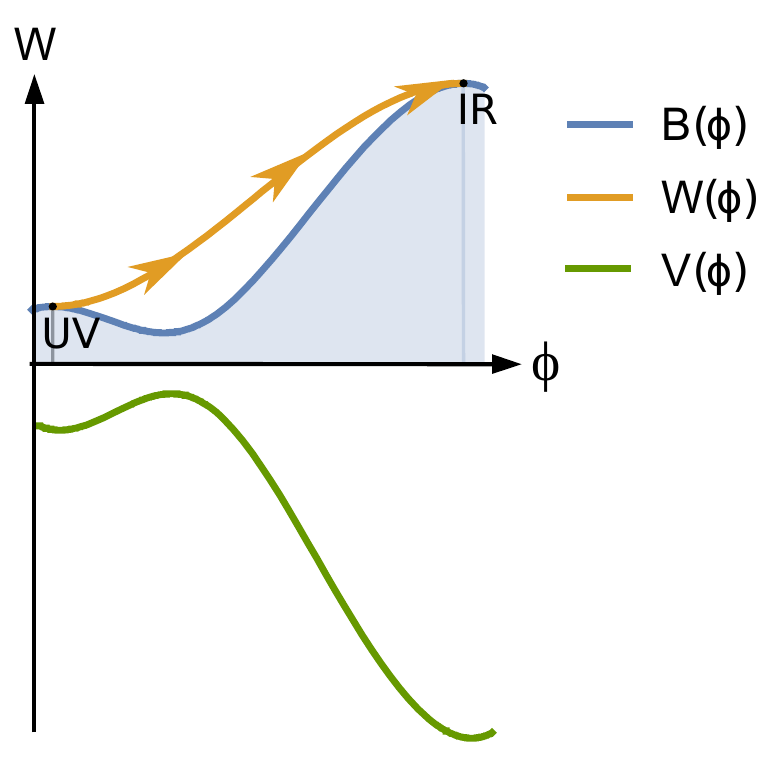}
\caption{
Sketch of a flow connecting two minima of the potential.}
\label{fig:W1_intro}
\end{figure}

\paragraph{Flows between minima} (Figure \ref{fig:W1_intro}).  Usually, with our conventions (in which $V <0$), an UV fixed point corresponds to a local maximum of $V(\phi)$, and an IR fixed point corresponds to a local minimum. In this generic situation, the flow is driven by a relevant operator (corresponding to a bulk scalar with negative mass squared) away from the UV. However this is not the only possibility: one can also have a flow completely driven by the VEV of an {\em irrelevant} operator. In the holographic  dual this corresponds to flowing out of a {\em minimum} of the scalar potential to another minimum, as depicted in figure \ref{fig:W1_intro}.
These solutions do not exist in generic theories, but only if the potential is appropriately chosen. Here, we present one example \cite{Sibiryakov} and we show how one can construct bulk potentials which allow for these solutions.

\subsection{Discussion and Outlook}

Our findings pose many more questions that are interesting to consider further.

\begin{itemize}

\item The structure of Lorentz invariant saddle points is rather simple and no fancy techniques are needed, like those of \cite{net}, and no non-trivial constraints are envisaged as those discussed in \cite{sha}.

\item Our analysis was one-dimensional. It is clear that multi-dimensional analysis is necessary. The whole first order formalism giving rise to the holographic RG group is still operative, \cite{rg1} but now novel phenomena can happen, due to the fact that the generic extremum of the potential is a saddle. For example, one can have the interplay between extrema violating the BF bound in one direction and flows in the other directions, as seen in two scalar examples in \cite{vqcd}. Such an analysis in  currently in progress.

\item The exotic holographic examples of RG flows we found correspond to generic scalar potentials. It is interesting to identify whether such potentials can arise in effective theories of string theory, in particular maximal supergravities. Even if this is the case, of course, there is still the possibility that such behavior is an artifact of the large-N limit, and/or the strong coupling limit.

\item The example of skipping flows indicates that a single UV fixed point and a single perturbing operator may give multiple distinct flows with distinct IR fixed points. Such flows are different saddle points of the same QFT. Interestingly, the holographic dynamics indicates that the dominant saddle-point is the one whose IR CFT has the lowest $a$-anomaly. It would be interesting to understand this phenomenon from the QFT point of view.

\item A related issue concerns the fact that the superpotential is on the one hand an $a$-function for the holographic flows and second it satisfies an inequality, which  constrains it to be bounded by the square root of the bulk potential. This inequality seems intriguing from the QFT point and it would be interesting to understand its meaning better.

\item We have seen that flows with a finite number of bounces are generically allowed. Such flows share some similarities with limit cycles in QFT. In particular the coupling is not monotonic. On the other hand, they do not generate full cycles, which would involve an infinite number of bounces. The only flow that seems to have an infinite number of bounces (but does not correspond to a limit cycle) is a flow away from a BF-violating maximum of the potential. However as we know the UV CFT in that case violates unitarity and this is seen in the inherent instabilities of such flows.

    Demanding that a limit cycle is reproduced holographically requires that the bulk potential is infinitely multivalued. Multivalued potentials have been recently argued to exist in effective string actions, and they also exist in large N theories gauge theories like QCD(N). However at finite N, the multi-valuedness is at best finite-fold. These observations can produce an intimate link between the possibility of limit cycles in CFT and multivalued potentials in string theory that are interesting for cosmological applications.

\item A set of related questions associated with bouncing flows concerns their consistency as dual QFTs once other physical properties are examined. For example,  constructing  black-hole solutions associated with such flows is such a testing ground, and should be studied. Another potential observable may be the mass spectra associated with operators in these theories, especially the ones dual to the driving operators.

\item  Holography associates CFTs to the extrema of the scalar potential and defines local flows around them. This defines a graph with nodes and links, and it is the natural realm of application of Morse theory (which has been related to supersymmetric QM), \cite{wm}.  It would be interesting to investigate how Morse theory intertwines with the concept of possible holographic RG flows as determined by the bulk Einstein equations.

\item A related question involves Gukov's recent proposal to use bifurcation theory to study QFT RG flows, \cite{gukov}. Here of course we are in the holographic (large-N, strong coupling limit) but the interplay between holographic flows and bifurcation theory is intriguing.

\item In this paper we have been working exclusively in regions of field space where the potential is negative. Generic string theory potentials however have regimes where they are also positive. In this case there is a possibility for asymptotically de Sitter or asymptotically flat solutions. This will make the analysis of potential solutions richer, and the related phenomena even more interesting as they include real-time cosmology and its interplay with anti de Sitter regions. This might provide a larger arena where one can implement the de Sitter-anti de Sitter correspondence and the view on cosmological solutions along the lines developed of \cite{afim}.

\item The approach described here may provide a novel reorganization of the string landscape by giving it a new geometry based on regular flows rather than coordinates in field space. This novel organization may be physically more transparent when it comes to understand cosmological transitions in the string landscape.

\end{itemize}

This paper is organized as follows.

In Section 2 we set up the model, review the superpotential formalism and analyze solutions of the superpotential equation, specifically around critical points. We discuss the geometry close to  ``bounces'' where the superpotential becomes multi-valued and the holographic RG flow non-monotonic.

In Section 3 we present concrete  examples of exotic RG flows: we discuss solutions connecting minima of the bulk potential; multi-branched bouncing solutions ; infinitely cascading solutions emanating from a UV fixed point which violates the BF bound, and their instability; solutions interpolating between non-neighboring fixed points.

In Section 4 we extend the analysis to singular solutions reaching infinity in field space, and we discuss the criteria which make the singularity acceptable holographically. We also present concrete examples of solutions of this kind, which moreover present one or more of the exotic features described in Section 3.

In Section 5 we present the computation of the free energy associated with each flow and discuss criteria that select the  ground state  of the theory.

In Section 6 we discuss linear perturbations around holographic RG flow solutions and the conditions for stability.

Several technical details are left to the Appendices.

\section{Holographic RG Flows in Einstein-dilaton gravity}

We consider a gauge-gravity duality setup in which we will follow the dynamics of the bulk metric and a single scalar field, associated in the boundary QFT with the stress tensor and the coupling of a single-trace scalar operator, respectively.

The bulk theory  can be described by a two-derivative action that includes an Einstein-Hilbert term plus a minimally coupled scalar field.

In this section we review how  Einstein's equations can be transformed in a set of first order equations, in terms of a  {\em superpotential}, which allows one to make contact with the Holographic Renormalization Group, \cite{dVV,Bourdier13,rg1,rg2}. Next, we study the properties of the superpotential equation and its solutions.  These results will be used in later sections to construct several types of holographic RG flows.

\subsection{The setup}
\label{ssec:setup}

In our conventions,  the Einstein-scalar theory in $d+1$ dimensions, with signature $(-, +\ldots +)$,  has action:
\eql{S_E-S}{
			S\le[g,\f\ri]= \int d^{d+1}x \Vg \le(
			 R
			-\ha\p_a\phi \p^a\phi-V(\phi)  \ri)+S_{GHY}.
			}
where $S_{GHY}$ is the Gibbons-Hawking-York term.
We are interested in boundary field theories living in Minkowski space-time and we consider the most general solutions respecting d-dimensional Poincar\'e invariance. These solutions can always be put in the so-called domain-wall coordinate system:
\eql{Poinc}{
		 \phi=\phi(u),\qquad ds^2=du^2+e^{2A(u)}\eta_{\mn}dx^\m dx^\n
		}
where $u$ is  the holographic coordinate.

Our conventions are such that derivatives with respect to the holographic coordinate will be indicated by a dot and derivatives with respect to the scalar field with a prime:
\be
f=f(u),~~{df\over du} \equiv \dot{f}(u) ~~~~and~~~~g=g(\phi),~~{dg\over d\phi} \equiv g'(\phi)
\ee

Einstein's equations for field configurations of the form \eqref{Poinc} are:
\begin{subequations}\label{EE}
\begin{align}
	&2(d-1)\ddot{A}(u)+\dot{\phi}^2(u)=0,\label{EE1}\\
	&d(d-1)\dot{A}(u)^2-\ha\dot{\phi}^2(u)+V(\phi)=0.\label{EE2}
\end{align}
\end{subequations}
Equation \eqref{EE1} implies that $\dot A(u)$ does not increase. In  the holographic RG, this is  holographic c-theorem \cite{GPPZ,Freedman}.

The Klein-Gordon  equation is:
\eql{KG}{\ddot{\phi}+d\dot A\dot{\phi}-V'(\phi)=0,}
which can be deduced from \eqref{EE}.

We take the  QFT energy scale to be given by
\eql{mu}{
\m\equiv\m_0 e^{A(u)},
}
with some arbitrary $\m_0$. This choice gives the correct trace identities, once we identify $\phi(\mu)$ with the running coupling \cite{rg2}.

The fixed points of the RG flow correspond to asymptotically $AdS$ geometries,  For this reason we restrict ourselves to potentials such that all extrema have associated negative cosmological constants.

To make things simpler,  in this paper we will only consider strictly negative potentials:
\eql{negV}{V(\phi)<0\quad \forall~ \phi\in\{\mathbb{R}\cup(\pm\infty)\}.}
This simplification avoids the need to include time-dependent or asymptotically flat space-times in the landscape of solutions. Moreover, we assume for simplicity that $V(\phi)$ is analytic around any finite value of $\phi$.

Finally, we choose the radial coordinate such that $u$ increases along the flow, i.e. we take $A(u)$ to be monotonically decreasing. This choice is consistent throughout the solution, because  $\dot{A}$ cannot change sign: this follows from the fact that  $\ddot{A} \leq 0$ due to equation (\ref{EE1}).


\subsection{The Superpotential}
\label{ss:superp}

Here we present the formalism that makes contact between renormalization group flows in $d$ space-time dimensions and the Einstein equations in $d+1$ dimensions for geometries with $d$-dimensional Poincar\'e-invariance, \cite{ceresole,dVV,Bourdier13,rg1,rg2}. The goal is to convert the second order Einstein equation \eqref{EE1} into two first order differential equations and write them as a  gradient RG flow (as we will see, this is locally always possible except at special points corresponding to bounces).

In the standard quantization, the boundary conditions on the scalar field $\phi$ are given by the source $\phi_0(x)$ for an operator $\cO(x)$ at the boundary CFT \cite{Witten98,KlebWit,Witten02}. The generating functional of correlation functions of $\cO$ is of the form:
\eql{Schw}{
Z\le[\phi_0\ri]=\le<e^{\int d^d x~ \phi_0(x)\cO(x)}\ri>.
}
The boundary CFT is the UV theory. When the operator $\cal O$ is relevant it generates a RG flow away from the UV fixed point and the running coupling to this operator, $\phi(\mu)$, will flow according to the RG  equation.

In Quantum Field Theory, a RG flow for a single coupling $g$ is given by a first order differential equation:
\eql{QRG}{
		\m{d\over d\m}g(\m)=\b(g)
}
One very important feature of this type of equation is the possibility for fixed points. When $\b$ vanishes at a point $g_*$, a flow $g(\m)$ starting at $g_*$ will stay at $g_*$. If the $\b$ function has a regular series expansion around $g_*$, any flow $g(\m)$ that solves \eqref{QRG} and passes through $g_*$ will stop at this value of the coupling\footnote{On the other hand, if $\b(g)$ does not have a regular series expansion around a point $g_*$ where $\b(g_*)$ vanishes, a flow $g(\m)$ solving \eqref{QRG} does not necessarily stop at $g_*$. An example of a flow that continues after reaching a zero of the $\b$ function is shown in \cite{cycles}. We will consider this flow in more details in subsection \ref{sssec:bW} and appendix \ref{app:cycles}. }. In conclusion, {\em every flow $g(\mu)$ solving equation \eqref{QRG}  with a $\b$ function depending only on $g$ and admitting a regular series expansion around all its zeroes will be a flow that interpolates between nearest neighboring fixed points.}

In the present setting, we consider the case of a  single scalar coupling dual to the bulk scalar field.
Already in this simple setting, the situation in the gravitational dual appears different from the QFT side, since equations (\ref{EE1}-\ref{EE2}) are higher than first order.

To make contact with the QFT, we will rewrite our equations as a set of first order equations, following a well known procedure first introduced in the holographic setting by \cite{Freedman}. We define the function $W(\phi)$ such that\footnote{Such a function in supergravity theories is known as the superpotential, and the first order equations define the BPS flows. We will keep the same name in our non-supersymmetric case. Note that this has been also called the "fake" superpotential in \cite{townsend}.}:
\eql{defW}{W(\phi(u))=-2(d-1)\dot A(u) ,}
where the proportionality constant is chosen for future convenience.
Then, equation \eqref{EE1} is  automatically satisfied if $\phi$ obeys:
\eql{phiW}{W'(\phi)=\dot \phi (u).}
Replacing \eqref{defW} and \eqref{phiW} into \eqref{EE2} we obtain an equation for $W(\phi)$:
\eql{SuperP}{V(\phi)=\ha W'^2(\phi)-\frac{d}{4(d-1)}W^2(\phi)}
Even though there is no supersymmetry here, we will call $W(\phi)$ the superpotential and \eqref{SuperP} the superpotential equation.

One important consequence of \eqref{defW} and \eqref{phiW} is that we can automatically write the $\b$-function solely in terms of the coupling $\phi$:
\eql{betaW}{{d\phi\over d\log \m}=\b(\phi)=-2(d-1){W'(\phi)\over W(\phi)}.}

Once  a solution $W(\phi)$ of equation \eqref{SuperP} is found,  it is straightforward to solve for $\phi(u)$ and $A(u)$ using \eqref{defW} and \eqref{phiW}. {\it Our strategy from now on is, for a given $V(\phi)$, to solve the superpotential equation \eqref{SuperP} and find $W(\phi)$.}


\subsection{General properties of the Superpotential}
\label{ss:gen_prop}

In the following we list important properties of solutions $W(\phi)$ of equation \eqref{SuperP} for potentials $V(\phi)$ which are analytic\footnote{
By analytic we mean that around any finite $\phi$ the potential is given by a convergent power series.
}
over the entire real line and strictly negative.

We transformed one second order differential equation and one first order equation (\ref{EE1}-\ref{EE2}),   into three first order equations, namely (\ref{defW} - \ref{SuperP}), each one demanding one integration  constant. The first  consequence is that {\it for a given $V(\phi)$ there is a one parameter family of $W(\phi)$ that solve \eqref{SuperP}, parametrized by an integration constant}. As we only consider analytic and continuous potentials, the superpotential is real and continuous.

These solutions have the following general properties:

\begin{enumerate}

\item {\it W(u)$\equiv$W$(\phi(u))$ is monotonically increasing along the flow}
\label{it:Wu}

From equation \eqref{phiW},
\eql{Wu}{
	{dW(u)\over du}={d\phi(u)\over du}{d W(\phi)\over d\phi}=W'^2\geqslant0
}
Because the scale factor is  decreasing with $u$, as discussed at the end of subsection \ref{ssec:setup},   equation \eqref{Wu}  implies that $W(A)\equiv W(u(A))$ is monotonically decreasing, with $u(A)$ the inverse of $A(u)$. This is the holographic C-theorem, \cite{GPPZ,Freedman} and $1/W^{d-1}$ is, up to a constant, the  C-function for the flow.


\item {\it The $(W,\phi)$ plane has a forbidden region where no solution of \eqref{SuperP} exists.  This is the blue-shaded  area in the figures of this and the previous section, and it is the region bounded  by the critical curve:}
\label{it:bound}
\be
B(\phi) \equiv \sqrt{-\frac{4(d-1)}{d}V(\phi)}\label{B}
\ee
This statement  follows from the superpotential equation \eqref{SuperP}, which implies:
\eql{BoundW}{|W(\phi)|= \sqrt{\frac{4(d-1)}{d}\le(\ha W'^2-V(\phi)\ri)}\geqslant\sqrt{-\frac{4(d-1)}{d}V(\phi)}.}
Since we are assuming the  potential to be strictly negative,  $B(\phi)$ bounds $W(\phi)$ on the whole real axis.

As a consequence, $W(\phi)$ cannot change sign. Notice that, from the definition (\ref{BoundW}),   local minima of $V(\phi)$ are local maxima of $B(\phi)$ and vice-versa.

It is interesting to mention that the inequality (\ref{BoundW}) can be interpreted as a bound on the C-function associated with the bulk potential.

\item {\it We can assume $W(\phi)>0$ without loss of generality}.

Indeed, the superpotential equation \eqref{SuperP} is invariant under $W\to -W$. Also, the transformation:
\eql{inv}{
(u,W)\to-(u,W)
}
is a symmetry of the full set of equations (\ref{defW} - \ref{SuperP}).
Therefore, we can choose to work with one definite sign for $W$, which we take from now on to be positive:
\eql{Wpos}{W(\phi)>0}

This definition allows us to identify $W(\phi)$ with the effective potential of the dual field theory \cite{rg1,papa1}. Due to the symmetry (\ref{inv}),  $W<0$ can be interpreted as describing the same solution as $W>0$ but with $u$ increasing  towards higher energies (so we still have $dW/du >0$).

\item {\it For a generic point in the allowed region of the $(\phi,W)$ plane with $W> B$,
    there exist two and only two solutions, $W_{_{\uparrow}}(\phi)$ and $W_{_{\downarrow}}(\phi)$, passing through that point} (Figure \ref{fig:Crossing}).
\label{it:updown}

Indeed, equation \eqref{SuperP}  can be separated in two equations according  the sign of $W'$:
\begin{subequations}\label{W'B}
\begin{align}
W'_{\uparrow}(\phi)=+\sqrt{{d\over2(d-1)}\big(W_{\uparrow}^2(\phi)-B^2(\phi)\big)}\label{W'Ba}\\
W'_{\downarrow}(\phi)=-\sqrt{{d\over2(d-1)}\big(W_{\downarrow}^2(\phi)-B^2(\phi)\big)}\label{W'Bb}
\end{align}
\end{subequations}

Each of these two equations  has a unique solution  in a  neighborhood of a any point $\phi$ as long as the argument of the square root is non-vanishing. This follows from the existence  theorem for  differential equations of the form $y'(x) = F(x,y)$, which guarantees that a solution exist, and it is unique, in a neighborhood of any point where $F(x,y)$ is continuous and differentiable \cite{Jordan}.

Therefore, around a generic point $(\phi_*,W_*)$ with $W(\phi_*)>B(\phi_*)$ there will be a unique solution for each of the equations \eqref{W'B}. Since they have opposite derivative  the two  solutions cross at $\phi_*$ as shown in figure \ref{fig:Crossing}.
What happens on the critical curve $W(\phi)=B(\phi)$, where the above mentioned theorem fails,  will be discussed in detail in  subsection \ref{ssec:crit}.

\item{\it The geometry is regular if the potential and the superpotential are finite along the flow. This is equivalent to the scalar field staying finite along the flow.}
\label{it:reg}

The first statement follows from the fact that a generic curvature invariant can be written in terms of a polynomial in $W$ and $V$ and their derivatives \cite{Bourdier13}. 
For example,  the square of the Ricci tensor is given by:
\eql{Ricci2}{
R_\mn R^\mn=
{d\over 16(d-1)^2}
	\le[
		16(d+1)V^2
		+8dVW^2
		+dW^4
	\ri]
}
 The potential $V(\phi)$ is analytic on the real line and therefore finite for any finite $\phi$. As follows from equations \eqref{W'B}, no solution $W(\phi)$   can diverge at any finite  $\phi$, and it is analytic at any point in the interior of the allowed region $W>B$. As we will see in section \ref{ssec:crit},  regularity extends also  to points  on the critical line:  although $W$ may be non-analytic there, the geometry is regular.  Therefore, {\em regular solutions  are those where $\phi(u)$ is finite along the entire flow.}


\item {\it Extremal points of $W(\phi)$ lie on the critical line $B(\phi)$, } since there $W'=0$ by equation (\ref{W'B}).

As we will see, there are two kinds of critical points:
\begin{enumerate}
\item {\em fixed points}, i.e. points on the critical line where $B'=0$. These are the  extrema of $V(\phi)$, and correspond to the usual UV and IR  $AdS$ fixed point.
\item  {\em bounces}, i.e. generic points on the critical line where $V'\neq 0$.  These correspond to  points in the interior of the bulk solution, where the superpotential becomes multi-branched but the geometry is regular. They correspond to points where $\phi(u)$ reaches a maximum or a minimum and the  flow reverses its direction.
\end{enumerate}
Critical points will be the subject of the next section.

\end{enumerate}

\subsection{Critical points}
\label{ssec:crit}

Critical points are those where $W'(\phi)=0$. From equation \eqref{phiW},  they  correspond to extrema of the coupling $\phi(u)$. Therefore, if  $\phi_*=\phi(u_*)$ is a critical point, then:
\eql{beta}{
		\beta(\phi_*)=
		\le.{d\phi \over d\log \m}\ri|_{u=u_*}=
		\le.{d\phi \over d A}\ri|_{u=u_*}=0.
		}
If $\phi$ were a coupling in QFT, then equation \eqref{beta} would be a first order differential equation for the renormalization group flow of $\phi$. In QFT, therefore, the critical points would be the fixed points of the renormalization group flow. Here, however, equation \eqref{beta} has a different meaning:
because the gravitational equations are higher order, in general  a holographic RG flow does not necessarily  stop when $\dot\phi=0$.

The vanishing or not of the derivative of $V$  is what characterizes the behavior of $W$ at a critical point. Consider the derivative of equation \eqref{SuperP} with respect to $\phi$,
\eql{V'}{
V'(\phi)=W'(\phi)\le(W''(\phi)-\frac{d}{2(d-1)}W(\phi)\ri),
}
and suppose $W'(\phi_*)=0$. As $W$ is  finite for finite $\phi$,  there are only two possibilities:
\begin{enumerate}
\item $V'(\phi_*)\neq0$ with  $W''(\phi_*)$  divergent. \label{Wdiv}

{\bf or}

\item $V'(\phi_*)=0$; in this case  $W''(\phi_*)$ is finite.
\label{Wfin}
\end{enumerate}

The second statement is not entirely obvious,  and we prove it in appendix \ref{prel}.

Case \ref{Wdiv} is the generic case and, as seen in the previous subsection, item 6, it corresponds to bounces, i.e. points where two branches of $W(\phi)$ (one growing, one decreasing) are glued together. This generic situation will be treated in \ref{sssec:bW}.

Case 2 corresponds to  fixed points of the holographic RG flow. Below we will discuss separately maxima, minima and inflection points of $V$.

\subsubsection{Local maxima of the potential}
\label{sssec:max}

\begin{figure}[t]
\centering
\includegraphics[width=0.6\textwidth]{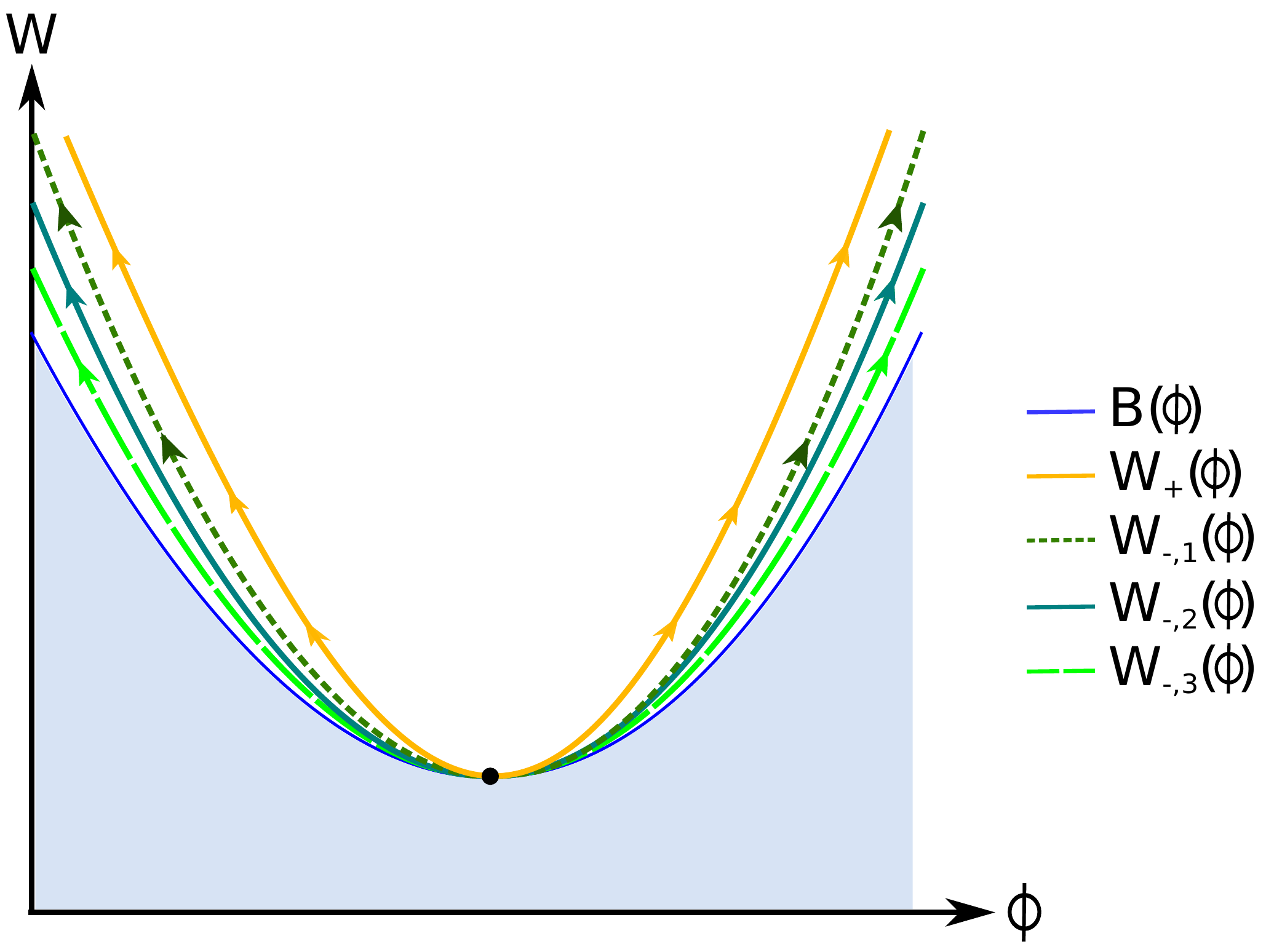}
\caption{Sketch of superpotential solutions  with critical point at a local maximum of the potential (local minimum of $B(\phi)$). There are two kinds of solutions, labelled $W_+(\phi)$ and $W_-(\phi)$, given in equations \protect\eqref{W_max} and \protect\eqref{W_max_BF}. The solution of the $W_+$ kind is unique, while there are infinitely many $W_-(\phi)$ solutions, parametrized by an integration constant of equation \protect\eqref{SuperP}.
  The arrows on $W_-(\phi)$ and  $W_+(\phi)$ indicate the direction of increasing holographic coordinate $u$, i.e. the direction of the RG flow.
}
\label{fig:max}
\end{figure}

Generically,  near a maximum  of the potential (that we place at  $\phi=0$ by an appropriate shift of $\phi$), the potential is generically expanded as:
\be
	V=-{d(d-1)\over \ell^2}+{m^2\over 2}\phi^2 + \cO(\phi^3), \qquad m^2 <0 \label{Vext}
\ee
As explained in Appendix \ref{crit},  equation \eqref{SuperP} has the following solutions \cite{papa1}:
\begin{subequations}			\label{W_max}
\begin{align}
	W_+(\phi)
			=&{1\over \ell}
			\le[
				{2(d-1)} + {\Delta_+\over2}\phi^2+\cO(\phi^3)
			\ri], 			\label{W_max_+}\\
	W_-(\phi)
			=&{1\over \ell}
			\le[
				{2(d-1)} + {\Delta_-\over2}\phi^2+\cO(\phi^3)
			\ri]+C|\phi|^{d/\Delta_-}\le[1 +\cO(\phi)\ri]
			+\cO(C^{2})
			\label{W_max_-},\\
	\Delta_\pm =&\ha \le( d \pm \sqrt{d^2 +4 m^2\ell^2}  \ri) \quad \text{with}\quad -{d^2\over 4\ell^2}<m^2<0,
	\label{D_max}
\end{align}
\end{subequations}
where $C$ is an integration constant. These solutions are  represented schematically in figure \ref{fig:max}. They are  of two types, characterized by the second derivative of $W(\phi)$ at the extremum: the $W_-$ branch is a continuous family of solutions parametrized  by the constant $C$; $W_+$ is instead an isolated $W_+$ solution (which however, as was observed in \cite{papa1},  can be obtained from the continous family in the limit  $C\to +\infty$, as we shall see below). The notation $\cO(C^{2})$  in equation (\ref{W_max_-}) does not mean that $C$ is small, but we use it to  mean schematically that the next terms in the expansion, which are proportional to powers of $C$, are accompanied by higher powers of $\phi$ and are therefore sub-leading. More precisely,   the general structure of the $W_-$ solution is a double series \cite{rg1,papa1}:
\be
W_-=\sum_{m=0}^{\infty}\sum_{n=0}^{\infty} ~A_{m,n} \left(C~|\phi|^{d/\Delta_-}\right)^m~\phi^n
\label{general}
\ee
The coefficients $A_{m,n}$ are constants that depend on the potential. This expansion is valid if $\Delta$ is irrational.  The expansion in (\ref{general}) has poles for rational values of $\Delta$. This implies that in such cases there are also $\log \phi$ terms entering into (\ref{general}), \cite{Martelli}. A detailed discussion of such phenomena is beyond the scope of this paper.

\begin{figure}[t]
\centering
\includegraphics[width=0.8\textwidth]{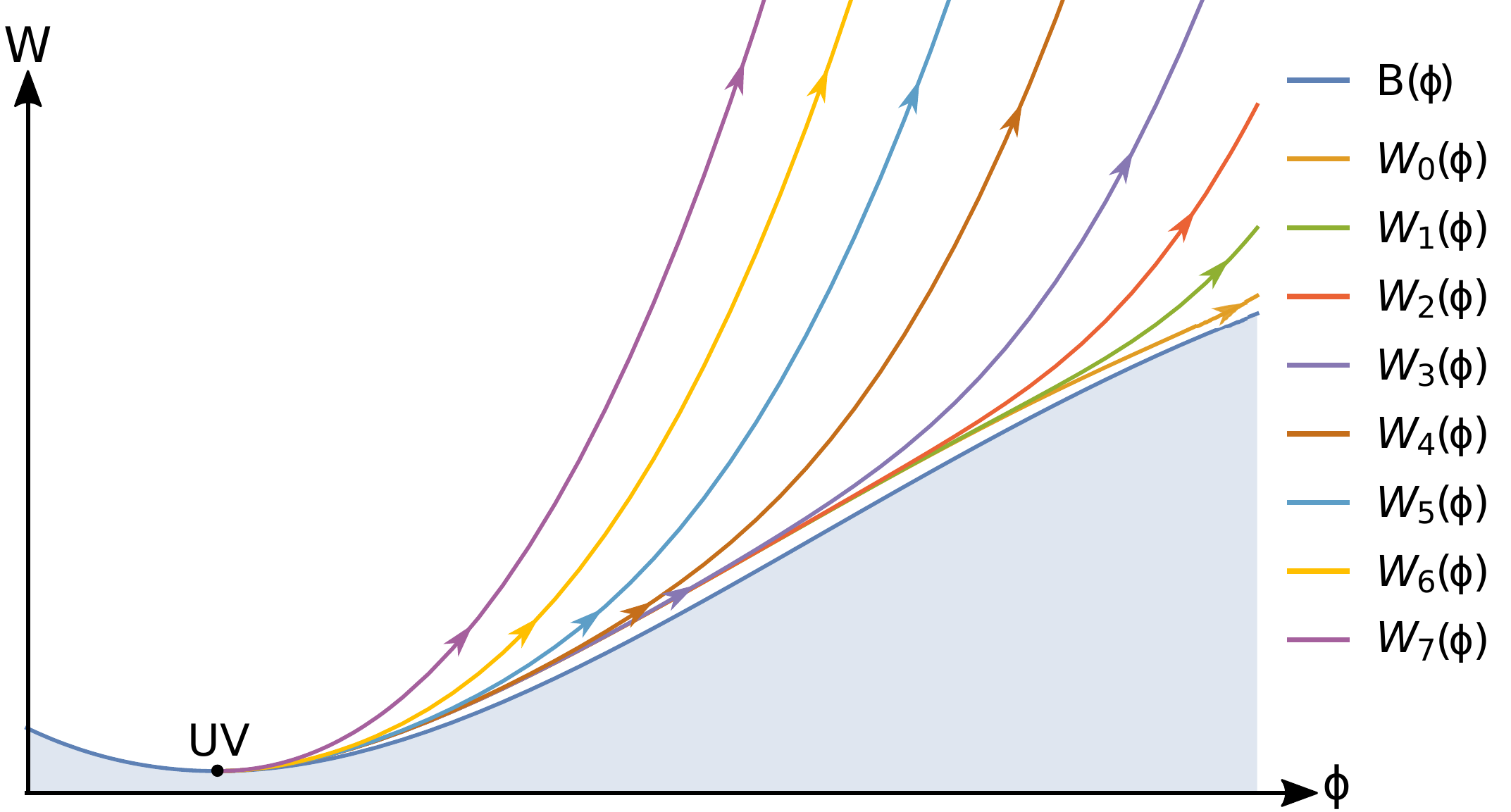}
\caption{These plots show  numerical solutions of the $W_-$ type (see equation \protect\eqref{W_max_-}),  with different values of $C$, obtained from the potential $V(\phi)=- 12 - 1/4 (\phi^2- 2)^2$, around the maximum at $\phi = -\sqrt{2}$.}
\label{fig:Delta}
\end{figure}

 In figure \ref{fig:Delta} we show actual numerical solutions corresponding to different values of the integration constant $C$ in \eqref{W_max_-} (shown on the right side of the critical point only).

We will now be  interested in what kind of geometry corresponds to the superpotentials (\ref{W_max_+}) and (\ref{W_max_-}). By solving equation \eqref{phiW} we obtain:
\begin{subequations}\label{a}
	\begin{align}\label{aa}
		&\dot\phi(u)=W_+'(\phi)
			\implies
		\phi(u)=\phi_+ e^{\Delta_+ u/\ell}+\cdots\\
		&\dot\phi(u)=W_-'(\phi)
			\implies
		\phi(u)=\phi_- e^{\Delta_- u/\ell}
			+{
				d~C\ell
					\over
				\Delta_-(2\Delta_+-d)}~\phi_-^{\Delta_+/\Delta_-}~
			e^{\Delta_+u \ell}+\cdots\label{ab}
	\end{align}
\end{subequations}
where $\phi_+$ and $\phi_-$ are integration constants.

The scale factors are found by integrating equation (\ref{defW}):
\begin{subequations}\label{Amax}
	\begin{align}
		&\dot A(u)=-{1\over 2(d-1)}W_+'(\phi)
			\implies
		A(u)=-{u-u_*\over \ell}+{\phi^2_+ \over 8(d-1)} e^{2\Delta_+ u/\ell}
		+\cO\le( e^{3\Delta_+u/\ell}\ri)
		\label{Amax1}\\
		&\dot A(u)=-{1\over 2(d-1)}W_-'(\phi)
			\implies
		A(u)=-{u-u_*\over \ell}+ {\phi^2_- \over 8(d-1)} e^{2\Delta_- u/\ell}
		+\cO\le(C e^{ud/\ell}\ri)
		\label{Amax2}
	\end{align}
\end{subequations}
where $u_*$ is an integration constant and the sub-leading terms come from the $\cO(\phi^2)$ terms in $W_\pm$.

The expressions above are valid for small $\phi$, hence we must take $u\to -\infty$. Therefore, the scale factor diverges and these solutions describe the near-boundary regions of an   asymptotically $AdS$ space-time,  with  AdS length $\ell$.  As $u$ increases away from the boundary,  the scale factor decreases, therefore these solutions corresponds to a flow leaving a UV AdS fixed-point. We conclude that {\it local maxima of the potential $V(\phi)$ and, hence, local minima of $B(\phi)$, correspond to UV fixed points}.

Notice that  each superpotential solution corresponds to {\em two} disconnected geometries: one with $\phi>0$,  the other with $\phi<0$,  because as $\phi$ approaches the critical point from each side the geometry is geodesically complete.

On the boundary field theory side, the continuous, $W_-$, branch is interpreted by associating with $\phi(u)$ a source\footnote{
This choice of source is called the standard quantization. Other quantizations exist \cite{BF82,KlebWit,Witten02}, as for example, the so-called alternative quantization where the roles of the source and the VEV are reversed. Different quantizations, however, only exist when $\Delta_-$ is between $d/2-1$ and $d/2$.
} $J=\phi_-/\ell^{\Delta_-}$ of an operator $\cal O$. The scaling dimension of $\cO$ is $\Delta_+$ \cite{Witten98}.
The vacuum expectation value (VEV) of $\cal O$ is given by
\eql{VEV_st-}{\le<\cal O\ri>_{W_-} ={1\over \ell^{\Delta_+}}{
				 d
					\over
				\Delta_- }\, C\ell\, 
				\phi_-^{\Delta_+/\Delta_-}}
so, for a given source $J$, the integration constant $C$ fixes the VEV \footnote{If we want to  interpret \eqref{VEV_st-} as the renormalized VEV,  the appearance of the same $C$ in \eqref{W_max_-} and in \eqref{VEV_st-} corresponds to a specific choice of holographic renormalization scheme \cite{papa1}.}.  The terms of order $C^2$ and higher do not contribute to  equation \eqref{VEV_st-} as they vanish at the boundary \cite{rg2,papa1}.

The $W_+$ solution in equation (\ref{W_max_+})  also corresponds to a boundary operator $\cal O$ with scaling dimension $\Delta_+$ and  a non-vanishing VEV given by:
\eql{VEV_st+}{\le<\cal O\ri>_{W_+}={2\Delta_+-d\over \ell^{\Delta_+}}\, \phi_+.}
The source of $\cal O$ however is set to zero.

Once the sign of the source $\phi_-$ in equation \eqref{ab} is fixed, close to the UV fixed point, the solution will correspond either to $\phi(u)\geqslant0$ or to $\phi(u)\leqslant0$. For either choice of sign the geometry will approach an AdS boundary, meaning that the positive source and the negative source solutions correspond to disconnected geometries. Similarly, a $W_+$ solution associated with a positive VEV in \eqref{VEV_st+} is disconnected from the geometry corresponding to a negative VEV. This corresponds to the sign of $\phi_+$ in \eqref{aa}.

It is easy to see from (\ref{ab}) that if we take the limit $\phi_-\to 0$ and $C\to \infty$ with $C\phi_-^{\Delta_{+}/ \Delta_{-}}$ kept fixed we obtain the $W_+$ solution in (\ref{aa}).
Therefore, the $W_{+}$ solution is the upper envelope of all $W_-$ solutions.

When the dimensions $\Delta_\pm$ are equal, $\Delta_+=\Delta_-={d\over 2}$,  we have $m^2\ell^2 = -d^2/4$, saturating the BF bound.  The solutions in this case are \cite{papa1}:

\begin{subequations}			\label{W_max_BF}
\begin{align}
	W_+(\phi)
			=&{1\over \ell}
			\le[
				{2(d-1)} + {d\over4}\phi^2+{\mathcal{O}}(\phi^3)
			\ri], 			\label{W_max_BF+}\\
	W_-(\phi)
			=&{1\over \ell}
			\le\{
				{2(d-1)} + \phi^2
				\le[{d\over4}
				\le(1+
					{1\over \log\phi}
				\ri)
				+{C\ell\phi^{2}\over (\log\phi)^2}
				+{\mathcal{O}}\le(\phi^2\over (\log \phi)^3\ri)
				\ri]
			\ri\}
				+{\mathcal{O}}(C^2)
			\label{W_max_BF-},\\
	&\Delta_\pm ={d\over 2} \quad \iff\quad m^2\ell^2=-{d^2\over 4}.
	\label{D_max_BF}
\end{align}
\end{subequations}
 Solving \eqref{phiW} we obtain:
\begin{subequations} \label{geo_max}
	\begin{align}
		&\dot\phi(u)=W_+'(\phi)
			\implies
		\phi(u)=\phi_+ e^{d u/2\ell}+...\\
		&\dot\phi(u)=W_-'(\phi)
			\implies
		\phi(u)=\phi_- e^{d u/2 \ell}
			\le[
				u
				+{\ell\over d}
					\le(
						1
						-{8C\ell\over d}
					\ri)+\cO\le(1\over u\ri)
			\ri]
	\end{align}
\end{subequations}
where $\phi_\pm$ are integration constants. Again, the scale factor behaves as in \eqref{Amax} to leading order.

Equations \eqref{W_max} and \eqref{W_max_BF} exhaust the  possibilities for a solution stopping at a  local maximum of the form \eqref{Vext}. As the $W_+$ solution lies above all the other solutions, one  consequence is that every solution $W(\phi)>W_+(\phi)$ will have $W'(0)\neq0$, skipping the critical point.

The BF bound $m^2\geqslant-{d^2\over 4\ell^2}$ is required in order to have stability of small perturbations close to a maximum of $V$ \cite{BF82} (see Appendix \ref{BF} for a quick review). On the boundary CFT, it corresponds to a reality condition on the dimension of the operator dual to $\phi$.
The BF bound is also required  to have  a real superpotential.

A local maximum of the potential which violates the BF bound does not correspond to a critical point for the  superpotential.  To see this,  we can rewrite the squared mass and the AdS length in terms of the potential with maximum at $\phi=\phi_*$ and replace the results into the BF bound inequality:
\eql{BFV}{
1\geqslant {4(d-1)\over d}{V''(\phi_*)\over V(\phi_*)}
}
If there is a solution of \eqref{SuperP} with $W'(\phi_*)=0$, then \eqref{BFV} becomes an identity:
\eql{BFW}{\le(1-{4(d-1)\over d}{W''(\phi_*)\over W(\phi_*)}\ri)^2\geqslant0.}

Therefore, having a critical point  of the superpotential is incompatible with a violation of the BF bound.
Another way to see this is that the expansion \eqref{W_max} around a BF bound-violating point leads to a  complex superpotential, as the dimensions $\Delta_\pm$  of \eqref{D_max} become complex.  As we will see in subsection \ref{ssec:casc}, this  represents a breakdown of the first order formalism. On the other hand,  a UV-regular solution $(\phi(u), A(u))$ which reaches the  AdS boundary with vanishing $\dot \phi(u)$ does exist. However,  this solution, (similarly to the AdS solution at  maximum of $V$) is unstable against linear perturbations and corresponds to a non-unitary CFT (see section \ref{sec:stab}).


\subsubsection{Local  minima of the potential}
\label{sssec:min}

We now assume the potential has the form \eqref{Vext},  but  with $m^2>0$. Then, as shown in Appendix \ref{crit},  solutions of the superpotential equation close to the critical point $\phi=0$ have the  regular power series expansion:
\begin{subequations}\label{W_min}
\begin{align}
	W_\pm(\phi)
			=&{1\over \ell}
			\le[
				{2(d-1)} + {\Delta_\pm\over2}\phi^2+\cO(\phi^3)
			\ri],
			\label{W_min_pm}\\
\Delta_\pm =&\ha \le( d \pm \sqrt{d^2 +4 m^2\ell^2} \ri)\quad \text{with}\quad m^2>0. \label{D_min}
\end{align}
\end{subequations}
Note that now we have necessarily $\Delta_-<0$  and $\Delta_+>0$.
These solutions are schematically represented in figure \ref{fig:min}. They are analytic around $\phi=0$ and  admit no continuous deformation, as shown in detail in appendix \ref{app:def}.

\begin{figure}[t]
\centering
\includegraphics[width=0.6\textwidth]{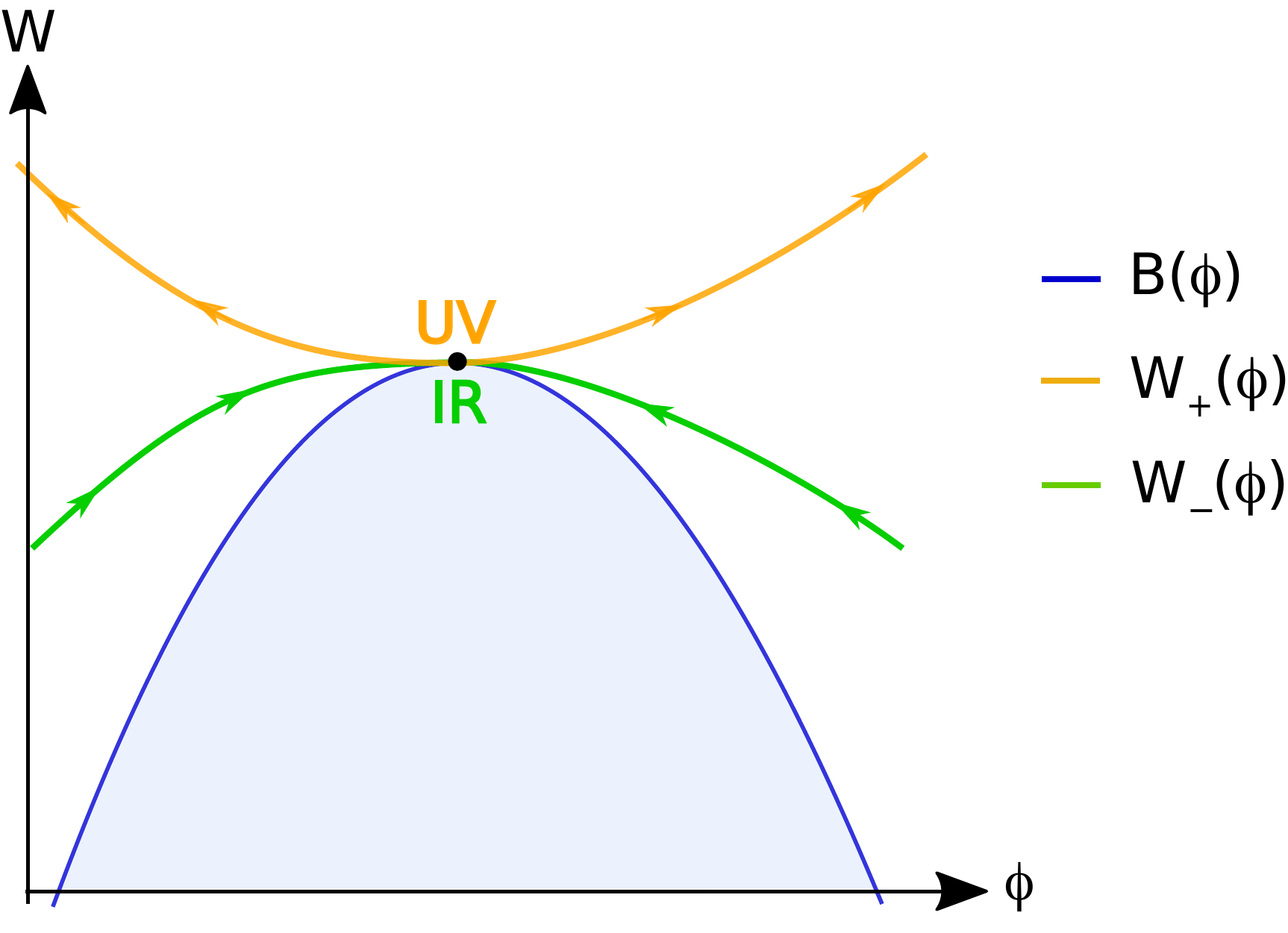}
\caption{Sketch of solutions of the superpotential equation with critical point at a local minimum of the potential (local maximum of $B(\phi)$). The $W_+(\phi)$ and $W_-(\phi)$   solutions correspond each to  two asymptotically AdS geometries in the UV and IR, respectively. The corresponding  geometries are not connected. The arrows on $W_-(\phi)$ and  $W_+(\phi)$ indicate the direction of increasing holographic RG flow.
}
\label{fig:min}
\end{figure}
The $W_+$ solution has a local minimum at $\phi=0$ while  $W_-$ has a local maximum. This implies a different geometrical and holographic interpretation. To see this, we solve for $\phi(u)$ using \eqref{phiW} and for $A(u)$ using \eqref{defW} with $W_\pm$ from \eqref{W_min} :
\begin{subequations} \label{geo_min}
\begin{align}
		&\dot\phi(u)=W_+'(\phi)
			\implies
		\phi(u)=\phi_+ e^{\Delta_+ u/\ell}+... \label{geo_mina}\\
		&\dot\phi(u)=W_-'(\phi)
			\implies
		\phi(u)=\phi_- e^{\Delta_- u/\ell}
			+...  \label{geo_minb}\\
	&A_\pm(u)=-{u-u_*\over \ell}
				- {1\over 8(d-1)}{\phi_\pm^2}e^{2\Delta_\pm u/ \ell}+...
\end{align}
\end{subequations}
where $\phi_\pm$ and $u_*$ are integration constants.
Equations \eqref{geo_min} are valid for small $\phi$ (near the critical point). Because $\Delta_-<0$, small $\phi$ in (\ref{geo_minb}) requires  $u \to +\infty$,  and $\m=\m_0 e^{A_-(u)} \to 0$. Therefore, {\em a $W_-(\phi)$ solution with a critical point at a local minimum of the potential corresponds to a flow that arrives at an infra-red (IR) fixed point}.

We can use again eq. \eqref{VEV_st-} with $C$ set to zero,  to see that the $W_-$ solution with positive $m^2$ represents a RG flow reaching an IR fixed point where the operator $\cal O$ of the dual field theory has dimension $\Delta_+$ and a vanishing VEV.

On the other hand,  because $\Delta_+>0$, small $\phi$ in (\ref{geo_mina})  requires $u \to -\infty$. In this case the scale $\m=\m_0 e^{A_+(u)}$ diverges, therefore the $W_+$ solution corresponds to a flow {\it leaving a UV fixed point}. Here however, the source for the operator $\cal O$ vanishes,  so it is a flow driven purely by a VEV.

The behavior of the solutions around a local minimum is summarized in  figure \ref{fig:min}. On each side ($\phi>0$ and $\phi<0$) there are two regular flows: one of them (with superpotential $W_-$)  is a flow towards the minimum, which then corresponds an IR fixed point.  The other one (with superpotential $W_+$)  leaves the minimum  driven by a VEV and it corresponds to a flow out of a UV fixed point. Similar remarks hold for the other side of the minimum, the difference residing in the sign of the source or of the VEV.


\subsubsection{Inflection points of the potential: marginal operators}
\label{sssec:marginal}
Close to an extremum of $V$,  the mass/dimension relation $m^2 = \Delta(\Delta -d) $ implies that $m^2=0$  corresponds to a (classically) marginal operator with $\Delta=d$. Here we discuss this case in detail.

Generically, if $V'(0)=V''(0)=0$, close to $\phi=0$  the potential has the form:
\be \label{marg1}
V(\phi) = -{d(d-1)\over \ell^2} - {\lambda \over 3} \phi^3 + {\cal O}\big(\phi^4\big),  \qquad \lambda>0,
\ee
where we have assumed, for definiteness, $\lambda>0$ (for $\lambda <0$ the results we present below are the same, but  the behaviors for  $\phi>0$ and $\phi<0$  are interchanged).
There are still two kinds of solutions  for the superpotential close to $\phi=0$ which, to order $\phi^2$,  have the same analytic expansion as in equations (\ref{W_max_+}-\ref{W_max_-}), where now $\Delta_-=0, \Delta_+ = d$
:
\begin{subequations}			\label{W_max_+massless}
\begin{align}
	W_+(\phi)
			=&{1\over \ell}
			\le[
				{2(d-1)} + {d\over2}\phi^2+\cO(\phi^3)
			\ri], 			\label{W_max_+massless}\\
	W_-(\phi)
			=&{1\over \ell}
			\le[
				{2(d-1)} + {\ell^2 \over d} {\lambda \over 3} \phi^{3}+\cO(\phi^{4})
			\ri], \label{W_max_-massless}
\end{align}
\end{subequations}
where we have included the first non-constant (cubic) term in the expression of $W_-$, but not in $W_+$, since there it does not play an important role.

\begin{figure}[t]
\centering
\includegraphics[width=0.6\textwidth]{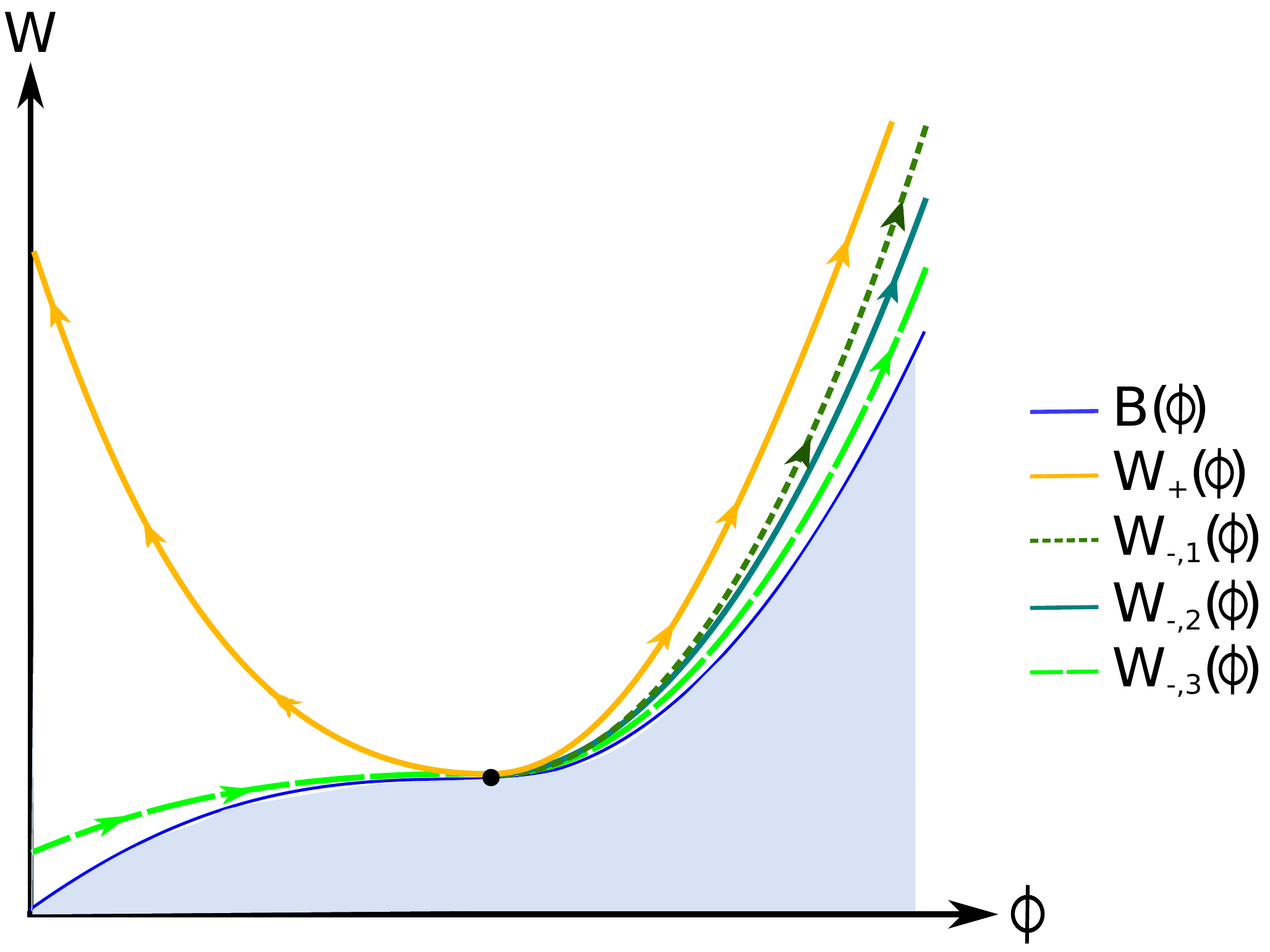}
\caption{Sketch of solutions with a critical point at an inflection point of $V$ (and of $B$). There are two kinds of solutions, labelled $W_+(\phi)$ and $W_-(\phi)$, as shown in equations \protect\eqref{W_max_+massless} and \protect\eqref{marg5}. The superpotential of the $W_+$ kind is unique, while the $W_-$ branch admits infinitely many solutions, parametrized by an integration constant, but  only on the  side of the critical point with $B''>0$. On the side  where $B''<0$, there is a unique solution of the $W_-$ type. Therefore, an inflection point behaves on one side as a maximum of the potential (figure \protect\ref{fig:max}) and on the other side as a minimum (figure \protect\ref{fig:min}). As in the case of minima and maxima, the geometries on either side are disconnected and locally geodesically complete.
  The arrows on the curves indicate the direction of increasing holographic coordinate $u$ and of the RG flow.
}
\label{fig:inf}
\end{figure}

As in the case $m^2\neq 0$,  $W_+$  corresponds to a VEV-driven  deformation of an $AdS$ UV fixed point, from which $\phi$ can equally flow towards negative or positive values. It does not admit continuous deformations.
For the $W_-$ solution on the other hand, the nature of the solution depends  on the sign of $\phi$, as can be seen by integrating the RG flow equations (\ref{defW}-\ref{phiW}) close to $\phi=0$, using (\ref{W_max_-massless})  as the superpotential. The solution is:
\be\label{marg2}
A(u) \simeq -{u\over \ell} , \qquad \phi(u) \simeq  -{d\over \ell \lambda} \,  {1\over u}.
\ee
From the above expression, it makes a difference if $\phi$ flows to the origin from the left or from the right:
\bea
&&\phi \to 0^+ \quad \Leftrightarrow  \quad u \to -\infty \quad \Leftrightarrow \quad A \to +\infty \qquad UV\\
&& \phi \to  0^- \quad \Leftrightarrow  \quad u \to +\infty \quad \Leftrightarrow \quad A \to -\infty \qquad IR
\eea
 Therefore, for  $\phi>0$, $W_-$  corresponds to a source-driven deformation of a UV fixed point at $\phi=0$;  instead, for $\phi<0$, it corresponds to a flow towards an IR endpoint at $\phi=0$.
This behavior is a generic feature of marginal operators in field theories: they are marginally relevant for one sign of the coupling, and marginally irrelevant for the opposite sign.  This analogy can be also seen by writing the holographic $\beta$-function corresponding to $W_-$, using equation (\ref{betaW}):
\be\label{marg3}
\beta(\phi)  = -\ell^2 \lambda \, \phi^2 + \cO(\phi^3).
\ee
Since the $\beta$-function does not change sign across the fixed point at $\phi=0$, the fixed point is attractive on the left and repulsive on the right. This is analogous to what happens in perturbative QFTs where the classical dimension of the operator vanishes and the running starts at one-loop, such as for example Yang-Mills theories (for $\alpha_{YM}$)  and scalar field theories for the quartic coupling. However, in these cases typically only one sign of the coupling gives a consistent field theory: in Yang-Mills, $\alpha_{YM} \propto g^2 >0$ and in scalar QFT a negative quartic coupling makes the potential unstable. Here on the other hand the coupling can be defined on both sides of the fixed point.

 As usual, the geometries on each side of the critical point are independent and  disconnected. Even though  the solutions in $W(\phi)$ in \eqref{W_max_-massless} has a series expansion valid for both signs of $\phi$, the geometry resulting from $W(\phi)$ with $\phi<0$ is independent from the geometry resulting from the same superpotential but with $\phi>0$.

We will now show that the $W_-$ solution admits a continuous one-parameter family of deformations only for $\phi>0$, whereas it is isolated for $\phi<0$. This is consistent with the fact that  only for a source-driven relevant deformation, the superpotential  needs to contain  an  integration constant parametrizing the VEV.  Following the procedure  in Appendix \ref{app:def},  from equation (\ref{def3}) we find the form of  a small deformation $\delta W_-$ around the  $W_-$ solution:
\be \label{marg4}
\delta W_- = C \exp \left[-{d^2\over \ell^2 \lambda} \, {1\over \phi}\right].
\ee
This is indeed a small deformation only as $\phi \to 0^+$, whereas it diverges for $\phi\to 0^-$. Therefore, only on the UV side (like around a maximum of $V$) is there a continous family of  solutions, in which the integration constant $C$ in equation (\ref{marg4}) parametrizes the VEV of the dual operator. The situation is sketched in figure \ref{fig:inf}.

We can briefly comment on cases when not only the quadratic, but also the cubic term in (\ref{marg1}) is vanishing, and the potential starts at quartic or higher order. It must be stressed  that this case is non-generic, and in perturbative field theories it corresponds to the special cases where  the one-loop $\beta$-function vanishes (e.g. QCD at the perturbative end of the conformal window). In this case  the fixed point  is symmetric  or not depending on whether the first non-trivial power in $V$ is even or odd, but the two situations will be qualitatively similar  to the quadratic and cubic cases respectively. If the leading  power in $V(\phi)$ is even, depending on the sign of its coefficient, the fixed point  is  attractive or repulsive on both sides.

For completeness, below we give the expression of the superpotential when the first non-trivial term in the potential is of order $n\geq 3$,
\eql{Vn}{
	V=-{d(d-1)\over \ell^2}-{\l_n\over n}\phi^{n} + \cO(\phi^{n+1}), \qquad \l_n >0, \; n\geqslant 3.
}
Then, the $W_+$ solution is the same as in equation (\ref{W_max_+massless}), whereas the $W_-$ solution (including the UV deformation) is given by:
\bea\label{marg5}
  	W_-(\phi)
			=&&{1\over \ell}
			\le[
				{2(d-1)} + {\ell^2 \l_n \over d \, n} \phi^{n}+\cO(\phi^{n+1})
			\ri]+\nonumber\\
			&&+C \, \phi^\omega\exp\left[-{d^2 \over (n-2) \ell^2 \l_n}{1\over \phi^{n-2}} \right]
			+\cO(C^{2})
\eea
where the value of $\omega$ controlling the power-like behavior in the second line depends on the coefficients of the sub-leading $\phi^{n+1}$ term  in equation  (\ref{Vn}).

\subsubsection{Bounces: $W'(\phi)=0$ and $V'(\phi)\neq0$}
\label{sssec:bW}

If we choose an arbitrary point $\phi_*$ on the critical curve $W(\phi)=B(\phi)$, $V'(\phi_*)$ will generically be non-zero. Despite this fact, $W(\phi)$ still has a critical point at $\phi_*$, as follows from equations \eqref{W'B}. In fact, generic critical points occurs where $V'(\phi_*)\neq 0$.

As we will see below,  these critical points are reached  at  finite values of the scale factor, unlike critical points  at  extrema of $V$, which instead are reached as $A \to \pm \infty$.

\begin{figure}[t]
\centering
\includegraphics[width=0.49\textwidth]{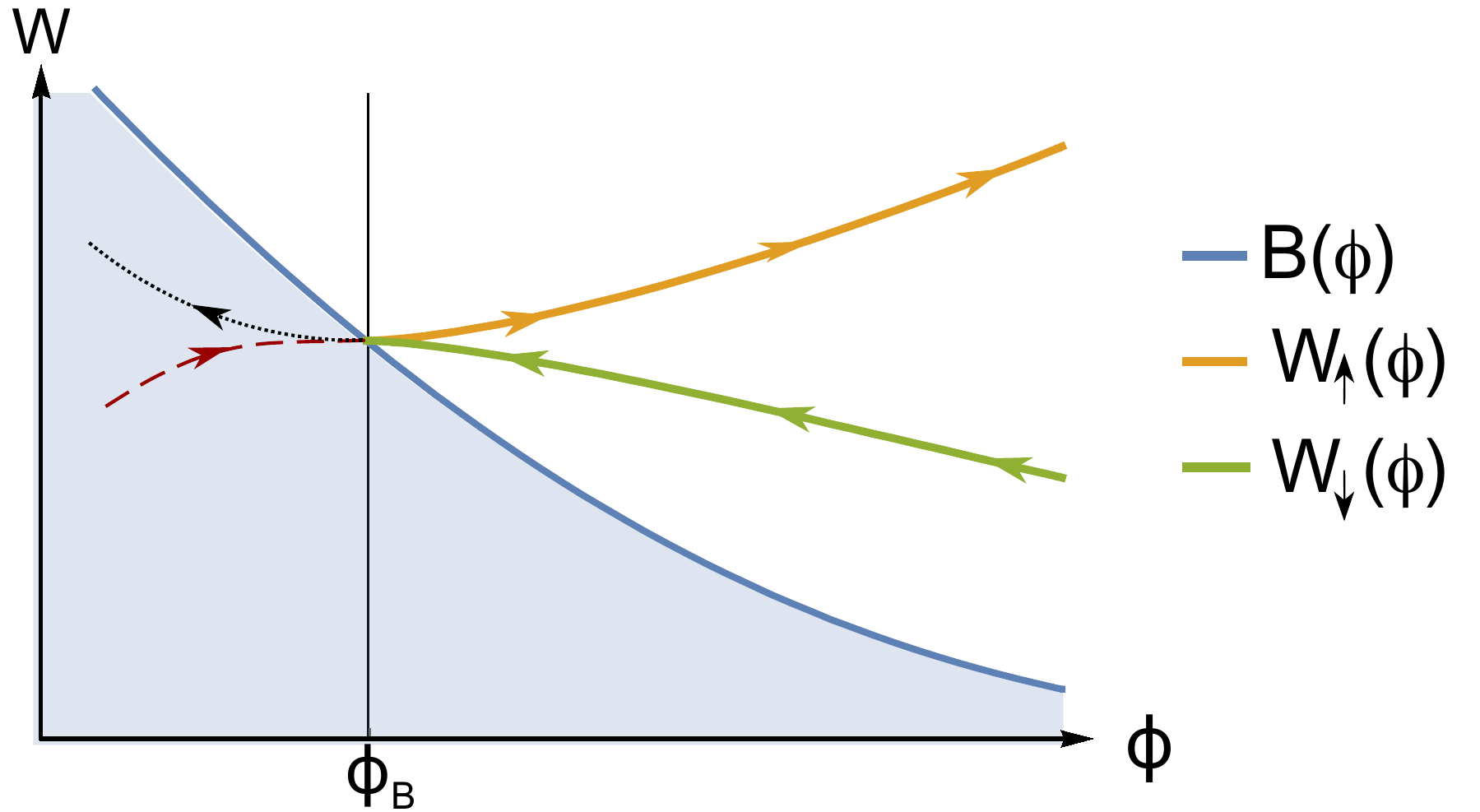}
\includegraphics[width=0.49\textwidth]{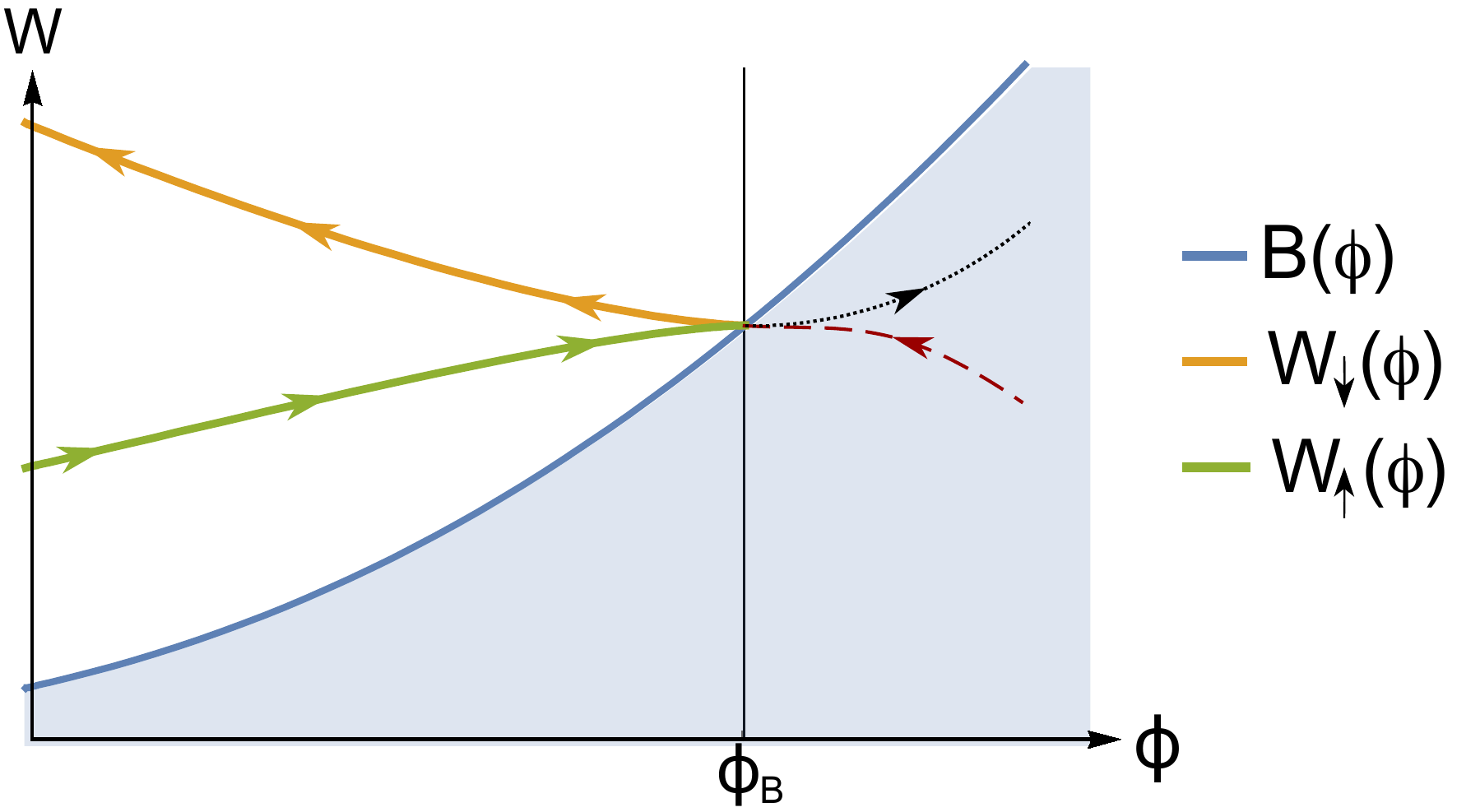}\\
(a) \hspace{0.35\textwidth} (b)
\caption{These figures illustrate a bounce at a critical point of $W$ with (a) $V '>0$  and (b) $V'<0$.  The functions  $W_{\uparrow}(\phi)$ and $W_{\downarrow}(\phi)$ together represent a bouncing solution, giving  a local minimum (a) or a local maximum (a) of the scalar field at $\phi=\phi_B$.%
The arrows
indicate the direction of the flow towards lower energies.
}
\label{fig:bounce2}
\end{figure}

We denote a generic critical point by $\phi_B$ where by definition $W'(\phi_B)=0$. The  superpotential equation also implies that $W(\phi_B)=B(\phi_B)$ and both are finite.
Then, equation (\ref{V'}) implies that  $W''$ must necessarily diverge at $\phi_B$. Therefore,  we can approximate equation \eqref{V'} by:
\eql{bW0}{
 W'(\phi)W''(\phi) \simeq V'(\phi) \qquad \phi \simeq \phi_B,
}
which can be integrated once to give, close to $\phi_B$:
\be\label{bW1}
W'(\phi) \simeq \pm \sqrt{2(\phi-\phi_B) V'(\phi_B)},
\ee
where we have used the assumption that $V$ has a regular power series expansion around $\phi_B$.
We conclude that a real superpotential exists only on the right (left) of $\phi_B$ for $V'(\phi_B)$ positive (negative). In either case, there are two solutions $W_\uparrow$ and $W_\downarrow$ terminating at $\phi=\phi_B$ from the right (left), which correspond to the two choices of sign in equation (\ref{bW1}).  After one more integration, we obtain the approximate form of the two solutions close to $\phi_B$.  For $V'(\phi_B)>0$,  they are:
\bea
 W_{\uparrow}(\phi)\simeq  W_B+\frac{2}{3} \sqrt{2V'(\phi_B)}(\phi-\phi_B)^{3/2}, \nonumber \\
&& \quad \phi > \phi_B, \label{bW2}\\
 W_{\downarrow}(\phi)\simeq  W_B-\frac{2}{3} \sqrt{2V'(\phi_B)}(\phi-\phi_B)^{3/2},  \nonumber \\
\eea	
where
\eql{W_B}{W_B=\sqrt{-{4(d-1)\over d}V(\phi_B)}.}
We call the two-branched  solution in equation (\ref{bW2}) a {\em bounce.} In  particular, equation  (\ref{bW2}  describes the increasing and decreasing branches at a left bounce,   as in figure \ref{fig:bounce2} (a). For $V'<0$  the solutions have the same  expression,   but now   $\phi<\phi_B$, with $W_\uparrow$ and $W_\downarrow$ interchanged, as in figure \ref{fig:bounce2} (b) (bounce on the right). At $\phi=\phi_B$, both solutions  (\ref{bW2}) have vanishing first derivative  and infinite second derivative,  as expected from equation (\ref{V'}). As shown in Appendix B, bounces do not admit continous deformations, i.e. there are only two solutions reaching the critical curve at a given generic point $\phi_B$ which is not an extremum of $V$.

From figure \ref{fig:bounce2} it may seem that one can have a separate solution which corresponds  to each single-valued branch of the superpotential, and which  terminates at $\phi_B$. As we will see shortly however, each separate branch gives rise to a geodesically incomplete geometry. To obtain a complete geometry  we must glue the two solutions $W_{\uparrow}(\phi)$ and $W_{\downarrow}(\phi)$ into a single solution with multi-valued superpotential. Although the superpotential is non-analytic at $\phi_B$, the resulting geometry  is smooth, as we will show below.


To obtain the solution in terms of $A(u)$ and $\phi(u)$, we first rewrite equation \eqref{phiW} for each of two  branches in equation (\ref{bW2}). Denoting by  $\phi_\uparrow(u)$ and $\phi_\downarrow(u)$ the two corresponding scalar field profiles, we have:
\bea
&& \dot\phi_\uparrow \simeq \sqrt{2\le(\phi_{\uparrow}-\phi_B\ri) V'(\phi_B)} \label{eqphiB1} \\
 &&  \nonumber \\
&& \dot\phi_\downarrow \simeq -\sqrt{2\le(\phi_{\downarrow}-\phi_B\ri) V'(\phi_B)} \label{eqphiB2}
\eea
where for definiteness we are considering the case a left bounce with $V'>0$ and $\phi>\phi_B$. (see fig. \ref{fig:bounce2} (a)).

Equations (\ref{eqphiB1}-\ref{eqphiB2})  integrate to the two halves $u>u_B$ and $u<u_B$ of a solution $\phi(u)$ which is analytic at the location of the bounce $u=u_B$ (which is an arbitrary integration constant):

\begin{align}
	\phi(u)=\phi_B+{V'(\phi_B)\over 2}(u-u_B)^2+\cO(u-u_B)^3=
		\begin{cases}
			\phi_\uparrow(u)\text{ for } u>u_B,\\
			\phi_\downarrow(u)\text{ for } u<u_B.
		\end{cases}
	\label{phiB}
\end{align}

Using equation  (\ref{phiB}) we can write  both superpotentials \eqref{bW2} as functions of the holographic coordinate $u$:
\eql{Wub}{W(u)=W_B+{\le(V'(\phi_B)\ri)^2\over 3}(u-u_B)^3+\cO(u-u_B)^4
=
		\begin{cases}
			W_\uparrow(\phi(u))\text{ if } u > u_B\\
			W_\downarrow(\phi(u))\text{ if } u< u_B
		\end{cases}
}
As for the scalar field, the two solutions $W_{\uparrow}(\phi)$ and $W_{\downarrow}(\phi)$  combine into a  a single-valued function of $u$. Finally,  we can integrate \eqref{defW} using \eqref{Wub} to obtain the scale factor:
\be \label{Ab}
A(u)=A_B-\sqrt{V(\phi_B) \over d(d-1)}(u-u_B)-{\le(V'(\phi_B)\ri)^2\over 4!(d-1)}(u-u_B)^4+\cO(u-u_B)^5,
\ee
which is also regular at the bounce. In particular, keeping only one branch of the bounce would result in an incomplete geometry which terminates at an ``artificial'' boundary at $u=u_B$\footnote{
The regularity at a bounce also has a simple alternative derivation without the superpotential. It consists in using the Klein-Gordon equation \eqref{KG} and using the fact that the product $\dot A\dot \phi$ vanishes at the bounce, at $u=u_B$, because $\dot A^2\propto V(\phi(u_B))$ is finite as a consequence of the Einstein equation \eqref{EE2}. This leads to $\ddot\phi(u_B)=V'(\phi_B)$ which, upon integration, results in \eqref{phiB}. Substituting the result in the Einstein equation \eqref{EE2}
we obtain a differential equation with solution \eqref{Ab}.
}.

\begin{figure}[t]
\centering
\includegraphics[width=0.7\textwidth]{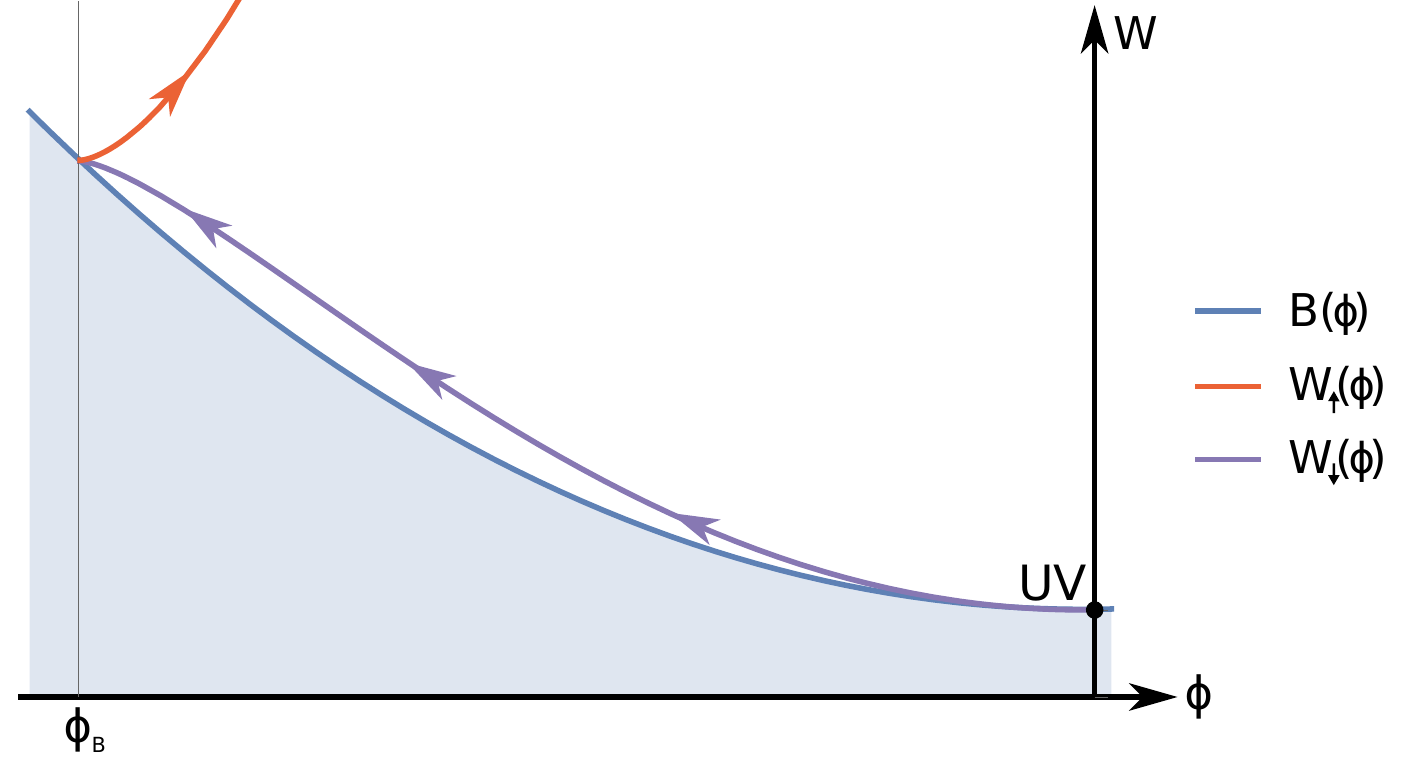}
\caption{The superpotential, obtained numerically, for a bouncing solution which starts at a UV fixed point.}
\label{fig:multi}
\end{figure}

\begin{figure}[t]
\centering
\includegraphics[width=0.48\textwidth]{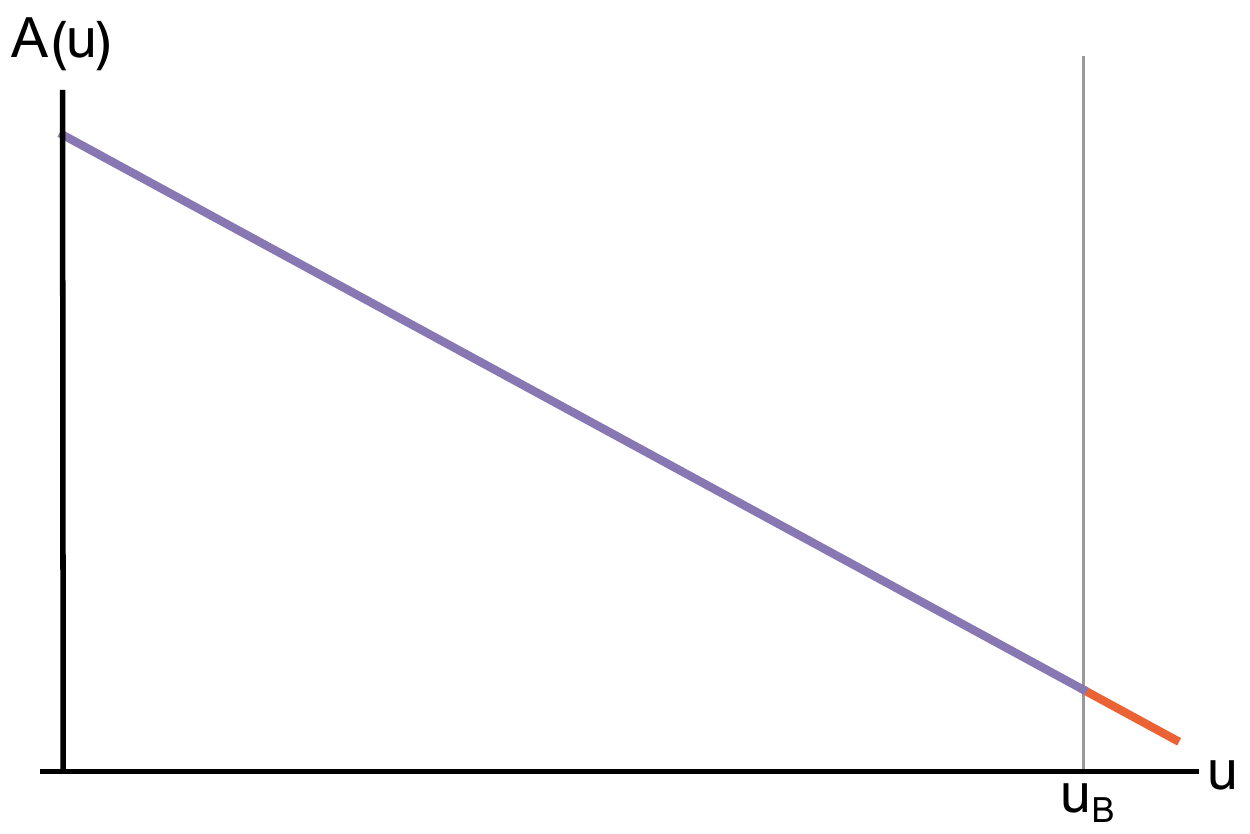}
\includegraphics[width=0.48\textwidth]{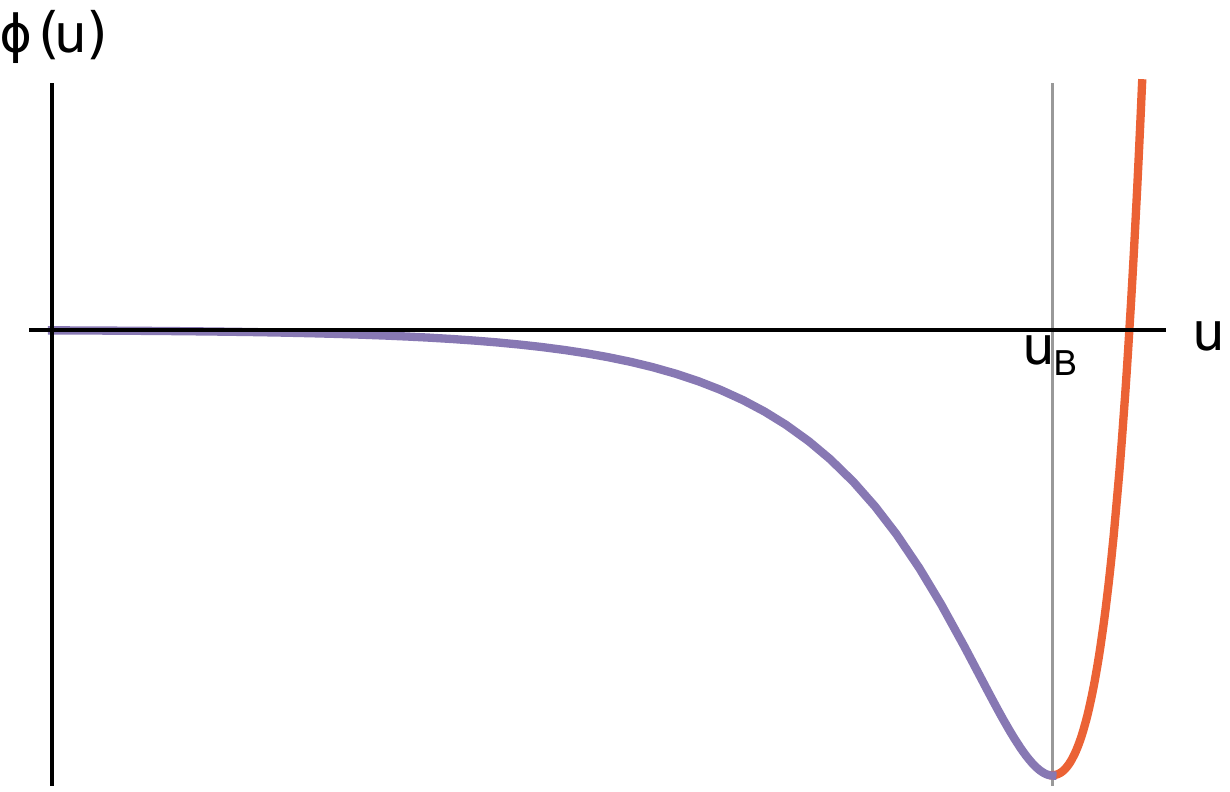}
\caption{The  profiles for the scale factor $A(u)$ and the scalar field $\phi(u)$ corresponding  to the solution represented in figure \protect\ref{fig:multi}. Both $\phi(u)$ and $A(u)$ are smooth at the bounce.}
\label{fig:phi8_00}
\end{figure}

In figures \ref{fig:multi} and \ref{fig:phi8_00} we present an example,  obtained numerically,  of a two-branched superpotential and the corresponding solution for the metric and scalar field as a function of $u$.  This is a detail of a full   numerical solution which will be presented and discussed in section \ref{ssec:bounce_inf}.

The geometry at the bounce is non-singular: all curvature invariants are finite . The curvature invariants of the geometry are functions of $V(\phi), W(\phi)$, and their derivatives \cite{Bourdier13}.  The fact that the geometry is regular at the bounce is clear when invariants are  expressed in terms of $(A(u),\phi(u))$ and their derivatives.
On the other hand it is not immediately obvious when we express them in coordinate-invariant form through $W(\phi)$, since $W''$ is singular at the bounce.  To explain this point, take for example the Ricci-squared invariant,  (\ref{Ricci2}). Taking two covariant derivatives, for example $\nabla^\mu\nabla_\mu (R^{\rho\sigma} R_{\rho\sigma})$  will lead to terms that contain $W''$. However, covariant derivatives are  with respect to  $u$, therefore:
\be
\ddot{W} =  2W''(W')^2~.
\ee
The right hand side is finite (in fact, it vanishes)  because the $W'$ factor cancels the divergence in $W''$, as one can observe by differentiating equations (\ref{bW1}).

To summarize, the full solution  is regular as it goes  through the bounce, which is characterized by a turning point for $\phi(u)$.  The non-analyticity (and multi-valuedness) of $W(\phi)$ are a consequence of the fact that $\phi(u)$ is not monotonic, therefore not a good coordinate in a neighborhood of $\phi_B$. This situation is represented in figures \ref{fig:multi} and \ref{fig:phi8_00}. The same considerations hold for a bounce to the right, i.e. such that $\phi<\phi_B$: in this case,  $V'(\phi_B)<0$ and $\phi(u)$ goes through  a relative maximum.

Finally, even if the geometry of the bounce is regular, one may worry that non-homogeneous perturbations in the metric and scalar field would introduce divergences in invariant quantities built using the space-time derivatives $\de/\de x^\mu$. If this were the case, the backgrounds would be holographically unacceptable as correlators would ill-defined.

We have performed a  detailed analysis of the   small perturbations around a bounce, which is presented in Appendix \ref{app:stab}. The conclusion is that fluctuations around the bounce are also regular, as are the perturbed curvature invariants.

 We now discuss the behavior of the holographic  $\b$-function at the bounce,  defined in \eqref{beta}. In terms of the superpotential $W(\phi)$, the holographic $\beta$-function is given by:
\be
	\b(\phi)\equiv {d\phi\over dA}=-2(d-1){d\over d\phi}\log W(\phi). \label{grad}
\ee
Therefore the resulting multi-branched $\b$-function near a bounce is
\be
	\b(\phi)=\pm \sqrt{-2d(d-1){V'(\phi_B)\over V(\phi_B)}(\phi-\phi_B)}
			 +{\mathcal{O}}\le(\phi-\phi_B\ri)\label{beta_bounce}
\ee
where we have assumed for concreteness a left bounce as in figures \ref{fig:multi} and \ref{fig:phi8_00}.  The sign in equation (\ref{beta_bounce}) must be taken to be negative for $u<u_B$ and positive for $u>u_B$.

The $\b$-function \eqref{beta_bounce} vanishes at the bounce, which is reached   at a finite energy scale. However the flow does not stop there, but rather it reverses its direction, so the bounce is a zero of the $\b$-function which is not a fixed point.

Even though to our knowledge there is no RG flow where a coupling has turning points (bounces) in Poincar\'e-invariant field theories, this behavior has been seen in condensed-matter effective field theories. In \cite{cycles} the example of the ``Russian doll superconductivity model'' \cite{doll} was presented and the similarity to the behavior \eqref{betaW} deserves some comments.

We consider more closely the example of the Russian Doll model \cite{doll}. The RG flow  is given by the following multi-valued $\b$-function for the coupling $g$ \cite{cycles}:
\eql{Russian_beta1}{{dg\over d\log\m}=\b_N(g)=(-1)^N\sqrt{1-g^2}, \quad N\in\mathbb{Z}}
This is a gradient flow that can be written in the form of equation \eqref{grad}. The corresponding superpotential is given in appendix \ref{app:cycles}.
Near any of the turning points, $g=\pm1$, the $\b$-function \eqref{Russian_beta1} has the same leading order behavior as our multi-branched $\b$-function \eqref{beta_bounce} near a bounce, up to a multiplicative constants: for example, as $g\simeq 1$,
\eql{Russian_beta2}{
	\b_{RD}(g)
	=
	\pm\sqrt{2(1-g)}+\cO({1-g})^{3/2}
}
Equation \eqref{Russian_beta1} has a solution analogous to a harmonic oscillator with ``time'' given by $\log\m$. This is an example of a first order RG flow for a single coupling that does not stop when the $\b$-function vanishes but changes direction instead.

In appendix \ref{app:cycles} we show that, using equation  \eqref{grad}, we can  obtain a multi-branched superpotential and therefore in principle a bulk geometry. The corresponding scalar potential $V(\phi)$  however, obtained via   equation \eqref{SuperP}, turns out to be itself multi-valued. On the other hand, the bulk potential should be a single valued function of $\phi$. This means that  {\em despite the same local behavior at the bounces, the $\b$-function \eqref{Russian_beta1} cannot have a (unitary) holographic realization}.


\section{Exotic holographic RG flows}

In this section we present examples of holographic RG flows with unusual properties.

 The first example, in subsection \ref{ssec:mm}, is a flow between two minima of the potential
   \cite{Sibiryakov,Libanov14} showing that a local minimum of the potential does
    not always corresponds to an IR fixed point.
More exotic flows, with no known counterpart in Poincar\'e-invariant field theories, are presented next.
In subsection \ref{ssec:bounces} we show an example of flows featuring   multiple  bounces: these are solutions that leave  a UV fixed point, reverse  direction several times,  before ending  at an IR fixed point. Next, in subsection \ref{skip} we present one example of a flow which starts at a UV fixed point, skips the nearest IR fixed point, and ends at a minimum of $V$ further away.

All these examples go  against the field theory intuition based on first-order RG flows, which we present for comparison in the language of superpotentials in the next subsection.

All such solutions are regular geometries, as they all asymptote to $AdS$ in the the IR, and bounces do not introduce singularities in the interior.  Moreover, they are all stable under small perturbations. This is a general feature of the regular geometries, as well as of a class of IR-singular solutions  which will be presented in the Section 4. The issue of stability will be discussed further in Section 6 and in Appendix \ref{app:reg}

\subsection{Standard Holographic RG flows}

\begin{figure}[t]
\centering
\includegraphics[width=0.5\textwidth]{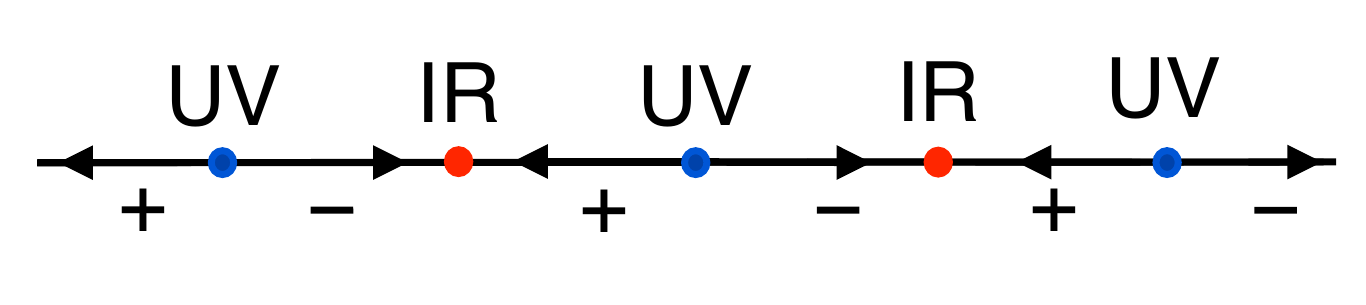}
\caption{Standard RG flow for a single coupling. The signs of the $\b$-function on the intervals between fixed points are indicated. The arrows correspond to the direction of decreasing energy along the RG flow.}
\label{fig:RG}
\end{figure}

Standard holographic RG flows are those which have a qualitative correspondence with RG flows from QFT: they start at one UV fixed point, end at an adjacent IR fixed point and the coupling is a monotonic function of the energy scale. This behavior is schematically represented in figure \ref{fig:RG}.  These flows can be obtained from a variety of potentials and here we present a numerical solution of the superpotential equation \eqref{SuperP} with this property.

\begin{figure}[t]
\centering
\includegraphics[width=0.7\textwidth]{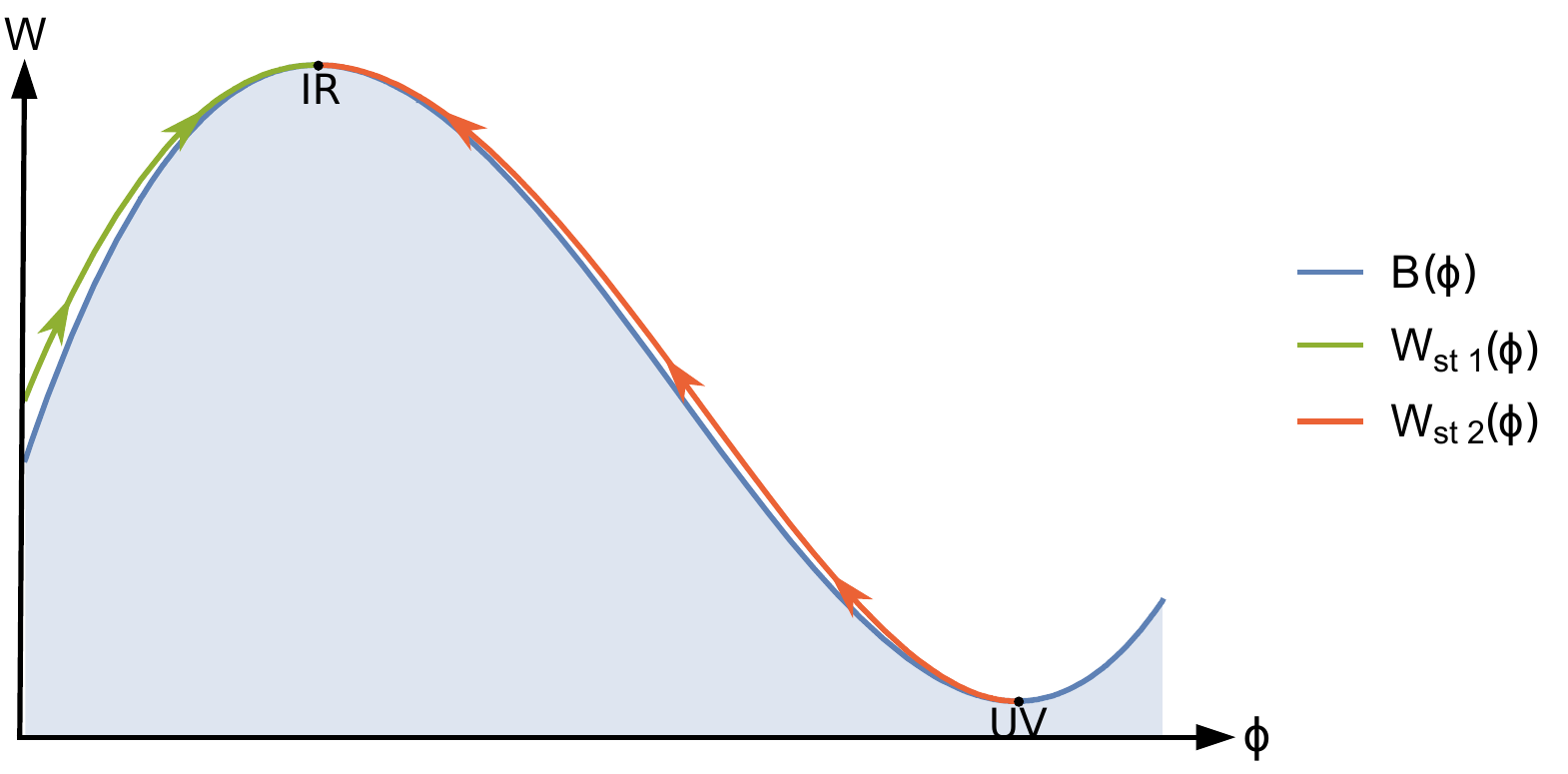}
\caption{Plot of the superpotential (red curve) interpolating between two fixed points in a model with polynomial $V(\phi)$, given in appendix \protect\ref{app:V12}. This solution represents a standard RG flow between two neighboring fixed points. The green curve represents the end of another standard RG flow that reaches the same IR fixed point, but with increasing $\phi$ (the UV starting point is not shown in the figure). The shaded area is the forbidden region $W<B$.  The arrows indicate the direction of the holographic coordinate (and of the flow).}
\label{fig:standard}
\end{figure}

In figure \ref{fig:standard} we plot an example of a complete regular RG flow, $W_{st 2}(\phi)$, with decreasing coupling $\phi$, starting at the UV fixed  point and ending at the IR point. On the same graph we also show the final part  of another standard flow with increasing $\phi$, described by $W_{st 1}(\phi)$, which approaches the same IR fixed point from the left.

 The green curve also has an asymptotically AdS region in the UV  (not appearing in the figure), with a different AdS length.  Both superpotentials increase with decreasing energy scale reflecting the holographic C-theorem.

From each side, the IR fixed point is reached by a unique solution, given by the $W_-(\phi)$ solution from \eqref{W_min}.

This means that among the infinitely many solutions leaving a UV fixed point at a local minimum of $B(\phi)$ (see figure \ref{fig:max}), at most one of them may end at a given IR fixed point. In other words, a flow ending at a given IR fixed point prevents any other flow to reach it from the same side (including flows from different UV starting points).

 When a holographic RG flow starts at a UV fixed point and ends at an IR fixed point, the full geometry is regular. Solutions lying above the superpotential of a standard flow like the one in figure \ref{fig:standard}  will either reach a different  fixed point further away (cf. section \ref{skip}), or flow to infinity in field space (cf. section \ref{Francesco}).



\subsection{Flows interpolating between two minima}
\label{ssec:mm}
Usually, we think of maxima of the potential as UV fixed points, and minima as IR fixed points. The reason is that only at a maximum is the operator corresponding to $\phi$ relevant. However, one can also flow out of a UV fixed point by giving a VEV to an {\em irrelevant} operator, while keeping the source absent. In the present setup, this correspond to a holographic RG flow between two minima of the potential.

As we discussed in section \ref{sssec:min}, the solution  arriving at  a minimum of $V(\phi)$ is unique and does not admit deformations. Therefore, to construct a solution that connects two minima, one needs to fine-tune the potential (generic solutions will have only one of their two endpoints at one of the minima). A similar  behavior was already discussed in \cite{Sibiryakov,Libanov14,Kakushadze00}.

Here we present a potential $V$ that generalizes the one from \cite{Kakushadze00} and allows for a flow between minima of V:
\eql{ir}{
V(\phi)={(kv)^2\over 2} \le[1-\le(\phi\over v\ri)^2\ri]^2
			-\frac{d}{4(d-1)}\le\{
							kv^2\le(\phi\over v\ri)\le[
								1-{1\over 3}\le(\phi\over v\ri)^2
							\ri]+W_0
						\ri\}^2.
}
Depending on the values of the parameters, the potential \eqref{ir} can have up to five extrema, among which two are always present, the ones at $\phi=\pm v$. The constant $W_0>0$ can be adjusted so that $V(\phi)$ is strictly negative between $\phi=\pm v$. This will allow us to obtain regular flows starting from $\phi=-v$ and ending at $\phi=v$.

The potential \eqref{ir} is such that there always exists a range of $\phi$ where $V(\phi)$ becomes positive  or zero, but for the purpose of this example this is not problematic,  as long as this does not occur in the region between $\phi=-v$ and $\phi=v$ to which the flow is confined.

\begin{figure}[t]
\centering
\includegraphics[width=0.45\textwidth]{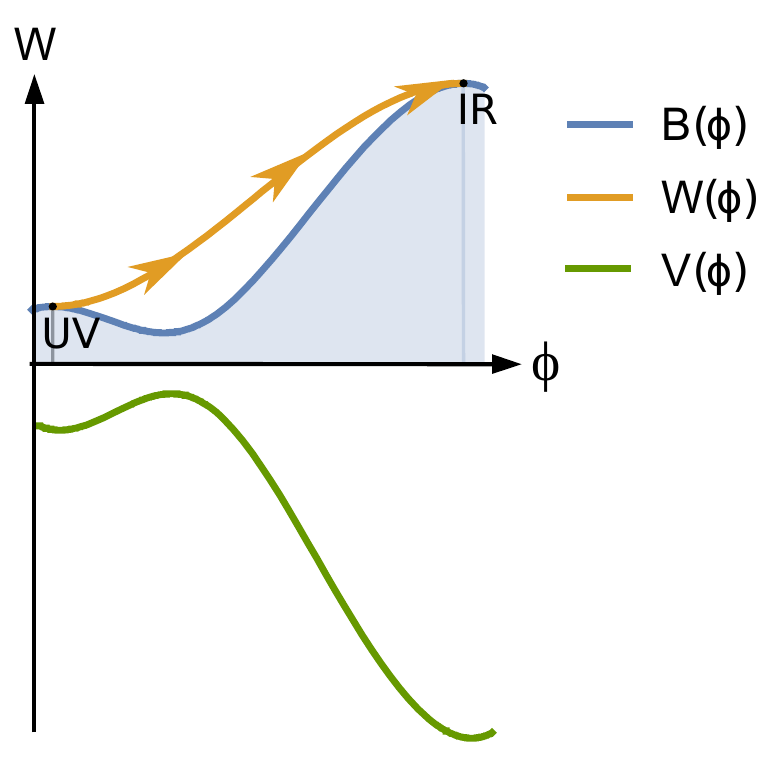}
\caption{
The potential \protect\eqref{ir}, the corresponding $B(\phi)$ and the superpotential \protect\eqref{ir2} interpolating between two minima are plotted for $v=k=1$, $d=4$ and $W_0=1.8$. In this particular case the local minimum of $B(\phi)$ (local maximum of $V(\phi)$)  between the two fixed points violates the BF bound and, therefore, does not correspond to a UV fixed point. This kind of solution, however, can also exist in cases when the BF bound is respected at the intermediate maximum of $V(\phi)$. }
\label{fig:W1}
\end{figure}

With the potential \eqref{ir},  equation \eqref{SuperP} admits as  its regular\footnote{More specifically, the resulting flow is asymptotically $AdS$ in the IR.} solution the following superpotential:
\eql{ir2}{
W(\phi)=kv^2\le(\phi\over v\ri)\le[
						1-{1\over 3}\le(\phi\over v\ri)^2
						\ri]+W_0
}
With a suitable range of parameters $W_0$, $k$ and $v$, the UV fixed-point and the IR fixed points are located at minima of $V(\phi)$, as shown in figure \ref{fig:W1}.
The corresponding scalar field profile is:
\eql{ir3}{
	\phi(u)=v \tanh(k~u.)
}

Other examples of this kind may be obtained constructively, using the inverse method (i.e. starting with the superpotential and then obtaining the  potential) described in Appendix \ref{app:inv}.

\begin{figure}[h!]
\centering
\subfigure[]{
\includegraphics[width=0.46\textwidth]{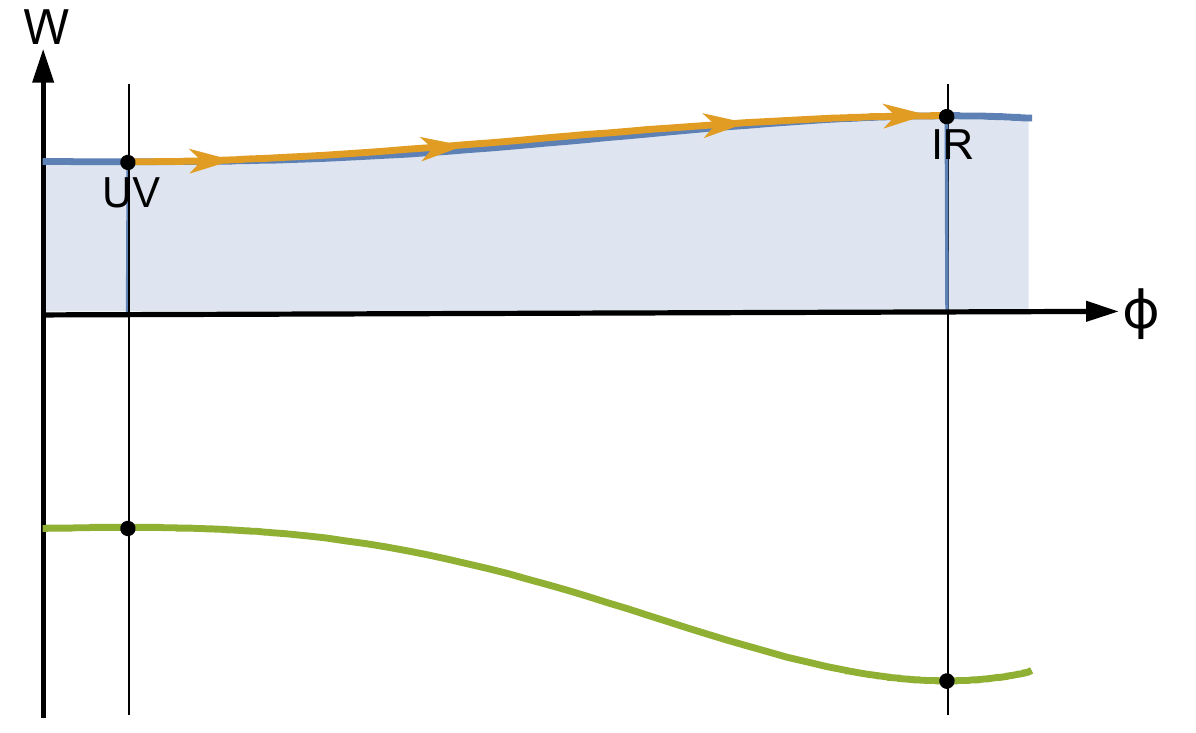}
}
\subfigure[]{
\includegraphics[width=0.49\textwidth]{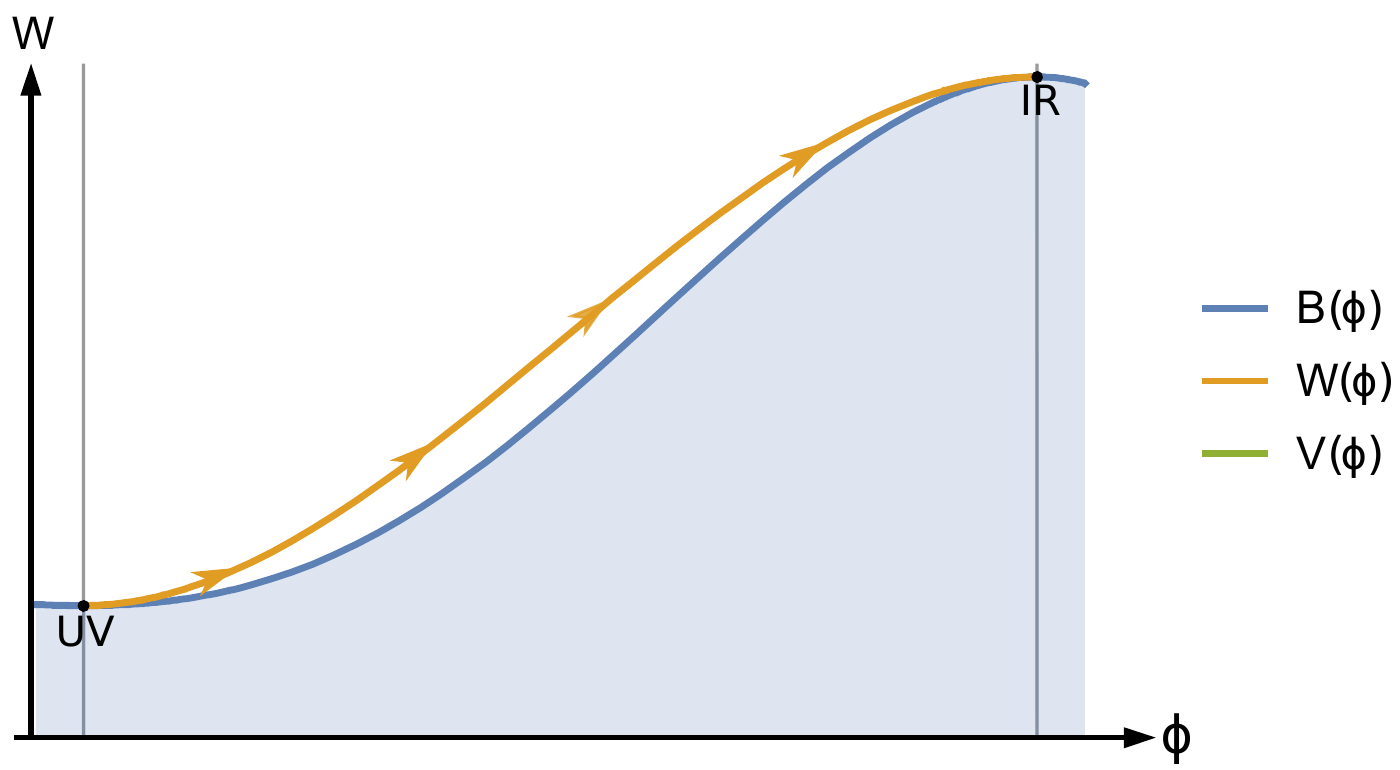}
}
\caption{
The potential \protect\eqref{ir}, the corresponding $B(\phi)$ and the superpotential \protect\eqref{ir2} are plotted for $v=k=1$, $d=4$ and $W_0=5$. Now $W(\phi)$ represents a standard holographic RG flow, interpolating between one local minimum of $B(\phi)$ and one local minimum.}
\label{fig:W2}
\end{figure}

We  have verified numerically that the local maximum which lies between $-v$ and $v$ violates the Breitenlohner-Freedman bound. We have not found explicit examples of flows between local minima of $V$ such that the intermediate  maximum respects the BF bound. However, there is no reason to believe they do not exist, since  all the basic elements elements  were shown  to exist separately (namely: a UV at a local minimum of $V(\phi)$ in the present section; flows skipping a local maximum of $V(\phi)$ which respects the BF bound, in section \ref{skip})  and one can imagine a flow displaying both features.

It is also possible to adjust $W_0$
so that the extremum of $V(\phi)$ at $\phi=-v$ becomes a local maximum. This is illustrated in figure \ref{fig:W2}.

\subsection{Bouncing solutions}
\label{ssec:bounces}

In subsection \ref{sssec:bW} we found that there are solutions of the superpotential equation \eqref{SuperP} where $W(\phi)$ is multi-valued. The points where the superpotential branches correspond to locations, which we called bounces, where $\dot \phi(u)$ changes sign. From the point of view of the RG flow, this means that the flow direction is reversed at a finite energy scale. As we have seen, the resulting geometry is regular at the turning point.  Bouncing solutions are those where $\phi(u)$ has at least one such a turning point.

In this subsection we construct   regular,  multi-branch solutions of equation \eqref{SuperP},  corresponding to bouncing  RG flows with endpoints at a UV and an IR fixed point.

In order to have a simple potential and sufficient freedom in choosing  parameters, we select a symmetric polynomial potential of order 8, with extrema at 0, $\pm\phi_0$, and $\pm\phi_1$, and mass at $\phi=0$ fixed so that  $\Delta(\Delta-d)$. The derivative of the potential is then fixed to be:
\eql{V8'}{
 V'(\phi):=-\phi\le(\phi^2-\phi^2_0\ri)\le(\phi^2-\phi^2_1\ri)\le(\phi^2-{\Delta(\Delta-d)\over\phi^2_0\phi^2_1}\ri),\quad 0<\phi_0<\phi_1.
}

We subsequently choose the integration constant that determines $V$ in such a way that the AdS curvature scale is set to one at the local maximum of V at $\phi=0$:
\begin{align}
		V(\phi)=&-d(d-1)+\Delta(\Delta-d){\phi^2\over2}
				-
				\le[
				\phi^2_0\phi^2_1+
				\Delta(\Delta-d) \le({1\over \phi^2_0}+{1\over\phi^2_1}\ri)
				\ri]{\phi^4\over 4}\nonumber \\
				&+\le[
				\phi^2_0+\phi^2_1-
				{\Delta(\Delta-d) \over \phi^2_0\phi^2_1}
				\ri]{\phi^6\over 6}-{\phi^8\over8}
		\label{V8}
\end{align}

The extrema are located at the zeroes of \eqref{V8'}: $0$, $\pm\phi_0$ and $\pm \phi_1$. The last term of equation \eqref{V8'} does not yield real roots because we are imposing $\Delta(\Delta-d)<0$ in order to have a local maximum of the potential at  $\phi=0$.
Specifically, we choose:

\be
	d = 4, \quad \Delta = 3,\quad \phi_0 =3.81261. \quad\phi_1=4\label{ParamV8}
\ee

\begin{figure}[h!]
\centering
\includegraphics[width=01\textwidth]{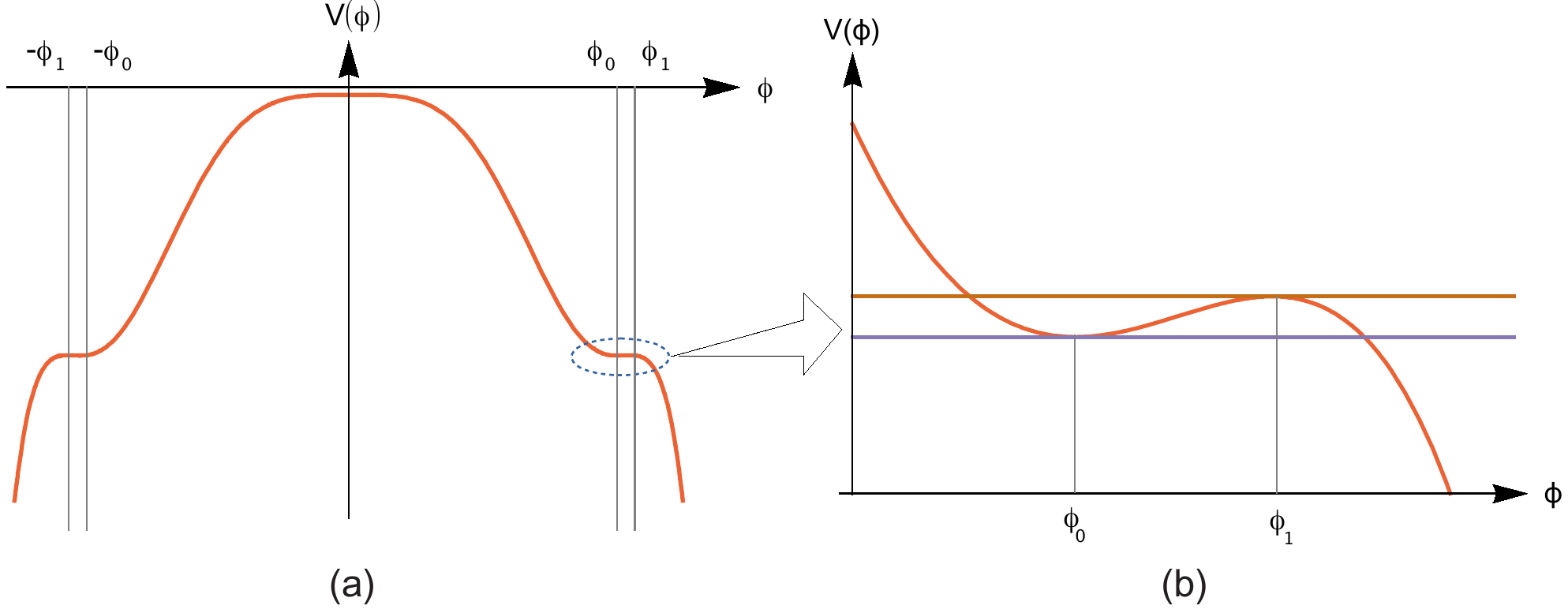}
\caption{
The potential V($\phi$) given by equation  \protect\eqref{V8} with parameters in equation \protect\ref{ParamV8}.}
\label{fig:phi8_00-2}.
\end{figure}

\begin{figure}[h!]
\centering
\includegraphics[width=0.6\textwidth]{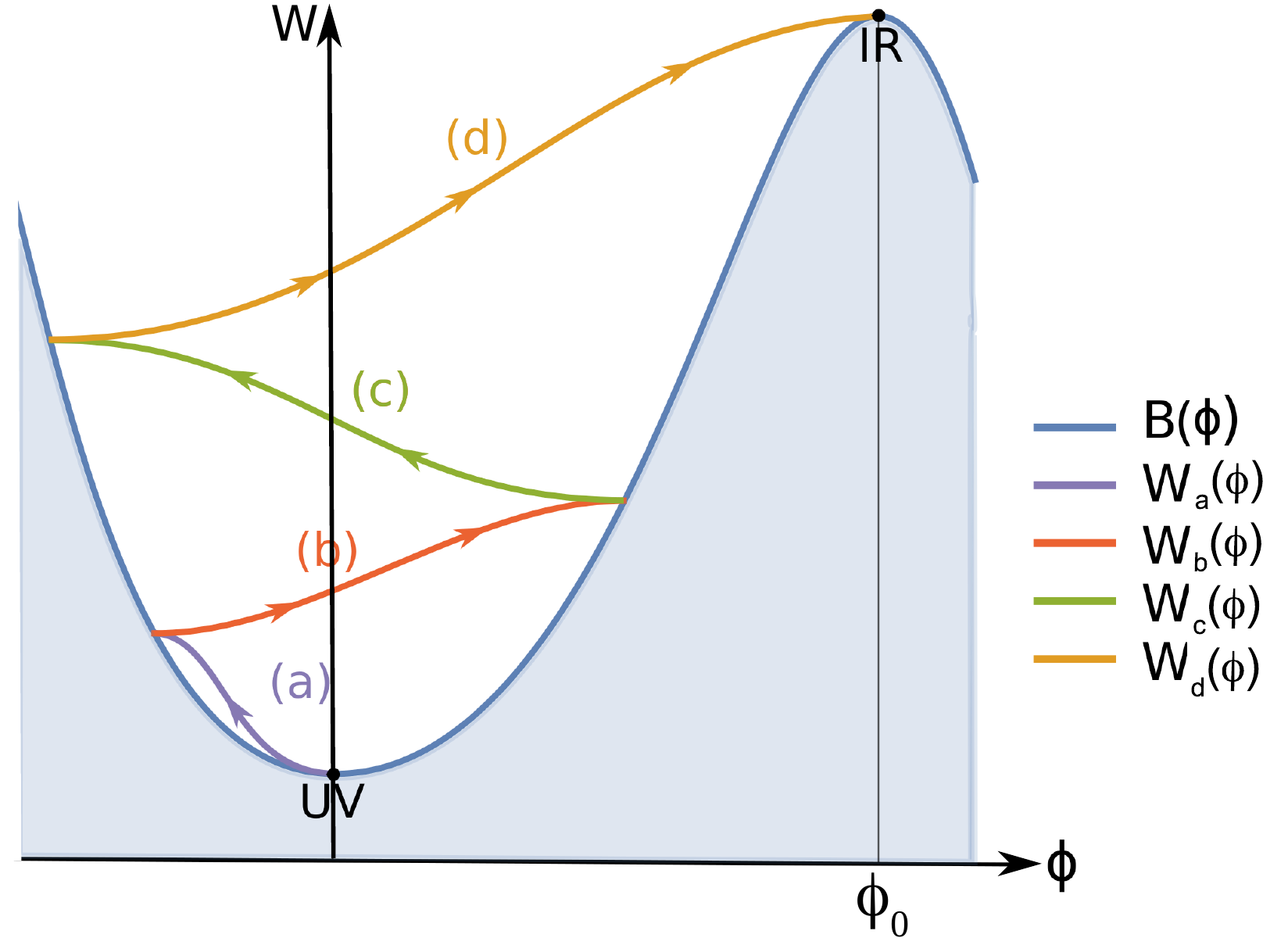}
\caption{Sketch of a bouncing solution where the superpotential has four branches. The arrows indicate the direction of increasing $u$. The actual numerical solution obtained using the potential (\protect\ref{V8}) is shown in figure \protect\ref{fig:phi8_01}.}
\label{fig:sketch}
\end{figure}

\begin{figure}[h!]
\centering
\includegraphics[width=0.46\textwidth]{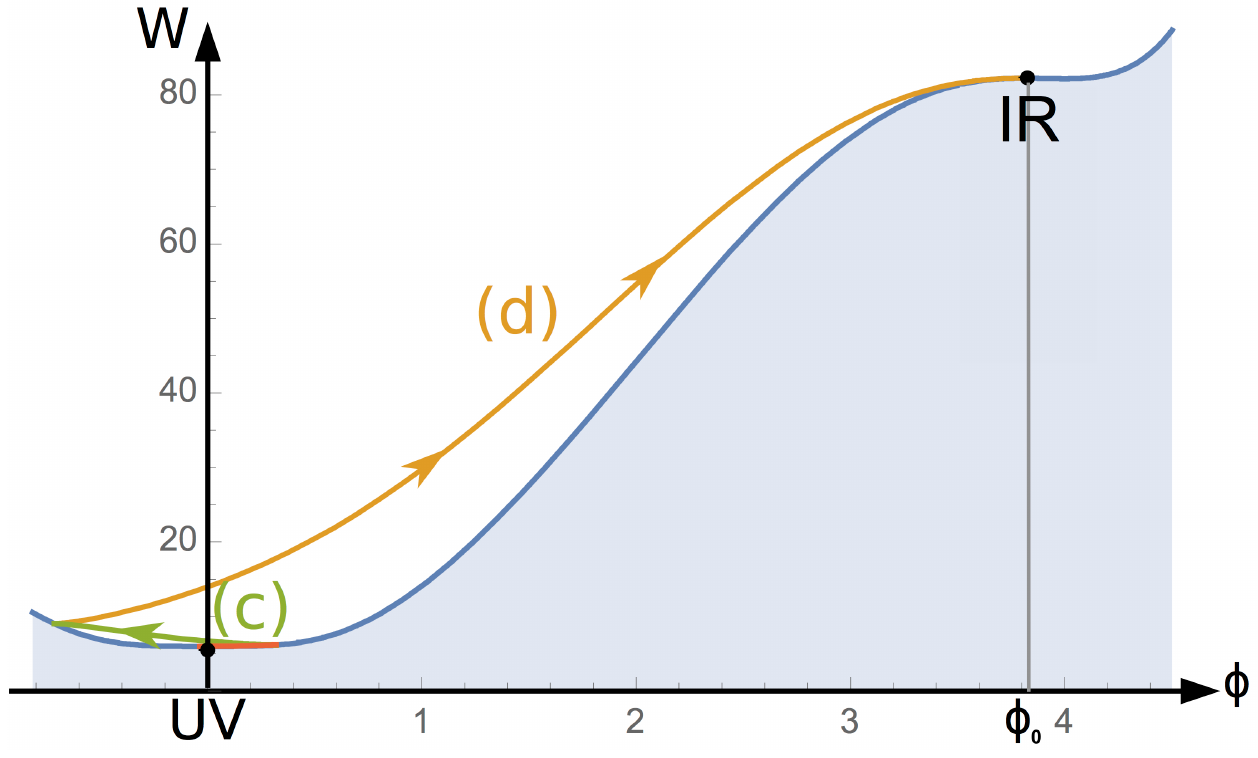}~~~~
\includegraphics[width=0.46\textwidth]{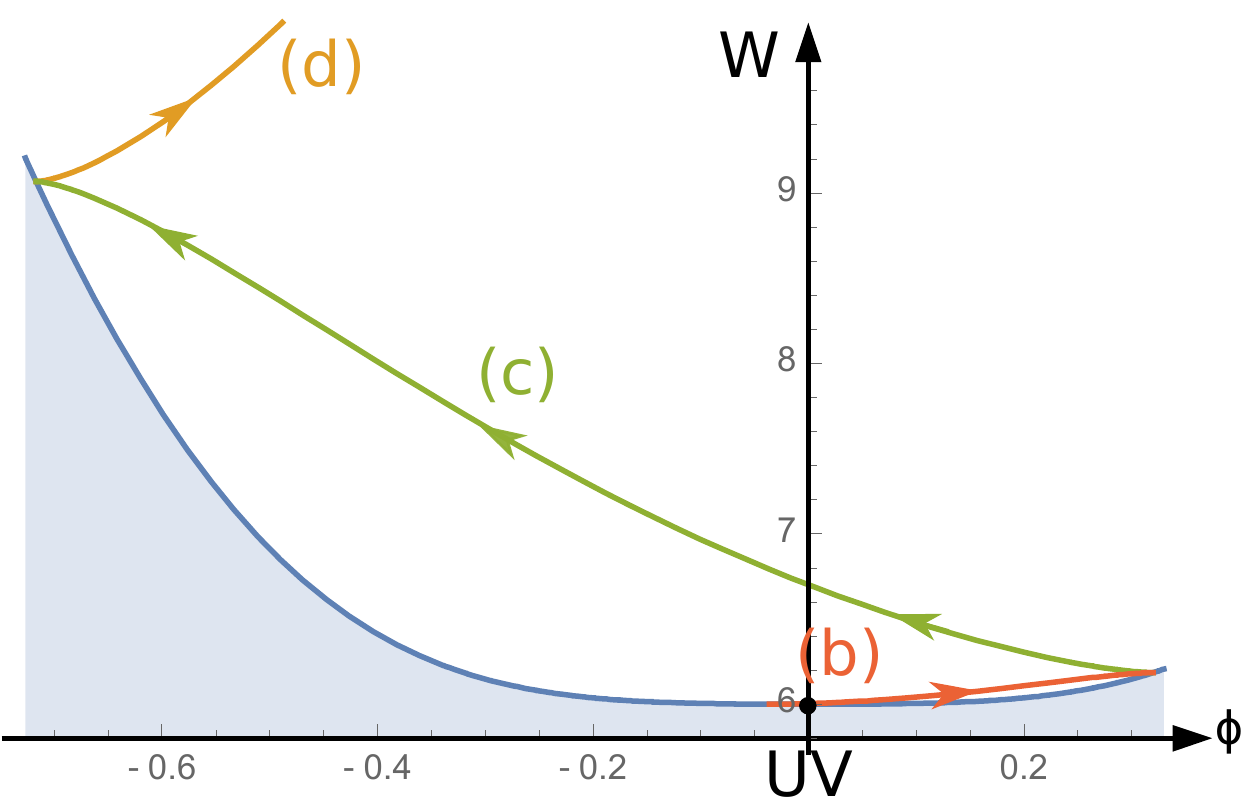}
\includegraphics[width=01.1\textwidth]{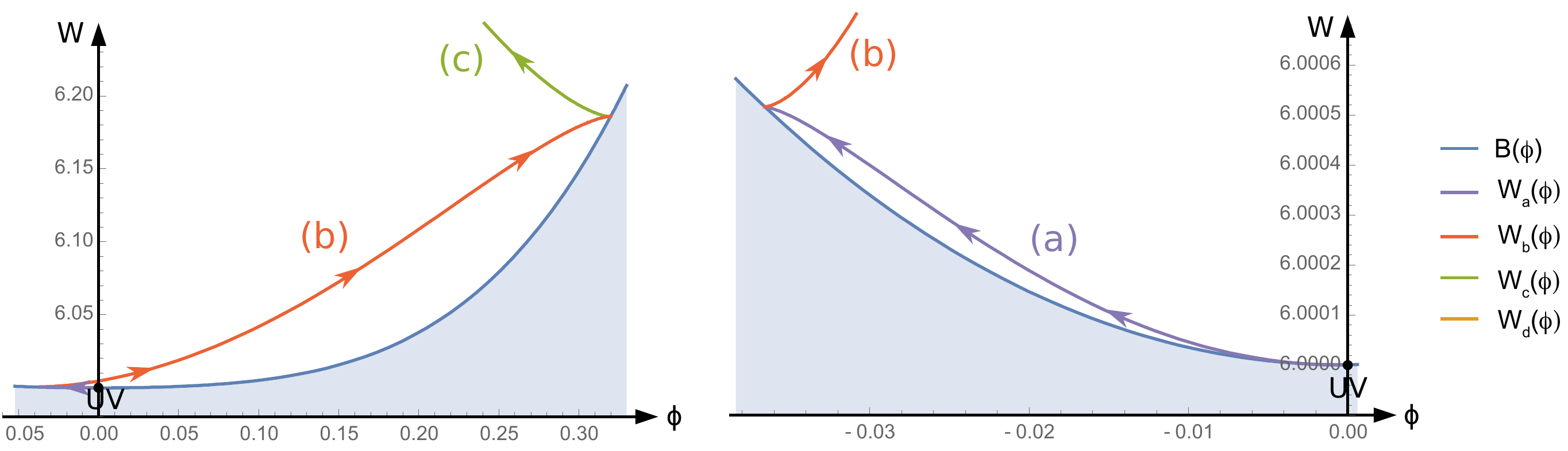}
\caption{
Here we plot a numerical solution which displays the same behavior as the sketch in figure \protect\ref{fig:sketch}. The potential V($\phi$) is given by equation \protect\eqref{V8} with parameters from equation (\protect\ref{ParamV8}). In the upper left figure we see the region near the infra-red fixed point. From  upper left to  lower right, the  plots  gradually zoom near the UV fixed point until we  see the last branch in the lower right figure. The curve in purple represents the branch of $W(\phi)$ that reaches the UV fixed point. For the scalar field profile and the scale factor see figures \protect\ref{fig:phi8_02} and \protect\ref{fig:phi8_03} respectively.}
\label{fig:phi8_01}
\end{figure}

\begin{figure}[h]
\centering
\includegraphics[width=0.6\textwidth]{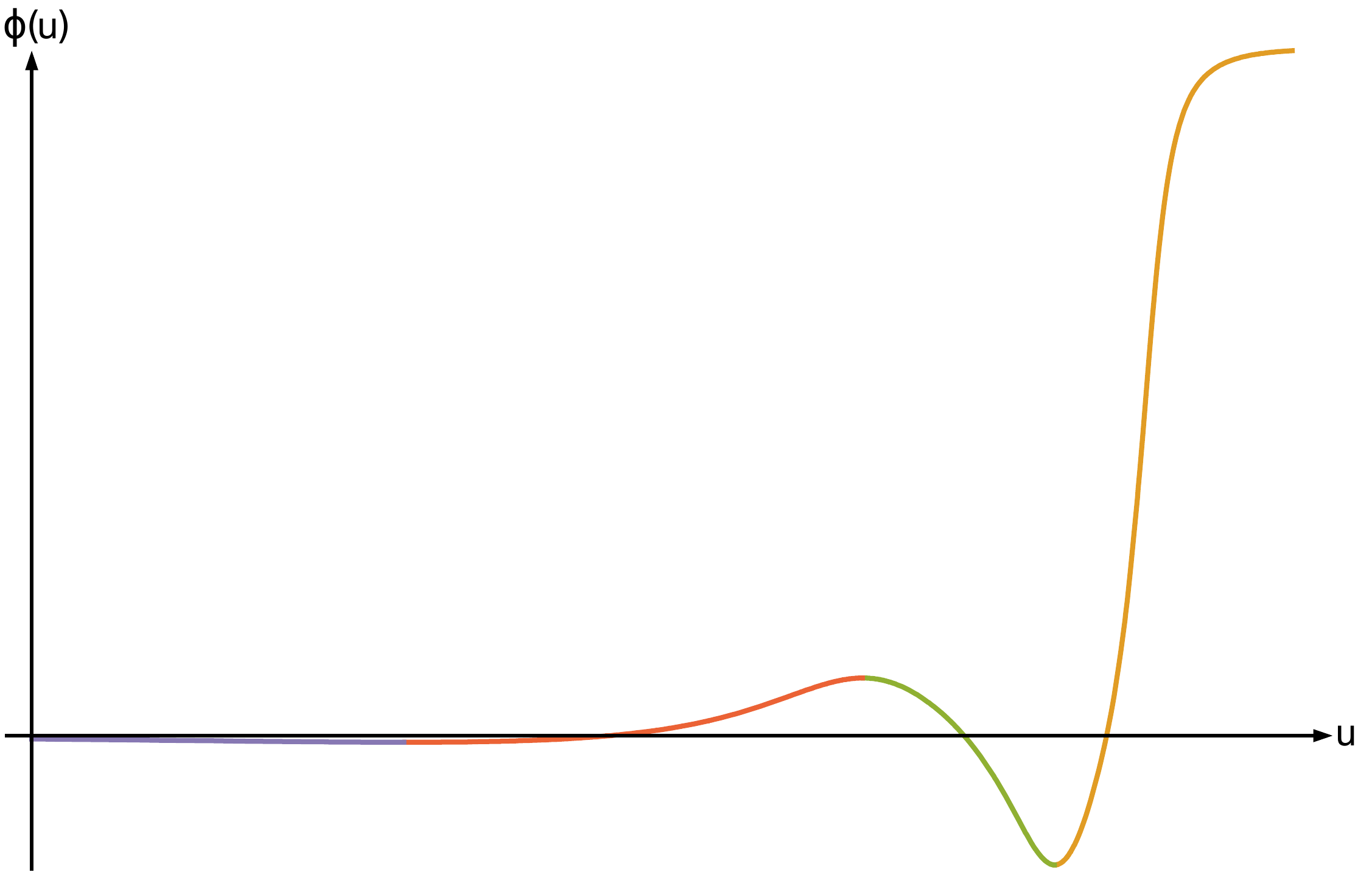}
\caption{The scalar field profile $\phi(u)$ corresponding to the holographic RG flow shown in figure \protect\ref{fig:phi8_01}. The solution leaves a UV fixed point where $u$ asymptotes to the negative infinity (purple curve), bounces three times, as indicated by the change of colors, and reaches an IR fixed point as $u$ asymptotes to the positive infinity (yellow curve). The corresponding scale factor is presented in figure \protect\ref{fig:phi8_03}.
}
\label{fig:phi8_02}
\end{figure}

\begin{figure}[h]
\centering
\includegraphics[width=0.6\textwidth]{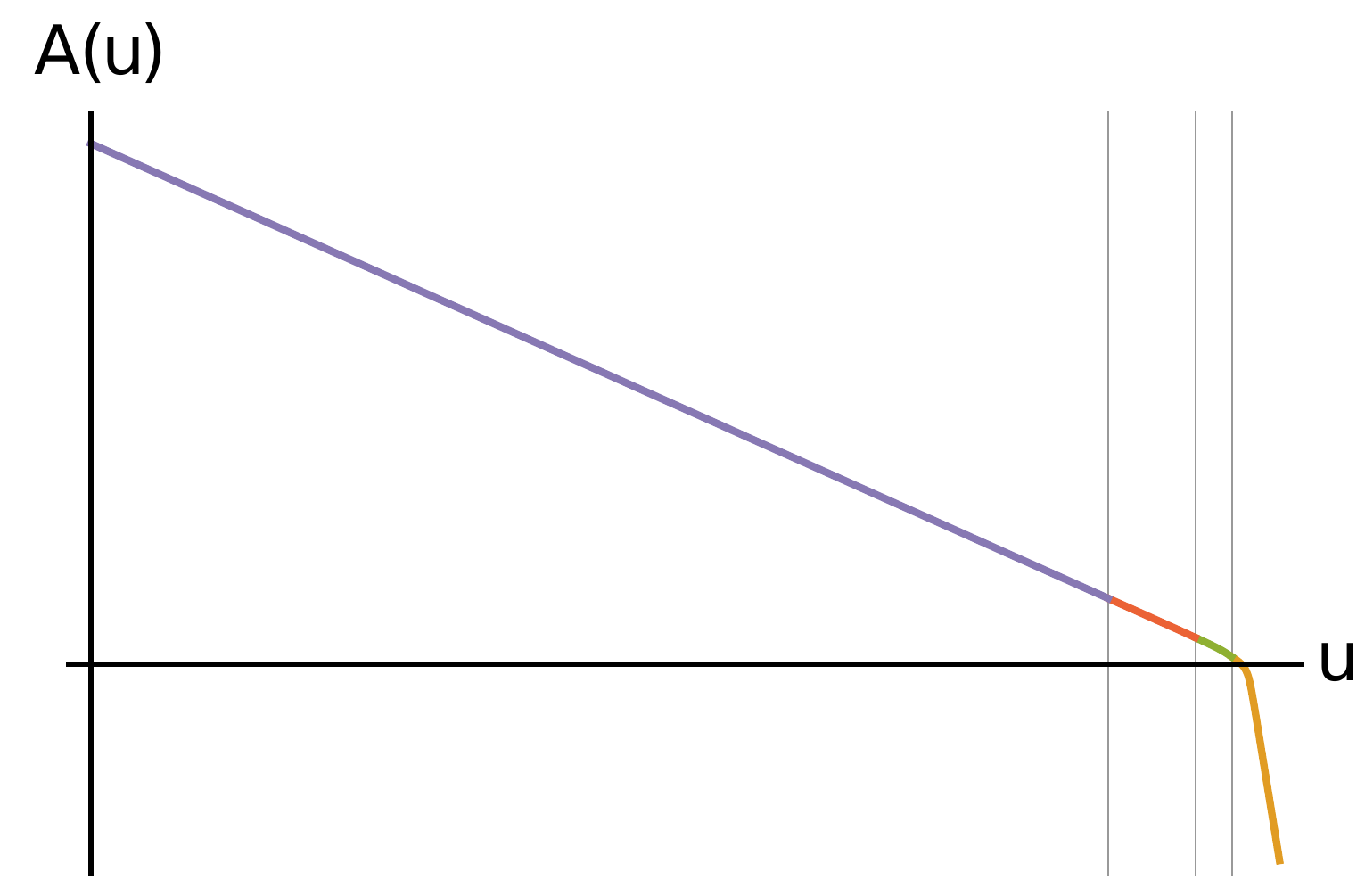}
\caption{The scale factor profile $A(u)$ corresponding to the solution shown in figure \protect\ref{fig:phi8_01}. The AdS length at the fixed points is inversely proportional to the asymptotic slope of $A(u)$, as $|u|$ asymptotes to infinity (see equations \protect\eqref{geo_max} and \protect\eqref{geo_min}). The asymptotic AdS length (i.e. the inverse of the slope of $A$) is smaller in the IR than in the UV, consistent with the holographic c-theorem.}
\label{fig:phi8_03}
\end{figure}

The resulting potential is plotted in figure  in figure \ref{fig:phi8_00-2}. With this choice, the model admits a regular flow  going through several  bounces.  More specifically,  the unique regular RG flow between the UV fixed point at $\phi=0$ and the IR fixed point at $\phi=\phi_0$  bounces three times,  and has four branches. A sketch of the complete flow is shown in figure \ref{fig:sketch} in order to make the four branches simultaneously visible, (unlike in the plot of the numerical solution). The actual superpotential   was obtained numerically and is  displayed in figure \ref{fig:phi8_01}. The corresponding scalar field and scale factor profiles are shown in figures \ref{fig:phi8_02} and \ref{fig:phi8_03}, respectively.

\subsection{Cascading solutions to BF bound-violating fixed points}
\label{ssec:casc}
As it is well-known, an extremum of the potential corresponds to a stable $AdS$ solution only if it satisfies the BF bound, \cite{BF82}:
\be
m^2\ell^2 > -{d^2 \over 4}.
\ee
This bound can be violated only at a local maximum of the potential and when this happens there are two consequences:
\begin{enumerate}
\item The UV AdS fixed point is unstable.
\item The RG flow does not admit a superpotential description with a fixed point at the extremum \footnote{Cf. equations  (\ref{BFV}-\ref{BFW}) and the related discussion.}.
\end{enumerate}

The existence of regular bouncing geometries however opens another interesting possibility: that  of {\em cascading} RG flows that bounce an infinite number of times towards a UV extremum violating the BF bound, without ever reaching it.

\begin{figure}[t]
\centering
\includegraphics[width=01\textwidth]{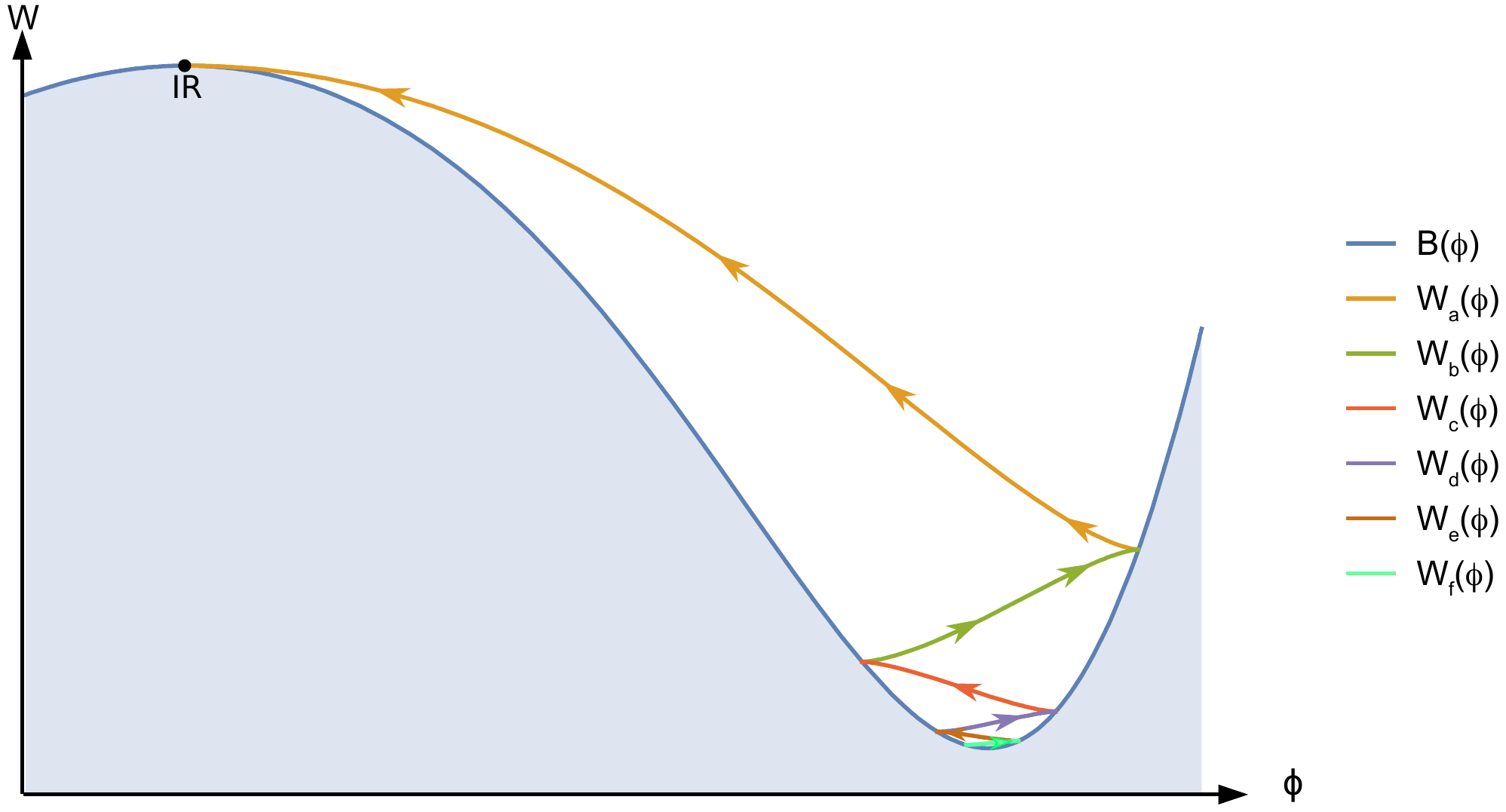}
\caption{The plot of a superpotential corresponding to  a cascading solution that reaches an IR fixed point (yellow curve) after bouncing an infinite number of times. Here only the first six branches are shown.
The potential used here is of the form \protect\eqref{V8} with
$d = 4$,
$\phi_0 = 7/4$,
$\phi_1 = 5/2$
and $\Delta(\Delta-d)=-4$.
The corresponding profiles of the scalar field and the scale factor are shown in figure \protect\ref{fig:cascfufA}.
}
\label{fig:cascW}
\end{figure}

\begin{figure}[t]
\centering
\includegraphics[width=0.5\textwidth]{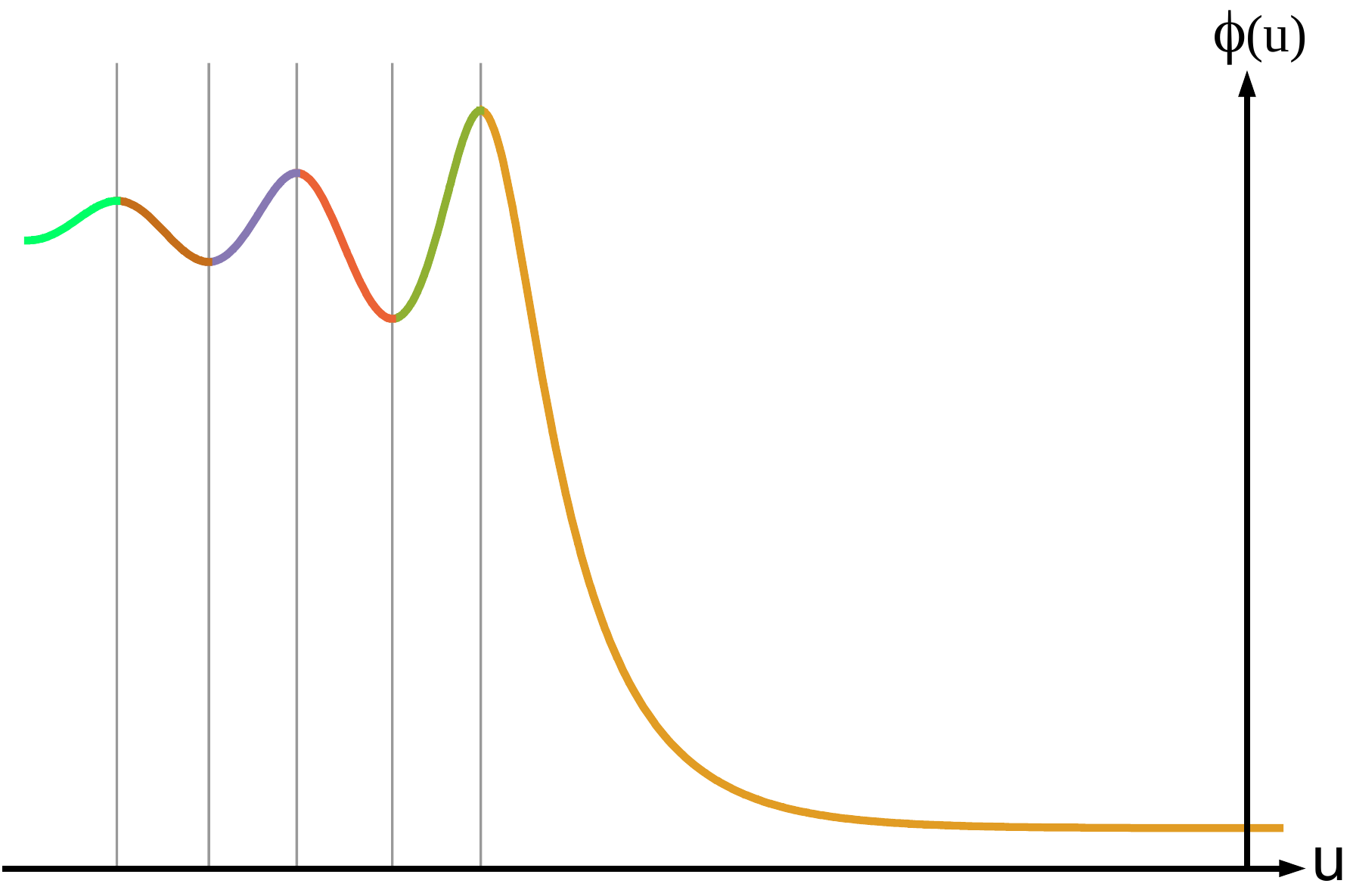}~~~~
\includegraphics[width=0.49\textwidth]{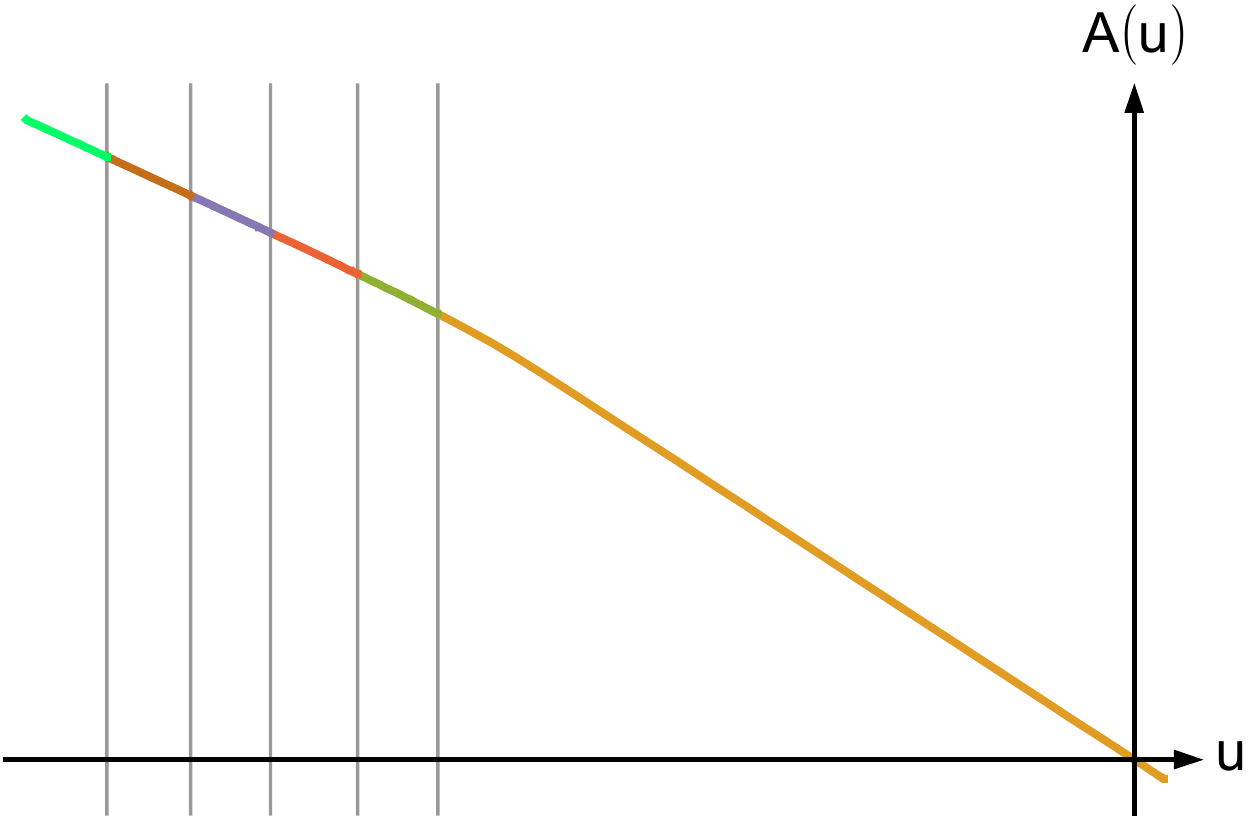}
\caption{The profiles of the scalar field $\phi(u)$ and the scale factor $A(u)$ corresponding to the first six branches of the cascading solution in figure \protect\ref{fig:cascW}. They  are both single-valued functions of $u$ that oscillate an infinite number of times as $u\to -\infty$. asymptotes to the negative infinity.}
\label{fig:cascfufA}
\end{figure}

The fact that  these solutions must exist is a consequence of the monotonicity of the superpotential as a function of $u$: if we start the superpotential equation with initial conditions in  the vicinity of a BF-violating maximum of $V(\phi)$, and we follow the solution backwards towards the UV, the superpotential will bounce off the boundaries of the forbidden region. However no solution $W(\phi)$ exists that ends at the fixed point, therefore it must continue bouncing indefinitely. This is confirmed by studying the solution for a scalar field close to a BF-bound-violating extremum, which exhibits an infinitely oscillating behavior:
\be
A(r) \simeq -\log {r\over \ell}, \quad \phi(r) \simeq \phi_0 r^{d/2} \cos \left[{|\nu |\over 2}  \log r + \varphi\right] \quad r\to 0.
\ee
where $\nu = \sqrt{4m^2\ell^2 + d^2}$ and $\phi_0$ and $\varphi$ are integration constants. This solution can extend to a full RG-flow away from $r=0$ and reach for example a regular IR fixed point. This is the case of the solution represented in figure \ref{fig:cascW}, where we show only six of the infinitely many branches of $W(\phi)$. The corresponding scalar field and scalar factor profiles are shown in figure \ref{fig:cascfufA}.

The existence of cascading solutions might suggest that one may, after all, make sense of BF-bound violating extrema in gauge/gravity duality, by excising them and replacing the UV fixed point by an infinite cascade, as in the Klebanov-Strassler  solution \cite{KS}: one could reach arbitrarily high energy by going up the cascade. Similarly to the AdS fixed point however, these cascading solutions are unstable against scalar fluctuations, as is shown in appendix \ref{BF}.

In conclusion, although they offer a glimpse of the behavior of the system near a BF-bound-violating extremum, infinitely cascading geometries must not be considered as part of the holographic landscape.


\subsection{Solutions skipping fixed points}
\label{skip}

\begin{figure}[h!]
\centering
\includegraphics[width=0.45\textwidth]{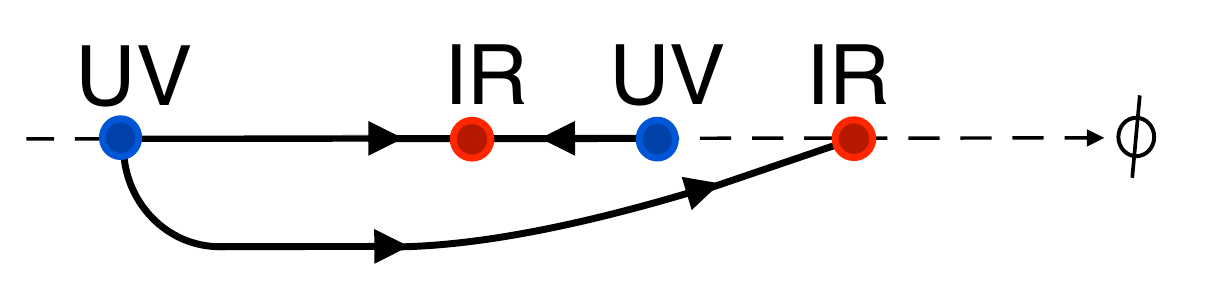}
\caption{Schematic structure an RG flow between non-neighboring fixed point. This structure is the same as  that of the  set of holographic solutions in figures  \protect\ref{fig:Skip2} and \protect\ref{fig:Skip3}.}
\label{fig:Skip}
\end{figure}

As a final example of ``exotic'' RG flows, in this subsection we present a solution interpolating between a UV fixed point and an IR fixed point, which skips an intermediate IR fixed point. This behavior, schematically represented in figure \ref{fig:Skip}, is normally not allowed in a first-order running of a single coupling as one has in standard perturbative field theories.

The existence of such solutions is suggested by the results of subsection \ref{ssec:bounces}. There, we showed  potentials which admits   solutions with bounces, where certain branches  skip an extremum of the potential without stopping there. As it is possible for a branch to skip a fixed point,  and this is a local property of the flow, one expects also situations in which an holographic RG flows interpolate between non-neighboring fixed points.

\begin{figure}[t]
\centering
\includegraphics[width=01\textwidth]{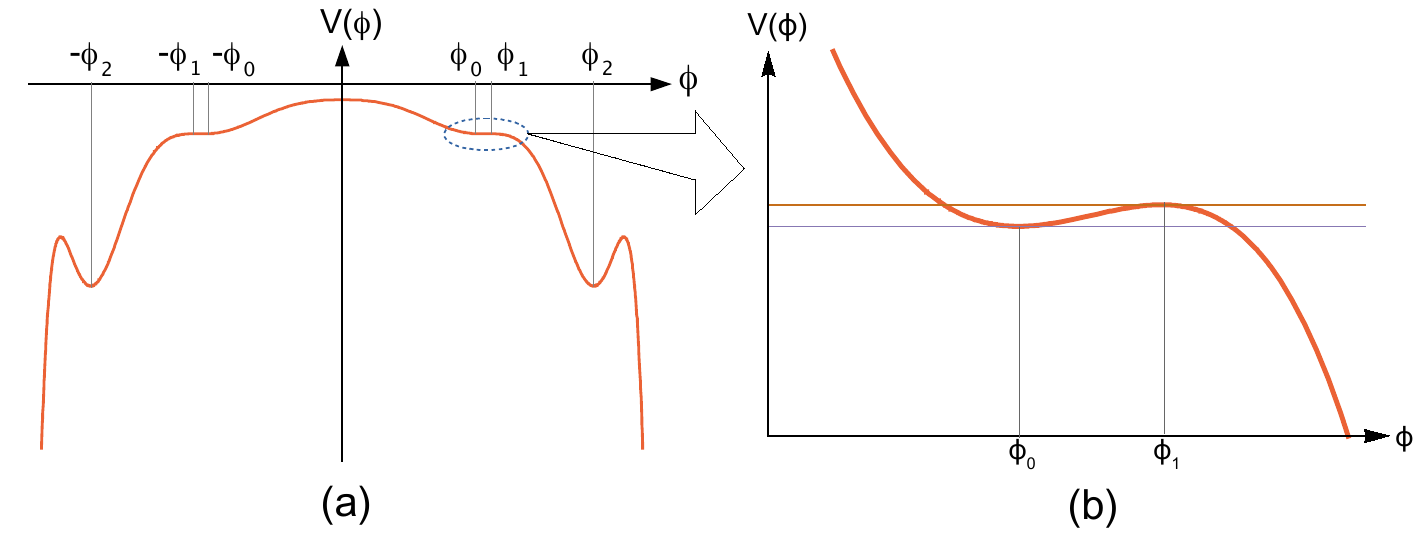}
\caption{Potential \protect\eqref{V12} with parameters from equation \protect\eqref{ParamSkip} and a zoom near the extrema $\phi_0$ and $\phi_1$, showing that there is indeed a local maximum of V at $\phi_1$.}
\label{fig:Skip0}
\end{figure}

As a  concrete example, we start with an $\cO(\phi^{12})$ potential and choose its extrema and its second derivative at $\phi=0$ through the factorized form of its first derivative:

\eql{V12'}{
	 V'(\phi):=-\phi\le(\phi^2-\phi^2_0\ri)\le(\phi^2-\phi^2_1\ri)\le(\phi^2-\phi^2_2\ri)\le(\phi^2-\phi^2_3\ri)\le(\phi^2-{\Delta(\Delta-d)\over\phi^2_0~\phi^2_1~\phi^2_2~\phi^2_3}\ri),
}
with $0<\phi_0<\phi_1<\phi_2<\phi_3$.
We choose the AdS length at the origin to be one:
\eql{V12}{
		V(\phi)=-d(d-1)+\int_0^\phi V'(x) dx.
}
The point  $\phi=0$ is a local maximum with fixed second derivative:
\eql{V12''}{
	 V''(0):=\Delta(\Delta-d)<0.
}
Because of equation \eqref{V12''}, the potential \eqref{V12} has extrema at $$\phi=0,\pm\phi_0,\pm\phi_1,\pm\phi_2,\pm\phi_3\;\;.$$
The explicit form of the potential in terms of $\phi_i$ and $\Delta$ is given in appendix \ref{app:V12}.
The parameters used in the numerical calculations are chosen to be:
\begin{align}
	&d = 4\quad \phi_0 = 1.0837\quad \phi_1= 1.1316 \nonumber\\
	&\Delta = 3  \quad \phi_2 =1.9200 \quad\phi_3 = 2.1500\label{ParamSkip}
\end{align}
and  the resulting potential is shown in figure \ref{fig:Skip0}.

\begin{figure}[t]
\centering
\includegraphics[width=1\textwidth]{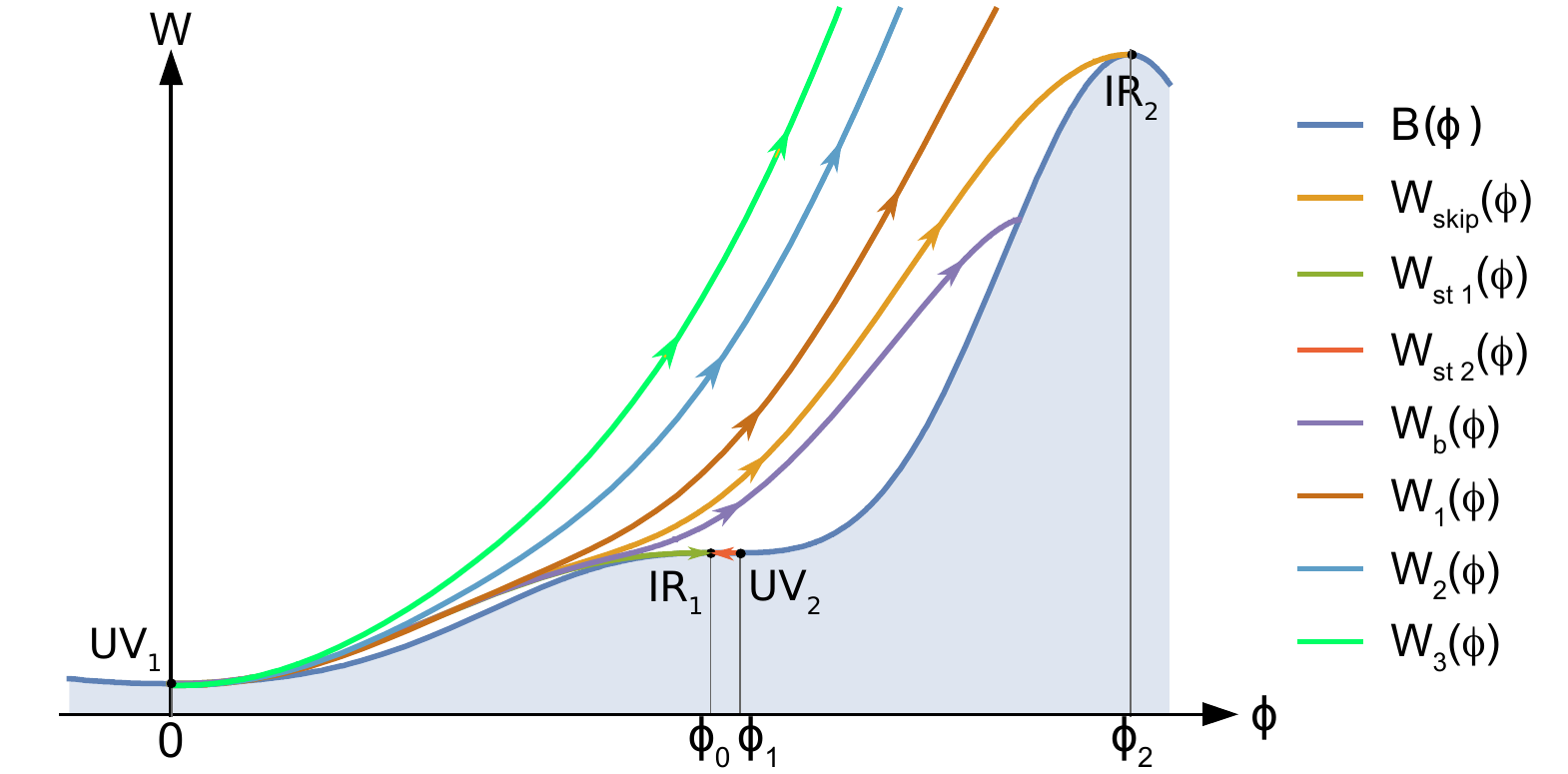}
\caption{Some of the  solutions $W(\phi)$ associated with the potential (\protect\ref{V12}),  leaving the same UV fixed-point at $\phi=0$. Most of them miss either fixed point (e.g. $W_1,W_2,W_3$) or hit the critical curve (e.g. $W_b$) leading to an IR singularity, or a  bounce. The three solutions $W_{st 1}$ and $W_{skip}$ reach two different IR fixed points, at $\phi_0$ and $\phi_2$ respectively. Also shown, is  the solution $W_{st 2}$, which leaves the  UV fixed point at $\phi_1$ and flows backwards to the IR fixed point at $\phi_0$ (displayed more in detail in figure \protect\ref{fig:Skip3}.)}
\label{fig:Skip1}
\end{figure}

\begin{figure}[t]
\centering
\includegraphics[width=01\textwidth]{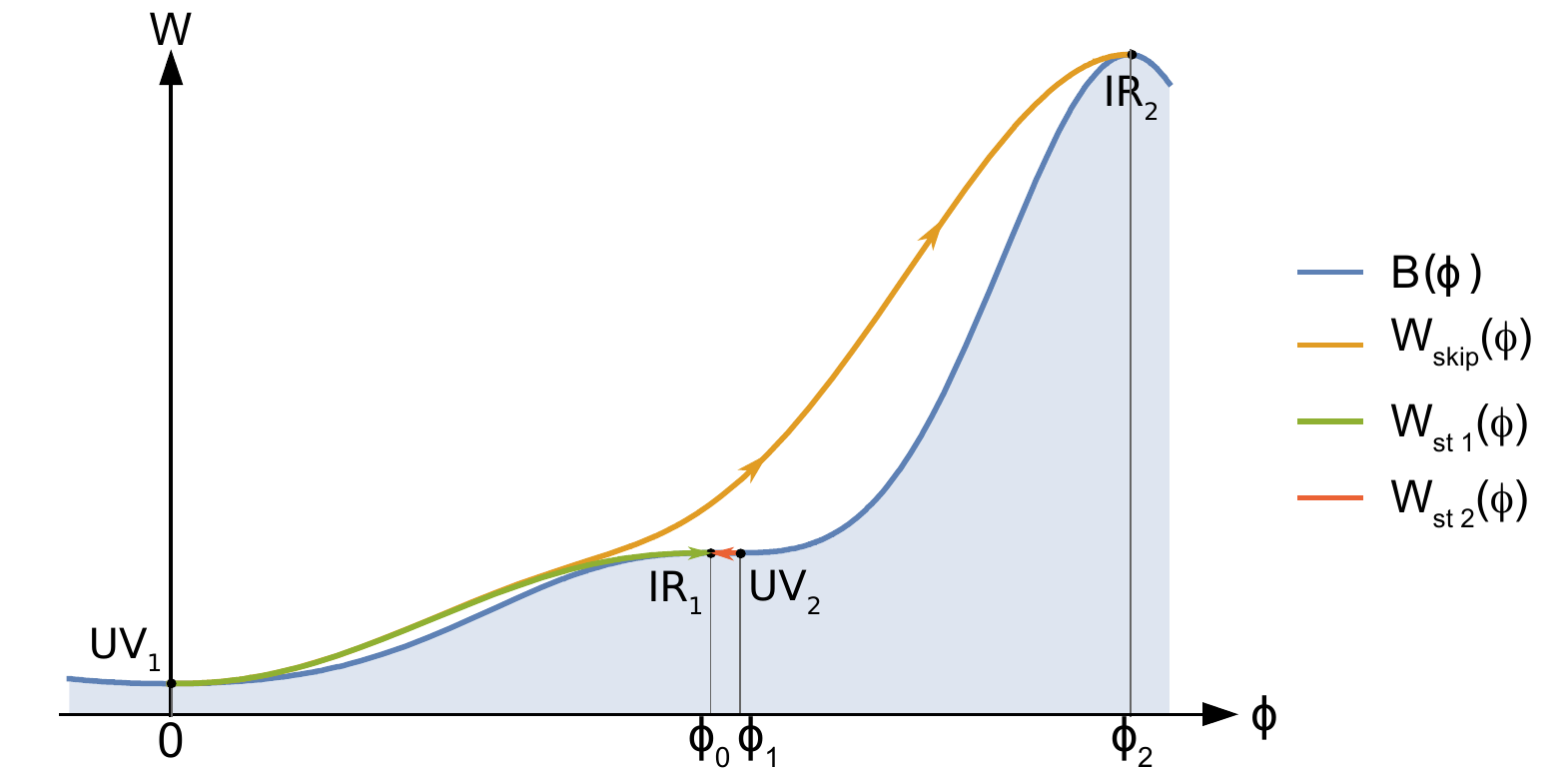}
\caption{Among the curves represented in figure \protect\ref{fig:Skip1} there are three which correspond to regular RG flows. The solution $W_{st 1}(\phi)$, represented by the green curve, goes to the nearby extremum, just as does  $W_{st 2}(\phi)$: these are standard RG flows. The yellow curve, $W_{skip}(\phi)$, skips the IR fixed point reached by the green curve and increases until it arrives at another IR fixed point. The potential V($\phi$) is given by \protect\eqref{V12} with parameters from equation \protect\eqref{ParamSkip}. }
\label{fig:Skip2}
\end{figure}

In figure \ref{fig:Skip1} we show some of the associated superpotentials.   Only three of them, i.e. those selected in figure \ref{fig:Skip2},  correspond to IR-regular flows, and one of them,  $W_{skip}(\phi)$,  skips the intermediate IR fixed point at $\phi_0$ and reaches to the next IR fixed point at $\phi=\phi_2$.

\begin{figure}[t]
\centering
\includegraphics[width=0.8\textwidth]{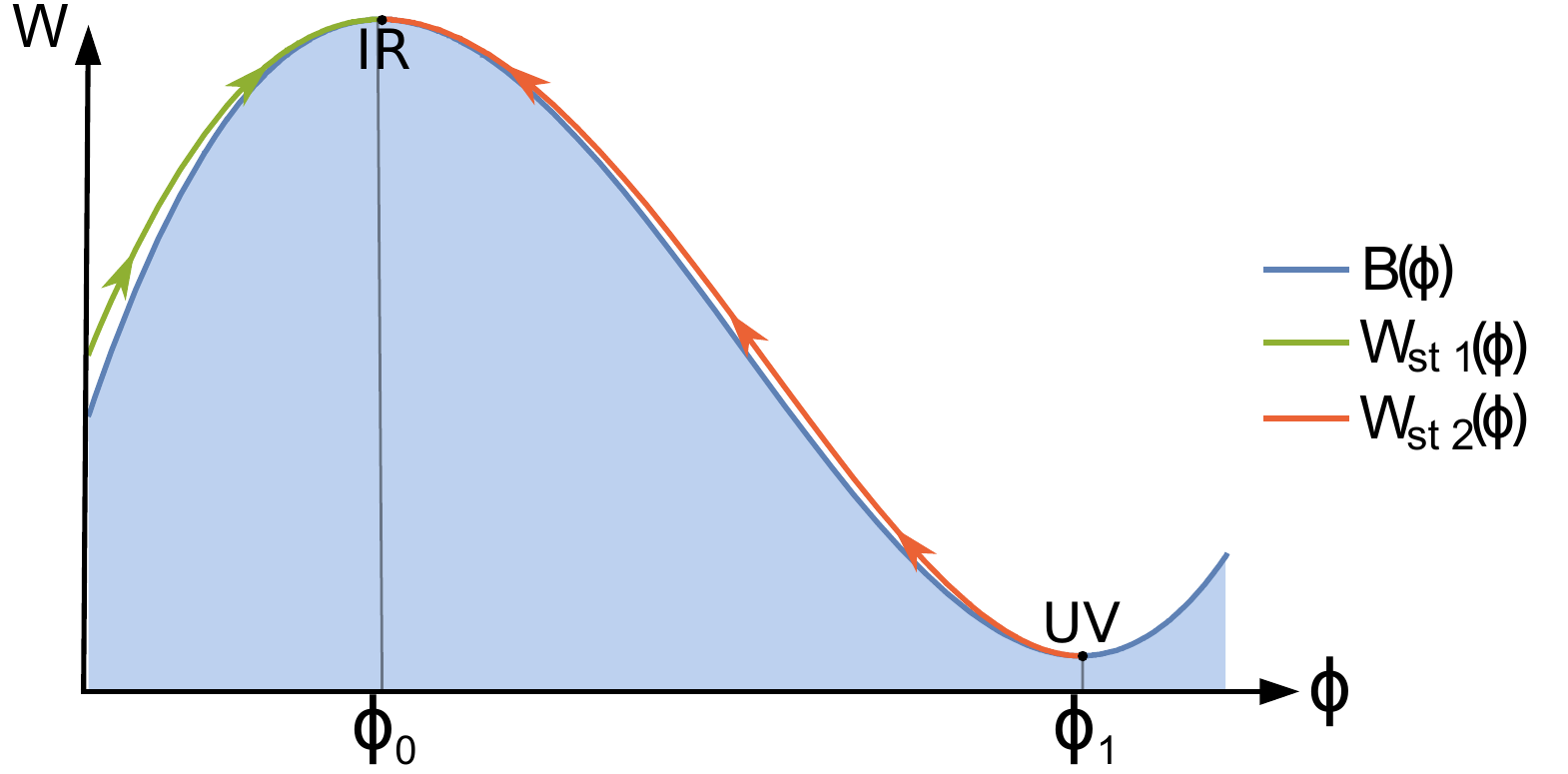}
\caption{Detail of figure \protect\ref{fig:Skip1}  showing the central IR fixed point $\phi_0$ and of the flows reaching it.}
\label{fig:Skip3}
\end{figure}

By choosing the extrema $\phi_0$ and $\phi_1$ to be very close,  it is possible for the extremum at $\phi_1$  to respect the BF bound.  Indeed, there exists a flow interpolating from $\phi_1$ in the UV to $\phi_0$ in the IR (solution $W_{st 2}$ in figure \ref{fig:Skip2},  shown more in detail in figure \ref{fig:Skip3}). However, there cannot exist a flow from $\phi_1$ to $\phi_2$, since the latter IR fixed point is already reached from the left by the flow $W_{skip}$. As we have seen in Section 2, an IR fixed point can be attained by only a single flow from each direction, therefore the existence of $W_{skip}$ forbids a flow from $\phi_1$ to $\phi_2$.

In general, given a bulk potential,  it is hard to pinpoint the exact conditions for the existence a solution skipping an intermediate  fixed point. However, some general statements can be made.   For concreteness, we consider the case $W'(\phi)\geqslant0$. To be relevant for the discussion,  the potential must allow for a first  UV fixed point, denoted by  UV$_1$ and two IR fixed points, which we denote by IR$_1$ and IR$_2$. This in turn implies the existence of another local maximum of $V$ between the two minima, which  we denote by UV$_2$.
We are interested in the existence of flows starting at UV$_1$, skipping IR$_1$,  and ending at IR$_2$ (therefore necessarily  skip UV$_2$ as well), as schematically shown in figure \ref{fig:Skip}.
This for example is the scenario with two UV and two IR fixed points  shown in figure \ref{fig:Skip2}.

Two approximate but general necessary conditions for the existence of fixed point-skipping solutions can be  extracted from the analysis of the previous sections:
\begin{itemize}
\item The first condition is that the IR fixed point where the flow ends, IR$_2$ is not reachable by any flow starting at UV$_2$.
This condition is necessary because a flow ending at a given IR fixed point is unique. When the condition above does not hold, there is either a standard or a bouncing RG flow from UV$_2$ to IR$_2$ and no solution skipping IR$_1$ will end at IR$_2$.

\item Another  condition is that the $W_+$ solution leaving UV$_1$  should   either skipping IR$_2$ or end at IR$_2$. If the $W_+$ solution skips IR$_2$, then there will be one of the infinitely many $W_-$ solutions that will reach IR$_2$. This is the case for  the numerical solution shown in figure \ref{fig:Skip1}. Experience with numerical solution shows that,  for this to happen, the local minimum of $V(\phi)$ corresponding to IR$_2$ should not be too deep compared to the neighboring maxima.
\end{itemize}


\section{Flows to the boundary of field space}
\label{Francesco}
In the previous sections we have considered holographic RG flows between two fixed points at finite $\phi$ (eventually going through a finite number of bounces). In these solutions, the range of field space explored is bounded, and the IR is regular (it  asymptotically approaches the interior of  $AdS$ ).

This is not the most general situation however: often, one finds solutions that extend all the way to $\phi \to \pm \infty$. The simplest such example is one in which the potential has only a maximum at the origin but no other extrema, for example
\be \label{inf000}
V(\phi) = -\cosh \gamma \phi,
\ee
as one often finds in truncations of higher dimensional supergravity theories. In this case, the running of $\phi$ away from the origin may be monotonic and $\phi$ may asymptote to infinity\footnote{This typically gives rise to scaling solutions which although not conformal, they have a scaling behavior. This behavior is known to correspond to violations of hyperscaling in condensed matter models, \cite{cgkkm,sa}.} .

The solutions of the type described above, in which $\phi$ diverges, are typically singular. As we observed in subsection \ref{ss:gen_prop}, item \ref{it:reg}, solutions without curvature singularities are the ones in which both $V$ and $W$ stay finite.
Therefore, if $\phi$ reaches infinity, the curvature generically diverges either because $V$ diverges or, even in the case where $V$ stays finite, because $W$ diverges, as we will see in the following subsection.

The fact that the solutions in this case are generically singular does not mean that they are all not meaningful in holography: indeed, it has been argued that under certain conditions, one can still make sense of the geometry in terms of a dual field theory, \cite{Gubcrit}. This includes the important case when the IR theory is confining with a mass gap (and therefore the flow  cannot end at a non-trivial  IR fixed point)

Here,  we will discuss two ``good singularity'' conditions:
\begin{enumerate}
\item {\bf Gubser's criterion:} A good singularity is such that it can be hidden behind an event horizon by an infinitesimally small deformation of the background geometry.
\item {\bf Spectral computability criterion:} A singularity is ``computable" if the spectral problem for the small fluctuations around the solution is well posed, i.e. it is uniquely determined  without the need of  extra boundary conditions at the singularity. This implies that the low-energy correlators are well defined and computable without the need to resolve the singularity, \cite{gkn,thermo}.
\end{enumerate}
We will review these two criteria below and translate them into the superpotential language.

Therefore, among the acceptable RG flows we should include those that, rather than ending at an  IR fixed point, reach infinity in field space, as long as the singularity satisfies one (or both) of the criteria stated above. In the following section we first classify the solutions in terms of the superpotential, then we translate the conditions above in the language of geometry.

\subsection{Superpotential equation for  ($\phi\to\pm \infty$)}
\label{scaling}

We now examine solutions of equation \eqref{SuperP} which reach the asymptotic region $\phi \to \pm \infty$. This section is essentially a generalization of the analysis performed in Appendix E of \cite{thermo} to arbitrary asymptotics of the potential at $\pm \infty$.

We remind the reader that we always suppose $W>0$. For definiteness, here it is sufficient to  discuss the case $W'>0$, i.e. the  $W_\uparrow$ branch. Indeed, the case $W'<0$ can be obtained by exchanging the two boundaries of field space i.e. $\phi \to +\infty$ and $\phi \to -\infty$. More precisely,  the properties of the growing branch of $W$ as $\phi\to +\infty$ for the potential $V(\phi)$  are the same as those of the decreasing branch for the potential $V(-\phi)$.

We have two distinct kinds of behavior as $\phi \to +\infty$ and as $\phi \to -\infty$. Furthermore we assume as usual the scalar potential $V(\phi)$ has a definite negative sign throughout field space.

It is convenient to rewrite the superpotential equation (\ref{SuperP}) as:
\be \label{inf00}
W^2 -  Q^2\left({dW \over d\phi} \right)^2 = B^2, \qquad B(\phi) \equiv \sqrt{-4{(d-1) \over d} V(\phi)},
\ee
where have introduced the notation:
\be \label{inf01}
Q \equiv\sqrt{{2(d-1) \over d}}.
\ee

\subsubsection{Behavior as $\phi \to +\infty$.}  \label{ssec:ir}

Assume  $\phi$ reaches $+\infty$. Recall that we are discussing the $W_\uparrow$ branch, so that $\phi \to +\infty$ corresponds to the IR of the geometry.

According to our general discussion in Section \ref{ss:gen_prop}, the superpotential  must obey the bound:
\be\label{inf1a}
W(\phi) > B(\phi),  \qquad \phi \to +\infty.
\ee
 There are two types  of solutions for the superpotential equation under these assumptions, depending on whether,  as $\phi \to +\infty$,  $W/B \to +\infty $ (continuous branch) or $W/B \to constant$ (special solution), which we discuss separately below (the third possibility, i.e. $W/B \to 0$, is inconsistent with the requirement (\ref{inf1a})).

\paragraph{Continuous branch.} These are the solutions for which $W \gg B$ as $\phi \to +\infty$. In this case, the superpotential equation can be approximated by:

\eql{LargeWsol}{
	W'(\phi)\approx {1\over Q} W(\phi),
}
whose solution takes the form:
\be \label{inf1b}
W(\phi)\approx C \exp\le[{1\over Q} \, \phi \ri] \qquad  \phi \to +\infty
\ee
where $Q = \sqrt{2(d-1)/d}$ as in (\ref{inf01}).
The solution  depends on an arbitrary integration constant $C$, and the scaling behavior does not depends on the details of the potential at $+\infty$, as long as the condition $W(\phi) > B(\phi)$ is satisfied. Since $B(\phi) \propto \sqrt{|V(\phi)|}$, these solutions exist only if the asymptotic behavior of $V(\phi)$ is slower than  $\exp\left[ \sqrt{2d/(d-1)}\phi\right]$ as $\phi \to +\infty$: if this is not the case, equation (\ref{inf1b}) is inconsistent with the condition that $W > B$ at infinity.

\paragraph{Special solution.}
Suppose instead that $W/B$ approaches a constant $\alpha \geq 1$ at infinity:
\be \label{inf2}
W(\phi) = \alpha B(\phi)  \, + \, sub-leading,  \qquad \phi \to \infty.
\ee
In this case,  equation (\ref{inf00}) fixes the value of $\alpha$ to be:
\be \label{inf3}
\alpha = \sqrt{1\over 1- k^2Q^2}, \qquad k\equiv \lim_{\phi\to +\infty} {1\over B}{dB \over d\phi}.
\ee

The  solution above is consistent only if $0\leqslant k < Q^{-1}$, i.e. again if the potential has a slower growth than  $\exp\left[ \sqrt{2d/(d-1)}\phi\right]$.

 If the potential stays finite  as $ \phi \to +\infty$, or it grows slower than an exponential, then $\alpha = 1$.

Finally, if the potential goes to a constant (including zero) in such a way that  $dB/d\phi <0$ asymptotically, then equation (\ref{inf2}) would imply that $W'<0$, contrary to our assumption that $+\infty$ corresponds to the IR. Indeed, it would correspond to a UV reached at infinity\footnote{To understand this case we can refer to the next subsection where we discuss the runaway UV as $\phi \to -\infty$.}. In this case the special IR solution (\ref{inf2}) does not exist.

The special property of the  solution (\ref{inf2}) is that it is isolated, i.e. it does not admit continuous deformations:

Following the procedure  in Appendix  \ref{app:def}, the deformation of a solution $W_0$, if it exists, is given by equation (\ref{def3}).   Under  the assumption (\ref{inf2}) for $W_0$, and the fact $B/B' \to k$, equation  (\ref{def3}) gives:
\be \label{inf5}
\delta W \simeq C \exp\left[{1\over Q^2 k} \phi\right]
\ee
Since $k < 1/Q$, we  see that  $\delta W/W_0 \to +\infty$, which is incompatible with the assumption that $\delta W$ is a small deformation.  Therefore, a solution with asymptotics (\ref{inf2}) does not admit continuous deformations, contrary to the solutions in the continuous branch.

We can summarize this discussion by concluding the following:

\begin{itemize}
\item If $V(\phi)$ grows {\em faster than $\exp (2Q^{-1} \phi)$} at infinity, there are no solutions reaching $\phi \to +\infty$: rather, all solutions bounce at a finite value of $\phi$ (which can be however arbitrarily large, in principle).
\item If  $V(\phi)$ stays finite or grows {\em slower than $\exp (2Q^{-1} \phi)$} at infinity, then there exist two types of solution that reach infinity: \begin{enumerate}
\item A continuous one-parameter family of superpotentials behaving as
\be\label{inf6}
W_C(\phi) \simeq C \exp {1\over Q} \phi \qquad Q \equiv \sqrt{2(d-1)\over d}
\ee
\item A {\em single special solution} $W_*$ behaving as
\be \label{inf7}
W_*(\phi) \simeq \alpha B(\phi) \propto \sqrt{-V(\phi)} ,
\ee
where $\alpha$ is completely fixed by the behavior of $B(\phi)$ at infinity, as in equation (\ref{inf3}).
\end{enumerate}
\item   From \eqref{inf7}, when the  $W_*$ solution exists, it  is sub-leading w.r.t. $\exp( Q^{-1}\phi)$,  therefore {\em the special solution lies below all the solutions in the continuous family}.
\item Finally, since all solutions reaching infinity are in either of the two classes above, {\em all solutions which lie below the special solution $W_*$  bounce before reaching infinity}
\end{itemize}

\vspace{0.5cm}

\subsubsection{Behavior at $\phi \to -\infty$} \label{ssec:uv}

We now consider  the opposite end of field space. We still assume $W >0$ and $W'>0$, so that now $\phi\to -\infty$ corresponds to the UV of the geometry.

Recall that  $W (\phi)$ is necessarily bounded from below by the curve $B(\phi)$, as follows from equation \eqref{BoundW}. Whether or not the solution extends to $-\infty$ depends on the asymptotic behavior of $V(\phi)$.
Suppose first that $B'$ stays strictly negative  asymptotically as $\phi \to -\infty$,  which happens if  $V(\phi) \to -\infty$. Then, the  growing solution $W_\uparrow(\phi)$ will necessarily hit the critical curve  at some finite $\phi$, and the solution will bounce before reaching infinity.

Therefore, if a $W_\uparrow$ solution is to extend to $\phi \to -\infty$, $V(\phi)$ (and therefore $B(\phi)$) must be bounded as $\phi\to-\infty$.

 Barring the case of infinite oscillations\footnote{One cannot exclude this a priori, but it would take a very special potential for this case to arise. We will neglect this possibility in this paper.}, $B(\phi)$ must therefore asymptote to a constant $B_0$ as $\phi\to -\infty$.  If $B_0\neq 0$,  we can neglect the derivative term in equation  (\ref{inf00}), and find:
\be
W \to W_0 = B_0 \qquad \phi \to -\infty.
\ee
This solution corresponds to a UV runaway $AdS$  fixed point,  like the one used in holographic duals of QCD\footnote{Upon analytic continuation this solution becomes the Starobinsky-like cosmological solutions, \cite{afim}.}  \cite{gkn}. The $AdS$ length is fixed by $W_0$ to be $\ell = 2(d-1)/W_0$. This solution also  admits continuous deformations: using the result from equation (\ref{def3}) we find a one-parameter deformation of the form:
\be
\delta W \simeq C \exp\left[{1\over Q} \int^\phi {W \over W'}\right]
\ee
Since the integrand is positive and divergent ($W\to W_0>0$ and  $W' \to 0^+$ as $\phi\to -\infty$),   $\delta W$ vanishes and it is indeed sub-leading with respect to the constant $W_0$. The detailed  behavior depends on the precise form of the reference solution, which in turn depends on exactly how the potential approaches  a constant\footnote{A concrete example is given by the asymptotically logarithmically $AdS$ behavior of Improved Holographic QCD models, see Appendix E of reference \cite{thermo}.}. Therefore, these solutions correspond to a UV fixed point deformed by a  marginally relevant operator, for which the parameter $C$ sets the value of the VEV.

To summarize:
\begin{itemize}
\item
If $V(\phi)$  diverges as $\phi \to -\infty$ there are only bouncing solutions with positive and growing $W$.
\item If $V(\phi) \to V_0$ as  $\phi \to -\infty$, then we have a continuous family of solutions all asymptoting a UV ``runaway''  $AdS$ fixed point.
\end{itemize}


\subsection{Good vs. bad singular geometries}
\label{ssec:goog}
We will now discuss which of the solutions that reach infinity have acceptable (i.e. ``good'' from the holographic standpoint) singularities.

The runaway UV solutions of section \ref{ssec:uv}  are mildly singular: indeed,  only the scalar $\phi$ is divergent, but all curvature invariants,  as well as all kinetic invariants built with derivatives of $\phi$,  are finite.  These geometries  satisfy both ``good singularity'' criteria we gave at the beginning of Section 4.

On the other hand, the IR singularity at $\phi\to +\infty$ may be more problematic.

For definiteness, we assume here that the potential is dominated by a simple exponential as $\phi$ asymptotes to $+\infty$, which for solutions with $W'>0$ correspond to the IR:
\be\label{good1}
V(\phi) \simeq V_0 \exp b\phi \qquad \phi \to \infty.
\ee
where $V_0$ is a negative constant and $b \geq 0$.  The discussion in the  previous subsection can be summarized by the statement that solutions that reach  $\phi \to +\infty$ exist only if $b<2Q^{-1}$, and that for such solutions the superpotential  has an exponential behavior:
\be \label{good2}
W \simeq W_0 \exp \gamma \phi,  \qquad \gamma = \left\{\begin{array}{ll} Q^{-1} & \quad \text{continuous branch}\\
b/2 & \quad \text{special solution} \end{array}\right.
\ee
For the continuous branch, $W_0$ is arbitrary; for the special solution it is fixed by equation (\ref{inf3}), which together with the definition (\ref{defB}) gives:
\be\label{good1-1}
W_0^{(spec)} = \sqrt{-{4(d-1)/d\over  1- (b Q/2)^2} V_0}
\ee

Since both the  superpotential and the potential diverge in the IR, there will be a naked curvature singularity as $\phi \to \infty$. In the case $b<0$ all the IR singularities fall into the continuous branch of \eqref{good2} because $B(\phi)$ becomes negligible compared to $W(\phi)$ as $\phi$ asymptotes to $+\infty$.

Integrating  the first order equations (\ref{defW}-\ref{phiW}) asymptotically, one  finds, for $\gamma\geq0$,  the metric and dilaton profiles:
\be\label{good3}
A(u) \simeq {1\over 2(d-1)\gamma^2} \log (u_* - u),  \qquad \phi(u) \simeq -{1\over \gamma} \log [\gamma^2 (u_*-u)]
\ee
where $u_*$ is an integration constant that sets the position of the singularity in the bulk.

Whether the singularity is good or not depends solely on the value of the parameter $\gamma$ in  equation  (\ref{good2}). Below we summarize separately, and without derivation, the conditions for  Gubser's criterion and for spectral computability, referring  the interest reader to previous work for details.

\subsubsection{Gubser's condition}
 To proceed with the analysis, one has to discuss black-hole solutions with running dilaton. These were discussed for a general potential in \cite{thermo}.  The black-hole scale factor and dilaton can also be derived from a superpotential $W_T(\phi)$, which satisfies a modified version  of equation (\ref{SuperP}), \cite{rg1}. Apart from this, the finite-temperature metric and dilaton profile are still given by first order equations:
\be
\dot{A}_T(u) = -{1\over 2(d-1)} W_T(\phi), \qquad \dot{\phi}(u) = W_T'(\phi).
\ee
For Gubser's criterion, we  must consider  small black-holes, i.e. solutions which are arbitrarily close to the vacuum (i.e. Poincar\'e invariant) solution. Since the latter extends to $\phi \to +\infty$, regular small black-holes must necessarily have a large value of the dilaton at the horizon. Therefore, small black-holes, if they exist, are those solutions in which the horizon value of the scalar field $\phi_h$ is arbitrarily large.

 Regularity (this time, in the strict sense) of  the black-hole horizon selects a single superpotential. One of the main results of \cite{thermo} can be summarized as follows: \\

{\em Regularity of small black-holes requires the finite temperature superpotential  $W_T$ to approximately match the  zero-temperature special solution $W_*$ as $\phi_h$ becomes large.}\\

This means that small the  black-hole metrics are an approximation to the zero-temperature {\em special} solution stemming from $W_*$, and at the same time they are far from any of the solution of the continuous branch, since the latter have a different exponential scaling at large $\phi$. Therefore, we reach the following statement:\\

{\bf Gubser's condition:} among  the solutions that extend to $\phi \to \infty$, only the special solution $W_*$ satisfies Gubser's criterion. \\

Since the special solution requires the scalar potential to grow at most as fast as in equation  (\ref{good1}) with  $b< 2Q^{-1}$  we arrive at what  is sometimes called {\em Gubser's bound,} which is a condition on the asymptotic growth of the potential: \\

{\bf Gubser's bound:} For a holographic Einstein-dilaton theory to admit solutions with IR-good singularities that reach $\phi \to \infty$, the potential must diverge slower\footnote{In theories that contain other bulk fields, like for an example an extra vector field the Gubser bound is modified, see \cite{cgkkm}.}than $\exp (2Q^{-1} \phi)$ at large $\phi$.\\

If this bound is not satisfied, then all solutions bounce before reaching infinity.

\subsubsection{Spectral computability}
\label{sssec:specc}
We now turn to the other condition we have adopted for a singularity to be useful: the computability of the spectrum/correlators.  Essentially, a singularity satisfies this criterion if the physical spectrum  is uniquely determined by normalizability (as e.g. in a quantum harmonic oscillator) without need to specify extra boundary conditions in the IR  (as it is the case with a hard wall, for example).

Although they are not inconsistent, solutions which display singularities that are ``hard wall''-like should  be handled with care in the holographic setting:  in these cases the model is not predictive given the  UV boundary conditions alone: rather, one needs to explicitly resolve the singularity in order to be able to compute physical correlators.

 Another way to frame this is that, in bad singularities, fluctuations with finite energy are in principle allowed to probe arbitrarily close  the high curvature region. This means that any resolution of the singularity will affect the energy spectrum, including the low-lying states, and results obtained using the singular solution are unreliable. On the other hand, for solutions satisfying the spectral computability criterion, the singularity is repulsive, wave-functions are  suppressed in the singular region,  and only highly energetic states can probe the high-curvature part of the geometry. Therefore,  the details of resolving the singularity will at most  affect the high energy part of the spectrum: the precise physics of the singularity decouples from the low-energy observables, which can then be reliably  computed using the singular geometry.

To state the criterion precisely in mathematical terms, we have to  translate the fluctuation spectral  problem  into an equivalent quantum mechanical problem. First, we change to conformal coordinates:
\be \label{spec1}
e^A dr = du
\ee
In this frame, the IR  singularity may be at finite or infinite $r$-coordinate, depending on the value of $\gamma$. Specifically:
\be\label{spec2}
A(r) \simeq \left\{ \begin{array}{lll}{1\over 2(d-1) \gamma^2 - 1}\log(r_*-r) & \quad r\to r_* & \qquad \gamma^2 > {1\over 2(d-1)} \\
& & \\
{1\over2 (d-1) \gamma^2 - 1}\log r & \quad r\to +\infty & \qquad \gamma^2 < {1\over 2(d-1)}\end{array}\right.
\ee
Notice that in both cases $e^A \to 0$ at the singularity.

Next, we consider the linearized fluctuation equation, in conformal coordinates, around the singularity. For simplicity, we discuss the massless Klein-Gordon equation
\be\label{spec3}
\de_r ( e^{(d-1)A} \de_r h) + \de_\mu\de^\mu h =0.
\ee
This is all we need to consider,  since
\begin{enumerate}
\item this is the equation satisfied by tensor modes in the full geometry;

\item     the equation for diff-invariant scalar fluctuations, i.e. equation (\ref{zeta1}) in  Appendix \ref{ssec:back},   reduces  to (\ref{spec3}) close to the singularity when $\phi \to \infty$ (the terms involving derivatives of $\phi$ turn out to be sub-leading).
\end{enumerate}

To find the d-dimensional  spectrum of mass excitations of the dual field theory, one looks for solutions with a fixed $d$-dimensional mass, $\de_\mu\de^\mu h = \m^2 h$. Moreover, a standard  change of  variables turns equation (\ref{spec3})  into  a one-dimensional time-independent Schr\"odinger equation, where $\m^2$ plays the role of the energy:
\be \label{spec4}
-\de_r^2 \psi + V_s(r) \psi = \m^2 \psi, \qquad \psi(r) = e^{{d-1\over 2}A(r)}h(r) ,
\ee
and the Schr\"odinger potential is given by:
\be
V_s(r) = \left({d-1 \over 2}\right)^2 \left({d A\over dr}\right)^2  +  {d-1 \over 2} {d^2 A\over dr^2}.
\ee
Close to the singularity, the warp factor is approximated as in equation (\ref{spec2}), and  the potential behaves as:
\be\label{spec5}
V_s(r) \simeq \left\{ \begin{array}{lll}\displaystyle{{q(q-1)\over (r_*-r)^2}} & \quad r\to r_* & \qquad \gamma^2 > {1\over 2(d-1)} \\
& & \\
\displaystyle{{q(1-q)\over r^2}}  & \quad r\to +\infty & \qquad \gamma^2 < {1\over 2(d-1)}\end{array}\right.
\ee
where we have introduced the notation:
\be\label{spec6}
q \equiv {(d-1) \over 2}{1\over 2(d-1) \gamma^2 -1},
\ee
with $\g$ defined in equation \eqref{good2}.
For $\gamma^2< (2(d-1))^{-1}$ the potential vanishes asymptotically as $1/r^2$ at infinity and the spectrum is continuous. We will comment on this case briefly later.

On  the other hand, for $\gamma^2 > (2(d-1))^{-1}$, which corresponds to $q>0$,  the singularity is at finite $r=r_*$. Intuitively, one may suspect that the spectrum will be discrete, as with a hard wall placed at $r_*$.  This intuition is correct, but depending on the value of $\gamma$, the spectrum may or may not be uniquely fixed. In fact, consider the two independent asymptotic solutions close to $r\simeq r_*$, where the $\m^2$ term in equation (\ref{spec4}) can be neglected:
\be \label{spec7}
\psi \simeq A (r_*-r)^q  + B (r_*-r)^{1-q}, \qquad r \to r_*.
\ee
Usually the spectrum is fixed by requiring normalizability of the wave-function, which {\em usually} picks only one out of the two independent solutions. This (together with the same requirement at the other asymptotic end of the $r$-direction) imposes a condition which  results in quantization of the eigenvalue $\m^2$. However, when $1-q > -1/2$, {\em both} solutions are normalizable and  so is any linear combination of them: therefore normalizability does not impose any condition. On the other hand, imposing any arbitrary linear relation between $A$ and $B$ in equation (\ref{spec7}) results in  a different (discrete) spectrum.   Technically, one says  the Hamiltonian operator in equation (\ref{spec4}) admits an infinite number of self-adjoint extensions, each parametrized by a different boundary condition at the singularity and each having a different discrete spectrum of eigenstates.

At this point it should be clear why this situation is unacceptable if we want the model to make sense from the holographic perspective: in holography, the definition of the theory should be completely specified in terms of a bulk theory plus a set of boundary conditions in the UV. For $1-q > -1/2$  on the other hand, we need an extra set of IR boundary conditions (one for each physical bulk excitation) to define the model. Therefore, the holographic model is not ``IR-complete'', and some information is missing\footnote{Notice that this is {\em not} the case for the positive $1/r^2$ potential at infinity (case $q<0$) since there we have only plane-wave-normalizable (i.e. improper) eigenfunction, and  the Hamiltonian has a single self-adjoint extension with continuous spectrum and (strictly speaking) empty {\em proper} spectrum. Which combination  of ingoing our outgoing plane waves one should pick depends on the physical question one wants to answer (e.g. ingoing b.c. for retarded correlators, etc).}

We are therefore able to summarize the condition for spectral computability (or IR-completeness). First, notice that the condition $1-q<-1/2$ translates into:
\be \label{spec8}
\gamma < \gamma_c \equiv  \sqrt{d+2 \over 6(d-1)}
\ee
Then, we can state the following criterion:\\

{\bf Spectral computability:} The singularity has a computable spectrum if the superpotential diverges as $\phi$ asymptotes to $\infty$ at most as fast as in equation (\ref{good2}), with  $\gamma < \gamma_c$ above. \\

A few comments are in order:
\begin{enumerate}
\item Since $\gamma_c <  Q^{-1}$ strictly for any $d>1$, the condition (\ref{spec8}) can only be satisfied by the special solution (cf.. equation (\ref{good2})).
\item  The resulting computability {\em bound} on  scalar potentials with asymptotic behavior (\ref{good1})   is
\be
b <  2 \sqrt{d+2 \over 6(d-1)}.\label{bcrit}
\ee
This is strictly stronger than Gubser's bound for any $d>1$.
\item Notice that we do {\em not} impose a similar requirement in the UV: in fact, the case when both independent perturbations  are UV-normalizable corresponds to the possibility of multiple quantizations: in this case, picking one specific quantization is a UV boundary condition and therefore part of the definition of the theory and of the holographic dictionary.
\end{enumerate}
\subsection{Multi-branch holographic RG flows from a regular UV to $\phi=\infty$}
\label{ssec:bounce_inf}

In this subsection we show  examples of  holographic RG flows that leave a UV fixed point, bounce and then reaches infinity asymptoting  the special solution $W_*(\phi)$, defined in equation \eqref{good2}.

This kind of solution is, to our knowledge, new in the literature. However, in \cite{bounce}, the authors studied black-hole solutions in the same model we consider below and some of the unusual properties they find can be explained by the fact that the zero-temperature solution, which we present here, bounces before reaching the IR.

The potential used in \cite{bounce} is:
\eql{Vcosh}{V(\phi)=\ha \Delta(\Delta-d)\phi^2 +d(d-1)\left(\ha  b^2\phi^2-\cosh( b \phi)\right).}
This expression is such that:
\be
V(0)=-d(d-1), \quad
V''(0)= \Delta(\Delta-d)
\ee
At large $\phi$,
\eql{Vcosh_asympt}{V(\phi)\approx-{d(d-1)\over2}e^{ b \phi}}
For appropriate choices of the parameters this potential does not have extrema, so the  only regular flows can end at infinity, provided  $b$ is below Gubser's bound.

\begin{figure}[h!]
\centering
\includegraphics[width=0.6\textwidth]{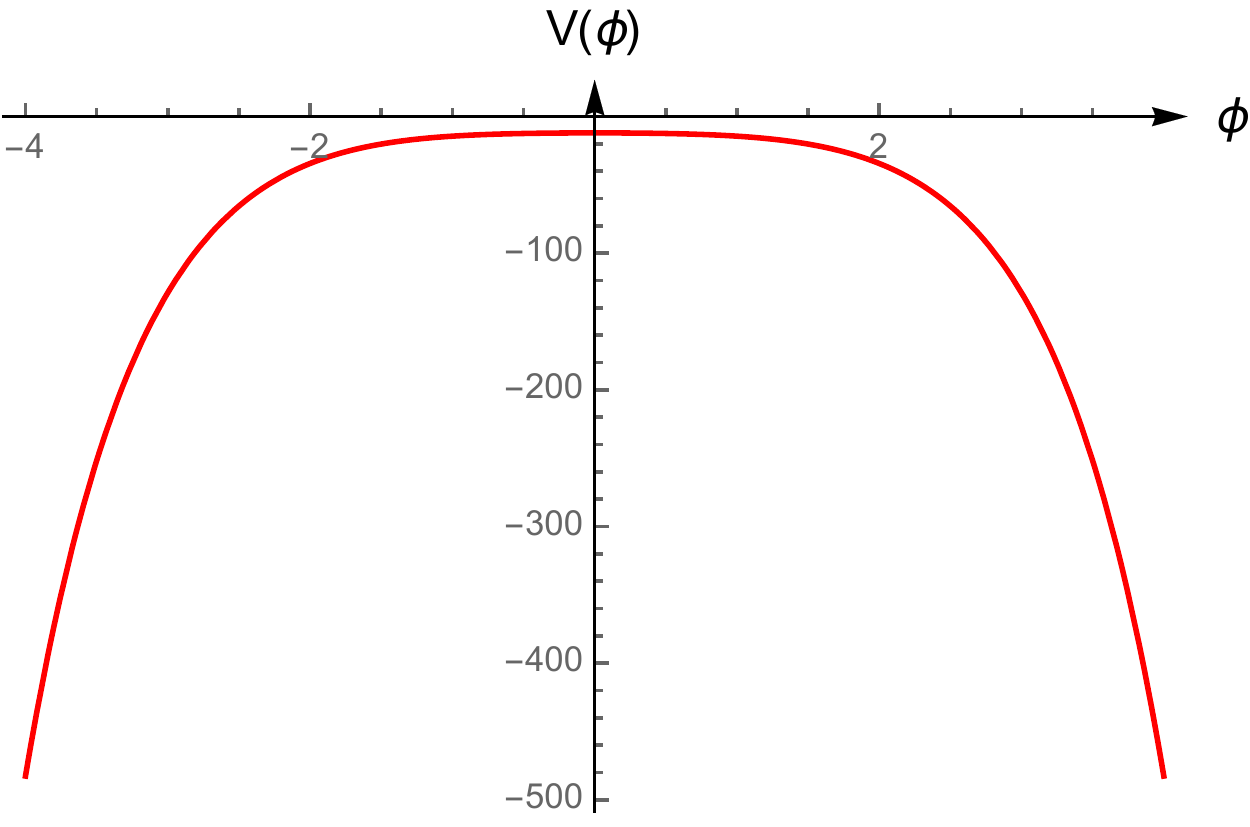}
\caption{
Potential in equation \protect\eqref{Vcosh} with parameters defined in equation  \protect\eqref{parGursoy0}.
}
\label{fig:Gur0}
\end{figure}

It follows from  equation \eqref{good2} and (\ref{Vcosh_asympt})  that the special solution has the asymptotic behavior:
\eql{Wcosh_1}{W(\phi) \simeq  W_0 e^{ b \phi/2},
\quad
\phi\to +\infty.}
The coefficient $W_0$ can is given by equation \eqref{good1-1}:
\be
W_0 ={(d-1)\sqrt{2}\over \sqrt{1-{(d-1)\over 2d} b^2}}\label{W0}
\ee
In \cite{bounce} the parameters are chosen to be:
\eql{parGursoy0}{\Delta=3,\quad d=4\quad \text{and} \quad  b={2\over \sqrt{3}}}
which, together with \eqref{W0}, yields $W_0=6$.  The resulting  potential   is plotted in figure \ref{fig:Gur0}.

\begin{figure}[h]
\centering
\includegraphics[width=1\textwidth]{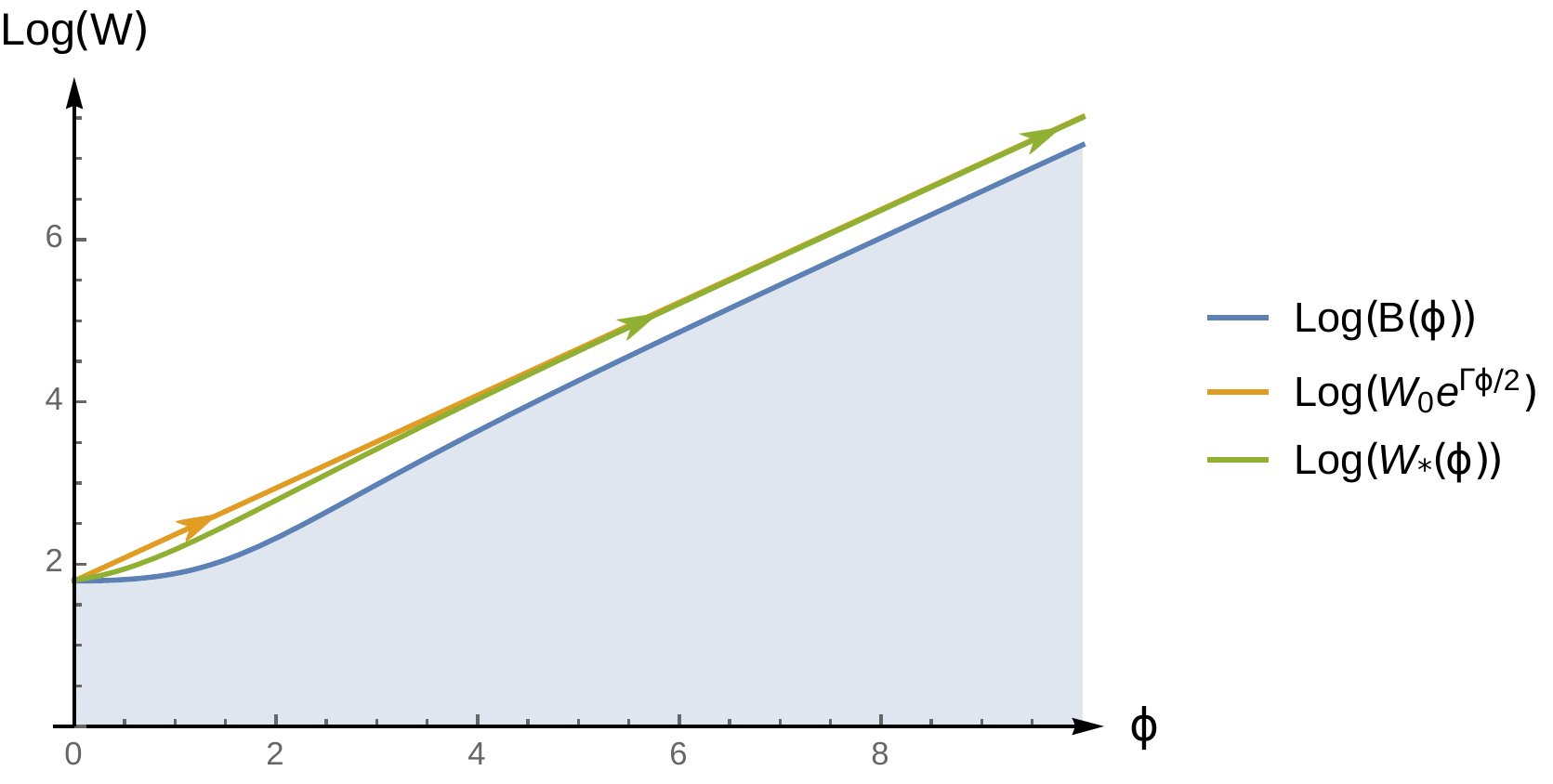}
\caption{
Plot of the superpotential obtained with the potential \protect\eqref{Vcosh} and parameters \protect\eqref{parGursoy0}. The green curve represents the special solution $W_*(\phi)$, obtained numerically. For comparison, the asymptotic behavior \protect\eqref{Wcosh_1} is represented by the yellow curve.
 The special solution is asymptotically parallel to the curve $B(\phi)$ as $\phi$ becomes much larger than $ b^{-1}=\sqrt{3}/2$. This can also be seen from  figure \protect\ref{fig:Gur2}.
\label{fig:Gur1}}
\end{figure}

\begin{figure}[t]
\centering
\includegraphics[width=0.8\textwidth]{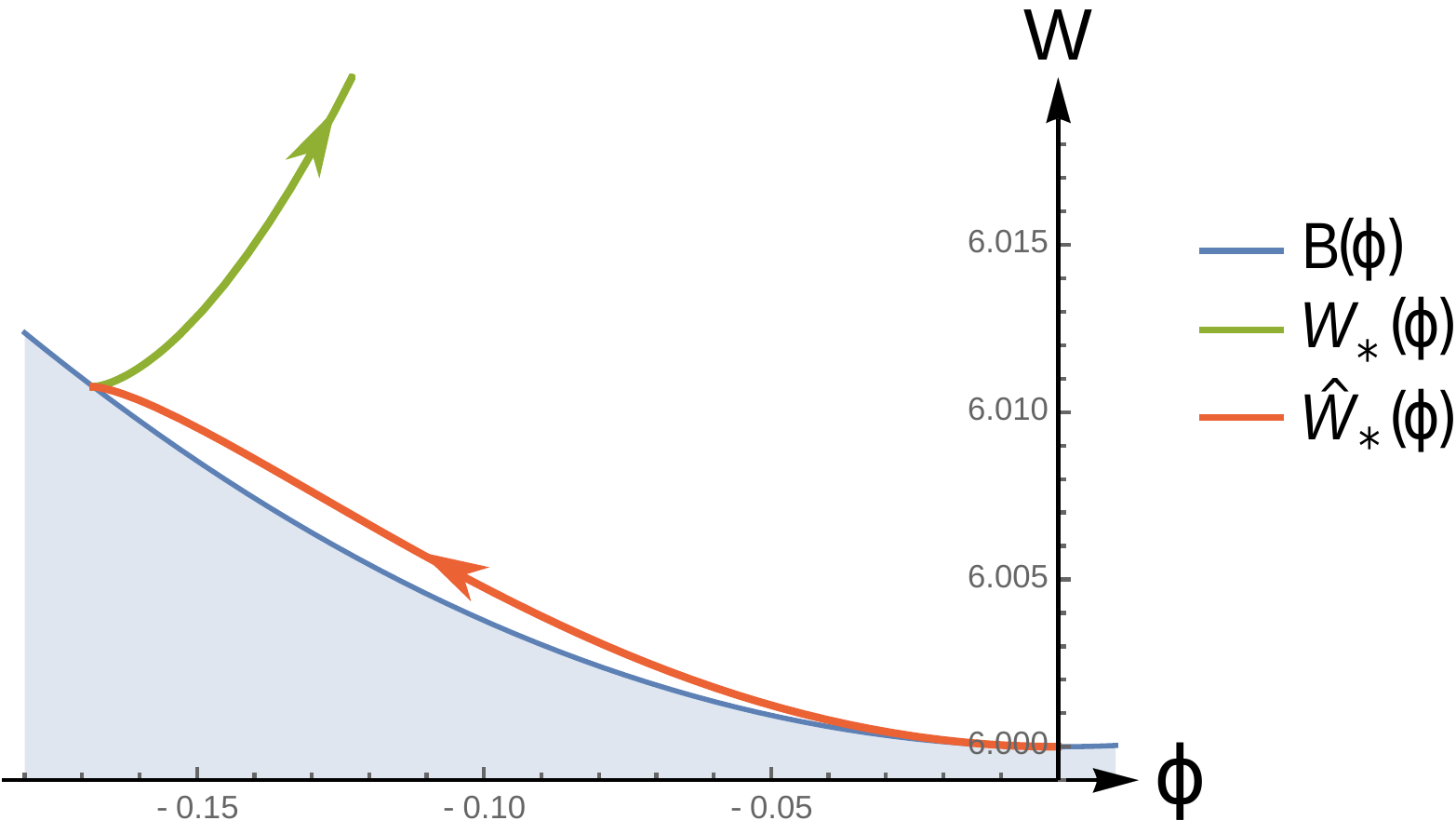}
\caption{
Near-UV and near-bounce region of the solution obtained with the potential \protect\eqref{Vcosh} and with parameters chosen as in \protect\eqref{parGursoy0}. The  solution is chosen in such a way that it matches the IR  special solution, $W_*(\phi)$.}
\label{fig:Gur3}
\end{figure}

 Using the potential described above, we have computed numerically the superpotential with special with asymptotic (\ref{Wcosh_1}) reaching
 $\phi \to +\infty$. The full flow is shown in figure \ref{fig:Gur1}. As we can see from the detail in  figure \ref{fig:Gur3}, if we follow the special solution  backwards  towards the UV, the flow overshoots the UV fixed point and bounce off the critical curve at a negative value of $\phi$. In other words, to flow towards $\phi\to +\infty$ in the IR, one must start with a {\em negative} coupling in the UV.

This explains the results of \cite{bounce}: there,  black-hole solutions were found only below a certain value of the scalar field at the horizon. Indeed, the authors of \cite{bounce} required that the solution reach the UV fixed point at $\phi=0$ from the positive $\phi$ direction. What happens is that,   beyond  a certain critical value of the horizon scalar field, the solution starts bouncing  and reaches the UV fixed point from the negative $\phi$ direction, as in figure \ref{fig:Gur3}. According to our analysis in Section 2,  the critical value should correspond to the topmost  solution that flows to the IR without bouncing, i.e. the VEV flow $W_+$. Indeed, in \cite{bounce} it was found that the VEV/source ratio diverges for this critical value.

\begin{figure}[t]
\centering
\includegraphics[width=0.6\textwidth]{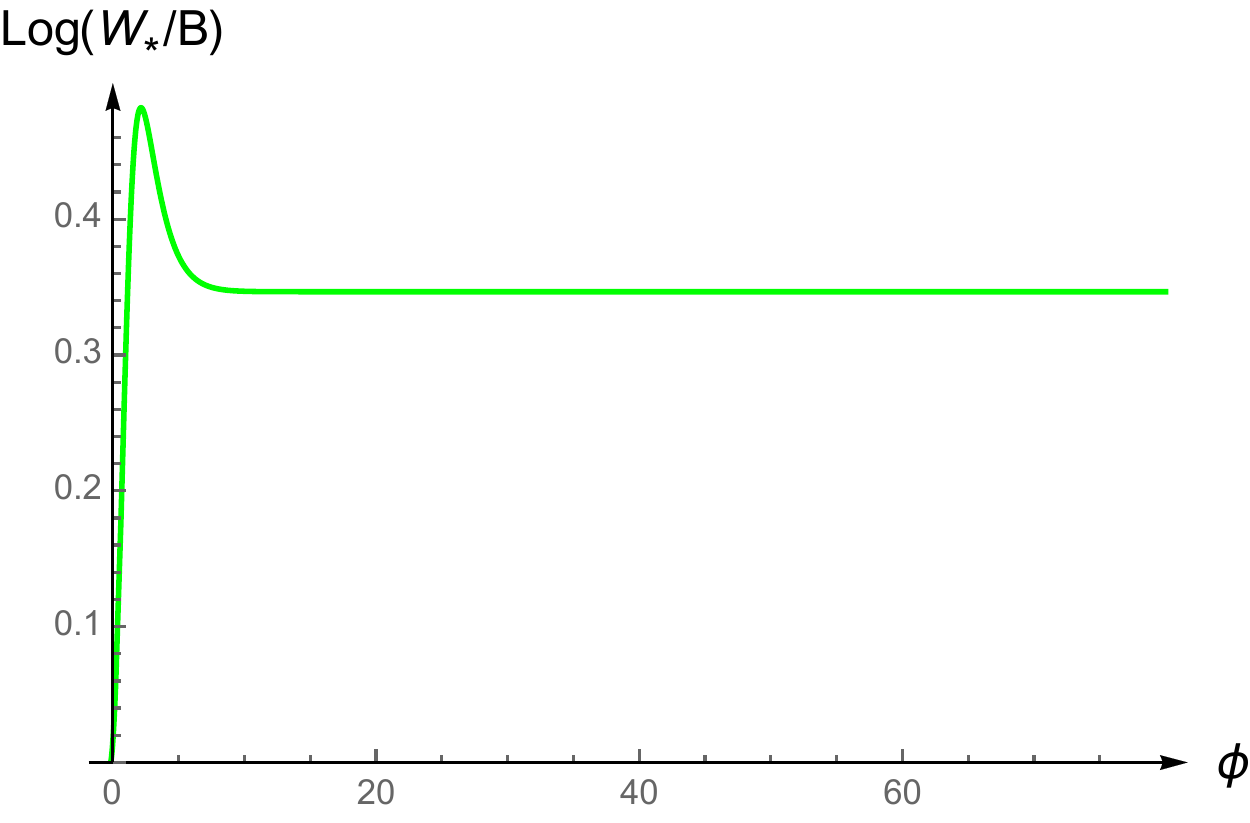}
\caption{
Evolution of the ratio betwewn $W_*(\phi)$ and $B(\phi)$ as a function of $\phi$ in the numerical solution plotted in figures \protect\ref{fig:Gur1} and \protect\ref{fig:Gur3}. The ratio approaches a constant, showing this is indeed a good approximation to the special IR-singular solution \protect\eqref{Wcosh_1} in this range.}
\label{fig:Gur2}
\end{figure}

From equation \eqref{bcrit},  we see that $b=2/\sqrt{3}$  is precisely the value that saturates the computability bound. This however is not related with the fact that the regular solution bounces: we checked that this feature persists  if we take $b=2/\sqrt{3}- 1/100$, for which the  bound is satisfied.

\begin{figure}[t]
\centering
\includegraphics[width=0.9\textwidth]{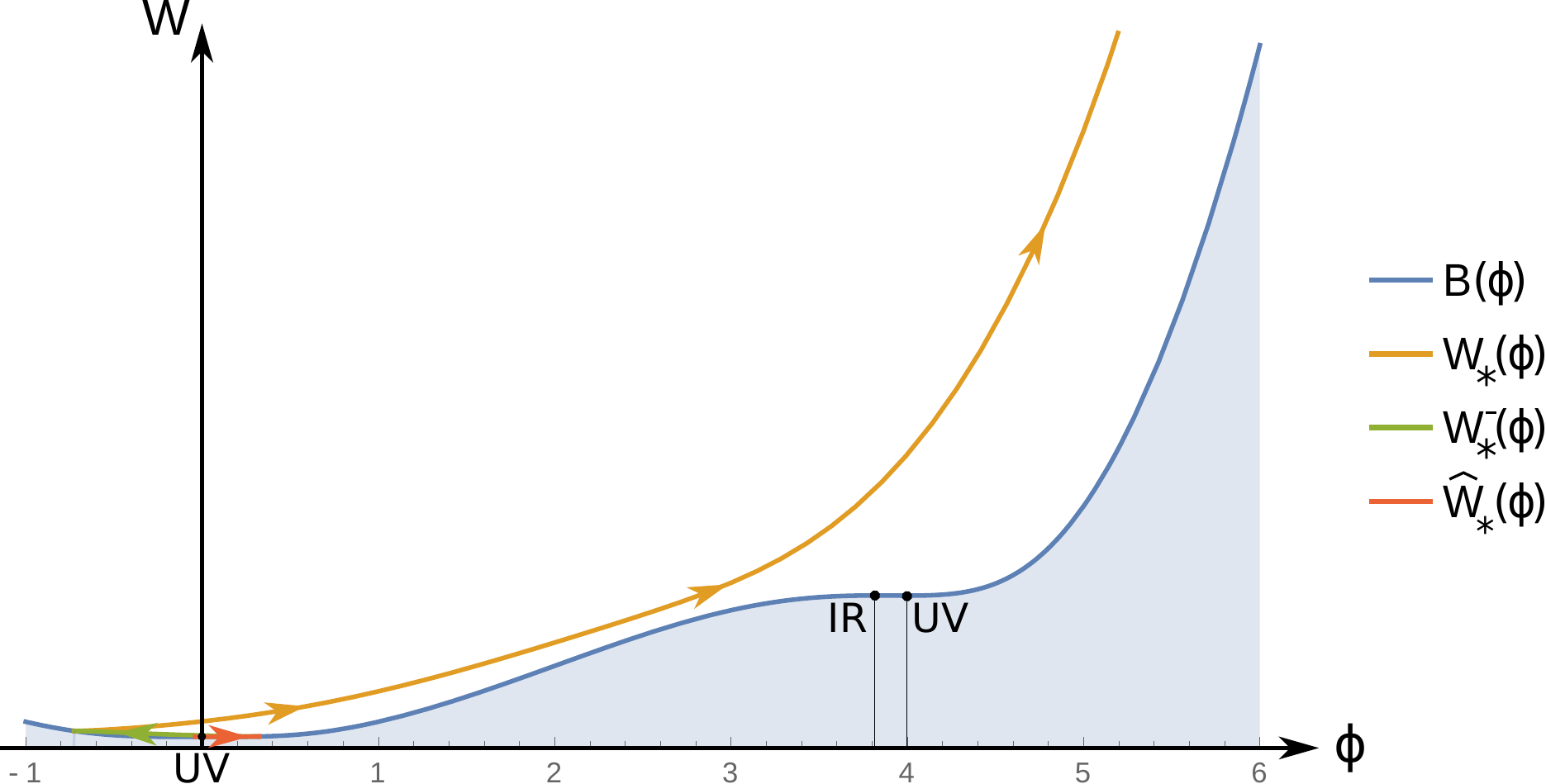}
\caption{The curves labeled $W_*(\phi)$, $W_*^-(\phi)$ and $\widehat W_*(\phi)$ correspond three branches of the same superpotential. In total there are four different branches but only three are visible in this figure (due to the small scale of the fourth one).  We use the potential (\protect\ref{V8}) with parameters \protect\eqref{ParamV8}. The near-UV behavior is very close to the one represented in figure \protect\ref{fig:phi8_01}.}
\label{fig:inf00}
\end{figure}

\begin{figure}[t]
\centering
\includegraphics[width=0.6\textwidth]{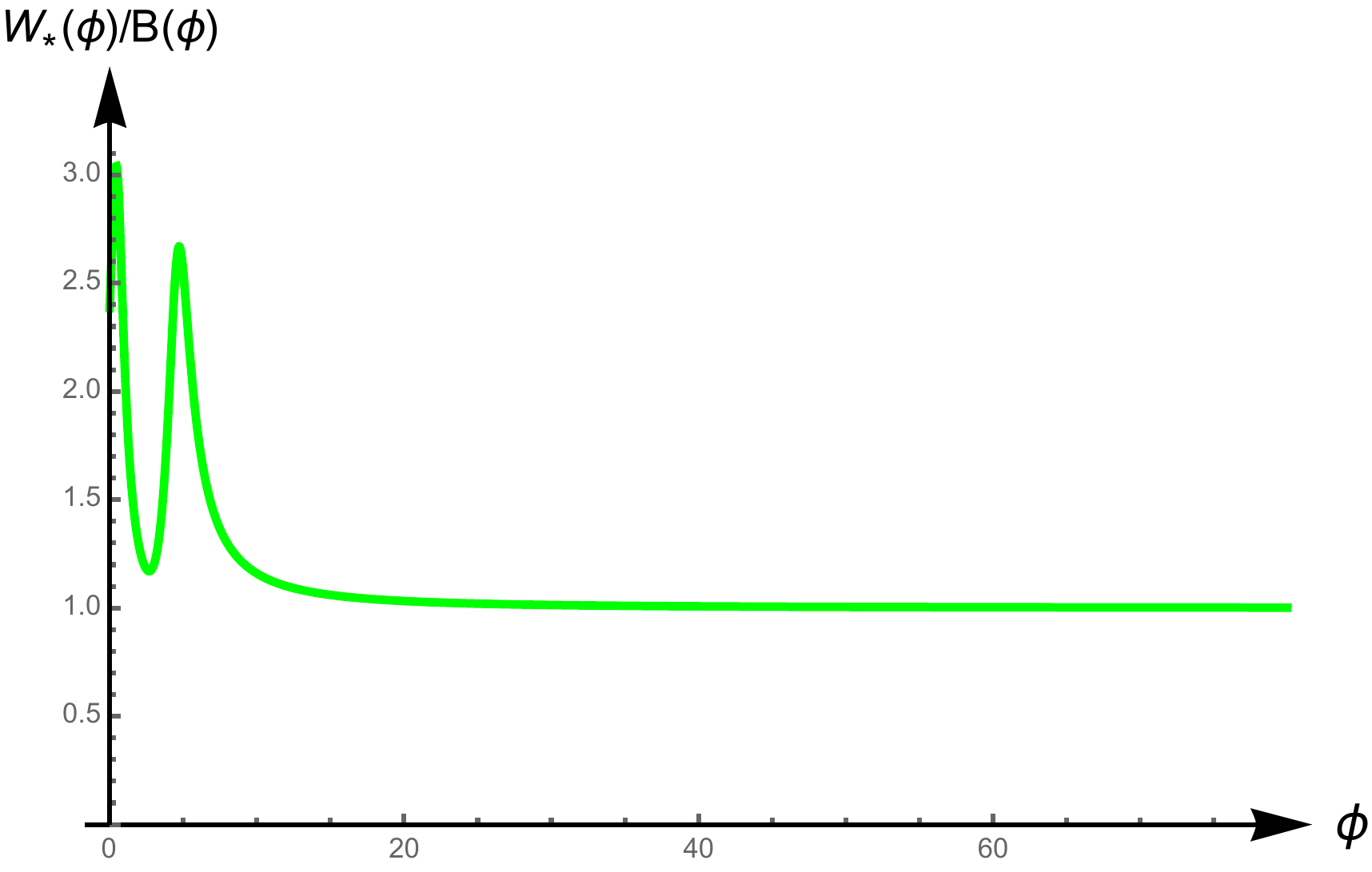}
\caption{
The ratio of $W_*(\phi)$ and $B(\phi)$ approaches a constant for large $\phi$ (cf. equation \protect\eqref{good2}). Here $W_*(\phi)$ corresponds to the upper branch of the superpotential plotted in figure \protect\eqref{fig:inf00}.}
\label{fig:All3}
\end{figure}

An even more exotic flow can be obtained with the same polynomial potential (\ref{V8}),  and the same parameters \eqref{ParamV8}, which we used in Section 3 to construct bouncing solutions. The  regular  solutions that extends to $+\infty$ is shown in figure \ref{fig:inf00}:  it starts at the UV fixed point at $\phi=0$,  bounces twice, skips the IR fixed point at  $\phi=\phi_0$ and finally, as $\phi$ asymptotes to infinity, locks onto the special asymptotics \eqref{good2}. The latter feature can be clearly seen  in figure \ref{fig:All3}. The difference between this solution and the one reaching the IR at $\phi=\phi_0$ (presented in Section \ref{ssec:bounces}) lies in the value of the coefficient of the sub-leading deformation $C$ in the UV, i.e. the two solutions correspond to different VEVs of the dual operator.

The yellow curve in figure \ref{fig:inf00} represents the upper of the four branches  of this bouncing solution $W(\phi)$ that corresponds to the special solution \eqref{Wcosh_1}.

Finally we make a technical remark about our numerical procedure to find the solution with special asymptotics: this being an isolated solution, obtaining it numerically is not straightforward, because the corresponding set of initial conditions has measure zero in the $(\phi,W)$ plane. To overcome this difficulty, we first notice that the special solution $W_*(\phi)$ is the one that has the minimum slope to reach infinity. Therefore, solutions with $W<W_*$ will not grow fast enough and will bounce for large but finite $\phi$.

 In order to numerically obtain a good approximation to $W_*$ one can choose a solution such that, for a given $\hat \phi\gg b^{-1}$, $W(\hat\phi)$ equals the asymptotic exponential \eqref{Wcosh_1}. For example, in the case of the potential   (\eqref{Vcosh}) with parameters \eqref{parGursoy0}, we chose  $\hat\phi=40$ to obtain the numerical solution (green curve in figure \ref{fig:Gur1}). This gives  a very good approximation to the  asymptotic exponential behavior expected for the special solution (yellow curve  in figure \ref{fig:Gur1}). In figure \ref{fig:Gur2} we show that this  matching continues for larger $\phi$, at least up to $\phi=80$ where we stopped the numerical calculations.

\section{The free energy}

As we have seen in the previous sections, when the potential admits several extrema one can have multiple regular solutions  leaving the same UV, each flowing to a different IR.  Each  solution is characterized by a different VEV (sub-leading term in the UV in equation (\ref{ab}) ) but fixed source. Therefore, in standard quantization (fixed source) the gravitational path integral  must  include all these saddle points, and the question arises of which flow dominates the path integral. This is equivalent to asking which among the possible solutions represents the true vacuum of the dual field theory. Notice that for a potential with a finite number of extrema,  there are always a finite number of regular solutions (those reaching one of the IR minima, plus the two special solution reaching $\pm\infty$.

This question  can be answered by comparing  the free energies of the  different vacua while keeping the source fixed. The free energy is given on the gravity side by the Euclidean on-shell action,
\be\label{free1}
{\cal F} =  S_{E}[\phi_-]
\ee
where we have made explicit the fact that the left hand side is a functional of the boundary data $\phi_-$, defined in equation (\ref{ab}).

We start from the action (\ref{S_E-S}). Evaluated on a given solution, it reduces to a boundary term equal to the corresponding superpotential \cite{papa1,papa2,rg1,rg2}:
\be \label{free2}
S_{on-shell} = \int d^d x \, \left(\left[e^{dA} W\right]_{UV} +  {1\over d-1}\left[e^{dA} W\right]_{IR} \right).
\ee
The difference between the UV and IR contributions is due to the fact that one should not  include the Gibbons-Hawking term in the IR, since this is not a boundary.

The IR contribution vanishes if the theory reaches a regular IR (either a fixed point, or the special solution in a flow to infinity). Indeed,  at an $AdS$ IR fixed point, $ e^{dA} \to 0$ while $W$ stays finite. For a flow extending to infinity with  special asymptotics (\ref{inf7}) this is less obvious, but it still holds: using equations (\ref{defW}-\ref{phiW}) the scale factor can be written as a function of $\phi$ as (see e.g. \cite{glueball}):
\be\label{free4}
e^{dA}   = \exp \left[A_0  -  {d\over 2(d-1)}\int d\phi \, {W\over W'}\right] \sim \exp\left[ -{d \over 2(d-1)} {1\over k} \phi \right], \quad \phi \to +\infty,
\ee
where we have used equation (\ref{inf7}), and where  $k$ is defined in equation (\ref{inf3}). Therefore, we can evaluate the IR contribution to (\ref{free2}) to be:
\be\label{free5}
\left[ e^{dA} W\right]^{IR} \sim \exp\left[ -{d \over 2(d-1)}{1\over k} \phi \right] e^{k\phi}  = \exp\left[-{1\over k}\left({1\over Q^2} - k^2 \right) \phi \right]
\ee
where $Q$ is defined in equation  (\ref{inf01}). The exponential in the equation above vanishes as $\phi \to +\infty$ in the IR, since $k < Q^{-1}$ as explained below equation (\ref{inf3}). Therefore, the IR contribution drops out for  regular RG-flows, as well as singular RG-flows with ``good'' singularities.

In the UV on the other hand we must introduce a cutoff and counter-terms, since there  $e^{dA}\to + \infty$ while $W \to 2(d-1)/\ell$.
The  appropriate counter-term to cancel the divergence is a reference solution $W_{ct}$ of the superpotential equation. This guarantees that all the UV-divergent terms cancel, and after removing the cutoff the surviving term is the one controlled by the parameter $C$ in equation (\ref{W_max_-}):
\be\label{free6}
S^{ren}_{on-shell} = \left(C-C_{ct}\right)\int d^d x \, \phi_-^{d/(d-\Delta)}
\ee
For a  fixed value of the UV coupling $\phi_-$, the expression above depends on the choice of the flow solution only through of the associated value of the parameter $C$ (which ultimately gives the VEV). On the other hand, the finite counter-term contribution $C_{ct}$ is fixed, since it is  part of the definition of the theory,  therefore it is the same for all solutions.

Equation (\ref{free6}) gives the value of the Lorentzian signature action. For homogeneous solutions,  this is a potential term. In going to the Euclidean signature, the action changes sign\footnote{Intuitively, one can understand this from  the fact that the potential energy enters with a negative sign in the action, but with a positive sign in the energy i.e. in the Euclidean action. More concretely, the Einstein-Hilbert term enters with a  negative sign in the action  for Euclidean gravity, i.e. the  opposite sign to the one in (\ref{S_E-S}),  which we used to derive (\ref{free6}). When we change the sign of the Einstein-Hilbert term, to make sure that the Euclidean version of  the same geometry solves the field equations we must change the sign for all terms in the action.}. The result for the free energy of each vacuum is then:
\be\label{free7}
{\cal F}_i = - \left(C_i - C_{ct}\right)\int d^4x \, \phi_-^{d/(d-\Delta)}
\ee
where the index $i$ labels the different RG flow solutions flowing from a fixed UV to different regular IR's, all having the same value of the UV coupling $\phi_-$.

The parameter  $C$ determines the growth rate of $W$ in the UV, as one can see from  equation (\ref{W_max_-}), all other leading terms being equal.  For single-branch solutions, this translates into a very simple statement about the IR:  since different  single-branch superpotentials (with a given sign of $W'$) can never cross, it follows that  the flows that reach further from the starting point must have larger  $C$.   from equation (\ref{free7}) we conclude that {\em the solution which dominates the path integral is the one that flows the longest distance to the IR.}

This statement can be rephrased  in the dual QFT in terms of the holographic $a$-theorem: since the superpotential is strictly increasing from the UV to the IR, the farther an IR fixed point in field space, the largest the IR value of $W$. Since the latter is the inverse of the central charge, we conclude that  {\em in a QFT with monotonic flows,  the vacuum with the lowest free energy is the one with the smallest IR central charge.} As a special case,  if the theory admits a confining vacuum, this will be the ground state even if there exist possible fixed points at intermediate finite values of the coupling.

\section{Stability}
\label{sec:stab}

In this section  we briefly address the question of stability under small perturbations of the solutions we have considered so far, leaving most of the details to appendix \ref{app:reg}. This problem has been well studied in the past, and here we add some details and generalizations to the results which can be  found,  for example in \cite{csaki1,Kofman04,Kiritsis06,gkn}).

 The problem of stability of the bulk solution at the linearized level is equivalent to the absence of tachyons in the particle spectrum of the dual QFT. This is a  standard result in holography, which follows from the fact that normalizable bulk fluctuations are dual to single particle excitations in the boundary theory. In this discussion, the key concept is that of normalizability: turning on a normalizable mode around a background solution  correspond to changing the {\em state}, but not the {\em theory}, i.e. does not introduce new sources in the UV. In the IR, normalizability is tied to the requirement of  regularity of the perturbation.
Here we will discuss the general features of the  problem, leaving all the details to appendix \ref{app:reg}.

The question whether there exist  unstable modes in the normalizable spectrum around the background translates into the positivity of a certain quantum mechanical Hamiltonian. To see this explicitly, it is convenient to write the metric in conformal coordinates,
\be
ds^2 = e^{2A(r)}\left(dr^2 + \eta_{\mu\nu} dx^\mu dx^\nu\right).
\ee
The bulk  equations for linear fluctuations $h(r,x^\mu)$ (which here may be tensor or scalar) can be typically put in the form (after a redefinition of the wave-function):
\be \label{stab1}
-\de_r^2  h  - \de^\mu\de_\mu h + V_s(r) h = 0
\ee
where $V_s$ is an appropriate radial potential. We can go to Fourier space in the space-time coordinates, i.e. look for modes of a given $d$-momentum $p_\mu$:
\be\label{satb3}
h(r, x^\mu) = \psi(r) \exp(i p_\mu x^\mu).
\ee
Equation  (\ref{stab1}) then becomes a Schr\"odinger equation for the radial wave-function $\psi(r)$, with energy given by the $d$-dimensional mass squared:
\be\label{stab4}
-\de_r^2  \psi + V_s(r) \psi= \mu^2 \psi ,\qquad \mu^2 = - p^\mu p_\mu.
\ee
Since a solution of this equation also satisfies  $(\Box + \mu^2 )h = 0$,  the existence of solutions   with $\mu^2<0$ would imply the presence of modes  growing exponentially in time, i.e. an instability (which in the dual theory translates to the presence of a  tachyon in the spectrum). Therefore, {\em  stability under small perturbations is equivalent to the positivity of the radial hamiltonian} in equation (\ref{stab4}).

We can now state a very general result: for any bulk potential $V(\phi)$,   perturbative stability of holography RG flow solutions  is guaranteed if:
\begin{enumerate}
\item The UV is asymptotically $AdS$, it satisfies the BF bound, and the boundary conditions are chosen according to standard quantization (i.e. the leading solution   corresponds to the source).
\item The IR is either a regular fixed point or an IR-computable singularity.
\end{enumerate}
Notice that normalizability and computability play a crucial role here: in particular, positivity is guaranteed whenever only one (but not both) independent solutions of equation (\ref{stab4}),   at both the UV and IR,  is normalizable. This implies that the Hamiltonian is of the form $P^\dagger P$ for a suitable operator $P$. If instead both  the asymptotic solutions are  normalizable, then the Hamiltonian is no longer self-adjoint. Rather,  it admits an infinite number of  self-adjoint extensions (corresponding to  the choice of boundary conditions), some of which have negative eigenvalues.

The two conditions layed above are sufficient, but not necessary: for example, one can relax the requirement about standard quantization in the UV and allow for mixed boundary conditions when the scalar admits  two quantizations in the UV ($-d^2/4 < m^2 \ell^2 <-d^2/4 +1$), provided the resulting additional $\delta$-function  boundary term in the potential in  (\ref{stab4})  has the correct sign, and  does not introduce negative energy bound states. This corresponds in the field theory to a double-trace deformation which does not make the energy unbounded below.

\section*{Note Added}
\addcontentsline{toc}{section}{Note Added}

After the publication of this work we were informed by Z. Komargodski that there are potential examples of multiple flows from the same UV fixed point distinguished by different scalar vevs. Such examples could be constructed by ``softly" modifying for example Seiberg-Witten theory by a scalar potential with discrete minima in the Coulomb phase\footnote{A similar strategy was adopted in \cite{sw}. For scales well below the Seiberg Witten scale the modification is solvable.}.

On the other hand flows that are driven by a a vev of an irrelevant operator are present, for example, in theories with a baryonic branch in the moduli space like in N=1 sQCD, \cite{ss}. A holographic baryonic branch was discussed in \cite{mm}.

E. Shaghoulian brought to our attention \cite{duality} where ``chaotic" QFT  RG flows were discussed. Seiberg duality was used and a linearization of supersymmetric QFT $\beta$-functions to investigate RG cascades, and duality walls. However none of these were visible on the holographic side where the RG flow is smooth, monotonic and regular.

\section*{Acknowledgements}
\addcontentsline{toc}{section}{Acknowledgements}

We would like to thank Sergei Gukov, Zohar Komargodski, David Langlois, Ioannis Papadimitriou, Krzysztof Pilch,  Edgar Shaghoulian, Sergey Sibiryakov and Nick Warner for discussions and suggestions. We   especially thank Vasja Susic who contributed to the initial stages of this work.

This work was supported in part by European Union's Seventh Framework Programme under grant agreements (FP7-REGPOT-2012-2013-1) no 316165 and the Advanced
ERC grant SM-grav, No 669288.



\newpage
\appendix
\renewcommand{\theequation}{\thesection.\arabic{equation}}
\addcontentsline{toc}{section}{Appendix}
\section*{Appendix}

\section{Skipping fixed points: complete expression for V($\phi$)}
\label{app:V12}

In this appendix we present the explicit form of the potential \eqref{V12} used to construct the RG flow, shown in section \ref{skip}, where fixed points are skipped, including one UV fixed point that does not violate the BF bound.
There we started defining the derivative of the potential:
\be
	 V'(\phi):=-\phi\le(\phi^2-\phi^2_0\ri)\le(\phi^2-\phi^2_1\ri)\le(\phi^2-\phi^2_2\ri)
\le(\phi^2-\phi^2_3\ri)\le(\phi^2-{\Delta(\Delta-d)\over\phi^2_0~\phi^2_1~\phi^2_2~\phi^2_3}\ri),
\ee
with $V''(0)=\Delta(\Delta-d)<0$, in order to fix the position of the extrema and to ensure that $\phi=0$ corresponds to a local maximum of $V$. The extrema are placed at $0$, $\pm\phi_0$, $\pm\phi_1$, $\pm\phi_2$ and $\pm\phi_3$. Choosing the  of $V(0)$ such that the corresponding AdS length is one, we obtain the full form:
\begin{align}
		V(\phi)=&-d(d-1)+\Delta(\Delta-d){\phi^2\over2}\nonumber\\
				&-
			\le[
				\phi^2_0~\phi^2_1~\phi^2_2~\phi^2_3+
				\Delta(\Delta-d) \le({1\over \phi^2_0}+{1\over\phi^2_1}+{1\over\phi^2_2}+{1\over\phi^2_3}\ri)
			\ri]
			{\phi^4\over 4}\nonumber \\
				&+
			\Big[
				 \phi^2_0~\phi^2_1~\phi^2_2+\phi^2_0~\phi^2_1~\phi^2_3+\phi^2_0~\phi^2_2~\phi^2_3+~\phi^2_1~\phi^2_2~\phi^2_3		 
				\nonumber\\
				&+\le(
				{1\over \phi^2_0~\phi^2_1}+{1\over\phi^2_0~\phi^2_2}
+{1\over\phi^2_0~\phi^2_3}+{1\over\phi^2_1~\phi^2_2}+{1\over\phi^2_1~\phi^2_3}+{1\over\phi^2_2~\phi^2_3}
				\ri)
				\Delta(\Delta-d)
			\Big]
			{\phi^6\over 6}\nonumber\\
				&-
			\Big[
				\phi^2_0~\phi^2_1+\phi^2_0~\phi^2_2
				+\phi^2_0~\phi^2_3+\phi^2_1~\phi^2_2
				+\phi^2_1~\phi^2_3+\phi^2_2~\phi^2_3
				\nonumber\\
				&+\le(
					{1\over\phi^2_0~\phi^2_1~\phi^2_2}
					+{1\over\phi^2_0~\phi^2_1~\phi^2_3}
					+{1\over\phi^2_0~\phi^2_2~\phi^2_3}
					+{1\over~\phi^2_1~\phi^2_2~\phi^2_3}
				\ri)
				\Delta(\Delta-d)
			\Big]
			{\phi^8\over8}\nonumber\\
			&+
			\le[
				\phi^2_0+\phi^2_1+\phi^2_2+\phi^2_3
				+{1\over\phi^2_0~\phi^2_1~\phi^2_2~\phi^2_3}
				\le(\Delta(\Delta-d)-{1\over 8}\ri)
			\ri]
			{\phi^{10}\over10}
			-{\phi^{12}\over12},\label{V12c}
\end{align}
with $0<\phi_0<\phi_1<\phi_2<\phi_3$ and $V''(0)=\Delta(\Delta-d)<0$.

The class of functions parametrized by \eqref{V12c} was chosen because it allows for simple potentials which are
symmetric, negative for large enough $\phi$,
have an absolute maximum of AdS length equal to unity at $\phi=0$,
present two local minima for $\phi>0$ (corresponding to two IR fixed points, one of which will be skipped)
and have enough adjustable parameters.

Symmetry is desirable feature but by no means necessary. The reason for this choice is that we are interested in monotonic superpotentials that have a UV fixed point at $\phi=0$ and a symmetric potential allows us to restrict ourselves to $\phi\geqslant0$ or to $\phi\leqslant0$ without loss of generality, therefore simplifying the analysis. We will henceforth drop the $\pm$ sign.

We consider that there are enough adjustable parameters when they allow us to ensure that:
\begin{itemize}
\item  $V(\phi)$ is negative for all $\phi$.
\item $\phi=\phi_1$ correspond to a local maximum, therefore constraining $V''(\phi_1)$ to be negative,
\item the BF bound is respected at $\phi_1$, imposing a limit on how negative $V''(\phi_1)$ can be, as follows from equation \eqref{BFV},
\item the local maximum of $V$ at $\phi=0$ is an absolute maximum,
\end{itemize}

If the local maximum at $\phi_1$ violates the BF bound, only unstable cascading solutions will leave this UV fixed point. The corresponding UV dual field theory would not be unitary.

If the BF bound is respected at $\phi_1$, it is in principle possible that a standard RG flow starts at this local maximum of $V(\phi)$ and ends at a local minimum at $\phi_2$. However, this is not the case  if the parameters are chosen to be, for example:
\begin{align}
	&d = 4\quad \phi_0 = 1.0837\quad \phi_1= 1.1316 \nonumber\\
	&\Delta = 3  \quad \phi_2 =1.9200 \quad\phi_3 = 2.1500. \label{p2}
\end{align}
For the choice above,  there is no regular RG flow from $\phi_1$ to $\phi_2$.
This happens because there is a monotonic flow from $\phi=0$ to $\phi_2$. Imposing that the absolute maximum is at the origin makes it easier to adjust the parameters and ensure that a monotonic flow from $\phi=0$ to $\phi_2$ exists.

\section{Solutions of the superpotential  equation near a critical point}
\label{app:def}

In this appendix  we will discuss in detail how to determine whether there are multiple solutions of the superpotential reaching  a critical point, where uniqueness is not guaranteed. In particular we are interested in the question whether there is a continuous family of solutions or a discrete number.

We recall the superpotential equation, which we write in the form:
\be
	B^2=W^2 - \frac{2(d-1)}{d}\big(W'\big)^2,  \qquad B(\phi)\equiv \sqrt{-{4(d-1)\over d}V(\phi)}. \label{SPA}
\ee

A regular point of this equation is a value of $\phi$ at which $W(\phi)\neq B(\phi)$.  At regular points, the solution in each branch (growing and decreasing) is unique. However this may not be true at a critical point $\phi_*$ where $W(\phi_*) = B(\phi_*)$: sometimes there is an infinite number of solutions  $W(\phi)$, all flowing to the same critical point (e.g if the latter coincide with a maximum of $V(\phi)$). Below we show how this question can be analyzed systematically.

The question can be rephrased as follows: given a solution $W_0(\phi)$ reaching the critical curve at $\phi_*$, is there one (or more) other solution $W_1$   in the same branch which also has a critical point at $\phi_*$?  If the answer is positive,  then  $W_0(\phi)$ and $W_1(\phi)$  must be very close to each other as we approach $\phi_*$, since by assumption $W_0(\phi_*) = W_1(\phi_*) = B(\phi_*)$. In other words, the quantity $\delta W = W_1-W_0$ must satisfy:
\be \label{def1}
{\delta W \over W_0} \to 0 \qquad \phi \to \phi_*
\ee
Therefore, the question of whether or not a solution $W_0$ is unique is equivalent to asking whether there exist a slightly  deformed  solution $W_0 + \delta W$ with the property (\ref{def1}).

The standard method to investigate whether a certain solution admits small deformations close to a fixed point was first discussed in \cite{papa1} and it works  as follows\footnote{This discussion holds for each of the two branches (growing and decreasing) separately. Therefore, when we say a certain solution is unique, we mean ``unique in each branch.''}:
\begin{enumerate}
\item we first find a solution $W_0(\phi)$ close to the critical point, e.g. by expanding all terms in the superpotential equation in a power series around $\phi_*$.
\item We then solve a linearized superpotential equation for $\delta W$ close to $\phi_*$: given a solution $W_0$, inserting a deformation   $W_0+\delta W$ in equation  (\ref{SPA}) and assuming (\ref{def1}) is valid, we obtain the linearized equation:
\be\label{def2}
{1\over2} W_0' \delta W' - {d\over 4(d-1)}W_0\delta W = 0
\ee
whose solution  is given by:
\be\label{def3}
\delta W(\phi) = C \exp\left[ {d\over 2(d-1)}\int^\phi  {W_0 \over W_0'}  \right] ,
\ee
where $C$ is an arbitrary  constant, implying  that  deformations of this kind come in a continuous family. Notice that to the  order at which we are working, the above expression is independent of which leading solution $W_0$ we use: if use a different one, the change in the approximate solution  (\ref{def3}) will be of order $(\delta W)^2$.

\item Finally, we  have to check that the expression in equation (\ref{def3}) actually  satisfies (\ref{def1}), i.e. it is really a small deformation. If this is not the case, then we have to conclude that there are {\em no} small deformations of the  original solution $W_0$,  and $W_0$ is unique (possibly up to discrete choices related to the existence of different branches) in reaching the critical point at $\phi_*$.
\end{enumerate}

Below we analyze the possible critical points of interest (for simplicity we suppose $\phi_*=0$ in all the examples below),  and (re-)derive the results presented in section 2. We always assume $V(\phi)$ has an analytic expansion around any of its points.

\subsection{Analytic expansion at an extremum of $V$.} \label{prel}
As a preliminary, we first derive the result that around an extremal point of the potential at $\phi=0$, any solution of equation (\ref{SPA}) must have a regular power-series expansion at least up to order $\phi^2$ included.
Close to $\phi=0$, the potential is given by:
\be
V=-{d(d-1)\over \ell^2}+{m^2\over 2}\phi^2 + \cO(\phi^3) \label{VA}
\ee
Any solution of equation (\ref{SPA}) reaching the critical curve at $\phi=0$ has to satisfy:
\be\label{prel1}
W(0) = B(0)= {2(d-1)\over \ell}, \qquad W'(0) = 0.
\ee
The statement that $W$ has a regular expansion up to order $\phi^2$ is equivalent to the statement that $W''(0)$ is finite. Below,  we will prove that  it is impossible to satisfy equation (\ref{SPA}) close to $\phi=0$ together with (\ref{prel1})  and  with divergent $W''(0)$.

Indeed, writing $W^2-B^2= (W+B)(W-B)$, we can rewrite equation (\ref{SPA}) close to $\phi=0$ as:
\be\label{prel2}
W' (\phi)\approx \sqrt {d \, W(0)\over d-1}\sqrt{W(\phi)-B(\phi)}, \qquad \phi \to 0.
\ee
Furthermore, we can approximate $B(\phi)$ as:
\be\label{prel4}
B(\phi) \simeq B(0) -  c \phi^2, \qquad c= {m^2 \ell\over 2d}.
\ee
Defining:
\be\label{prel5}
\omega(\phi) \equiv W(\phi)-W(0),
\ee
equation  (\ref{SPA}) can be rewritten, close to $\phi=0$, as:
\be\label{prel6}
\omega' \simeq \sqrt {d\,  W(0)\over  d-1} \sqrt{\omega + c\phi^2},  \quad \phi\to 0.
\ee
We now suppose that $W'\to 0$ and  $W'' \to \infty$ as $\phi\to 0$. Then, the $\phi^2$ term in equation (\ref{prel6}) is negligible, and we obtain by integrating:
\be
\omega(\phi) \simeq {d\,  W(0) \over  d-1} \, \phi^2, \qquad \phi \to 0.
\ee
This expression has {\em finite} second derivative at $\phi=0$, contradicting our assumption that $\omega''(0)$ diverges.

Therefore, we conclude that $W(\phi)$ has a regular power series expansion at least up to second order (and with vanishing first derivative) around an extremum of the potential.

Furthermore, we can compute the coefficient of the $\phi^2$ term by differentiating twice the superpotential equation and evaluating it at $\phi=0$:
 \be \label{w''}
m^2\ell^2 =  \ell W''( \ell W'' - d)
\ee
which has the two solutions:
\be\label{def6}
W''=\Delta_{\pm}/\ell, \qquad \Delta_{\pm} = {d\over 2} \pm {1\over 2} \sqrt{d^2 + 4m^2\ell^2}.
\ee

\subsection{Deformations near  critical points.} \label{crit}

We now apply the analysis of the deformations around a critical point along the lines explained at the beginning of this Appendix.  We must consider separately the cases of a critical point at a  maximum, a minimum, or a generic point of $V(\phi)$. Points where both the first and second derivative of $V$ vanish are discussed along the same lines in section \ref{sssec:marginal}.

\paragraph{Critical point at a maximum of the potential.} The potential has an expansion of the form:
\be \label{def4}
V=-{d(d-1)\over \ell^2}+{m^2\over 2}\phi^2 + \cO(\phi^3)  \qquad -{d^2\over 4}< m^2\ell^2 <0.
\ee

We first  look   for a (positive) superpotential $W_0$  with a power-law expansion  around $\phi=0$: using equations (\ref{prel1}) and  (\ref{w''}) we find,  up to order $\phi^2$:
\be\label{def5}
W_{0}(\phi) \simeq {1\over \ell}\left[{2(d-1)\over \ell} +{\Delta \over 2} \phi^2 + \ldots \right]
\ee
where $\Delta$ is one of the roots in  equation (\ref{def6}).

Therefore, up to this point,  we have two possible solutions differing at order $\phi^2$, which we denote $W_{\pm}$. Notice that both solutions belong to the growing branch for $\phi >0$ and to the decreasing branch for $\phi<0$. On the other hand there is no decreasing (growing) branch for $\phi >0$ ($\phi<0$) since they would violate the bound $W \geq B$.

We now look for deformations of these solutions. Inserting the expression (\ref{def5}) into equation (\ref{def3}) we obtain:
\be\label{def7}
\delta W_- = C  \phi^{d\over \Delta_-},  \quad \delta W_+ = C  \phi^{d\over \Delta_+}
\ee
Both deformations vanish as $\phi \to 0$. However, because $0<\Delta_-< d/2$ and  $\Delta_+> d/2$, we have:
\be
	{d\over \Delta_-}>2, \quad \quad 0< {d\over \Delta_+}<2,
\ee
therefore only the deformation $\delta W_-$ is sub-leading with respect to the undeformed solution $W_0$. Instead, $\delta W_+$ is leading with respect to  $\phi^2$,  which is inconsistent with the result in section \ref{prel}. Therefore, only $\delta W_-$ is allowed as a consistent deformation. Notice that, for generic $m^2$,  this is a non-analytic term, so we would not have found it using a power-law expansion.

\paragraph{Critical point at a minimum of the potential.}
We can repeat the above procedure, but now with $m^2 >0$ in equation (\ref{SPA}).  We arrive at the same result (\ref{def5})  and  (\ref{def7}), where now:

\be
	0<{d\over \Delta_+}<1,\quad  \quad{d\over \Delta_-}<0.
\ee
Neither solution $\delta W_{\pm}$  is sub-leading with respect to $\phi^2$, and moreover $\delta W_-$ diverges as $\phi\to 0$.  Therefore neither  solutions  (\ref{def5}) close to a minimum of $V(\phi)$  admits sub-leading continuous deformations.

\paragraph{Generic point on the critical curve.}

Close to a  generic point $\phi_B$  on the critical curve, as shown in Section \ref{sssec:bW},  the superpotential takes the form:
\be
W_0(\phi) \simeq  W(0) \pm {2\over 3}\sqrt{2V'(\phi_B)}(\phi - \phi_B)^{3/2},
\ee
where the $\pm$ refers to the growing/decreasing branches and we have assumed the critical curve is reached from the right (i.e. $\phi > \phi_B$ and  $V'(\phi_B) >0$).

To investigate whether these solutions admit continuous deformations or whether they are unique, we resort again to equation (\ref{def3}), which in this  case leads to:
\be
\delta W_{\pm} = C \exp  \left[{d\over 2(d-1)} {W(0) \over V'(\phi_B)}\sqrt{(\phi - \phi_B)}\right]
\ee
As $\phi \to \phi_B$ this expression has a finite limit, $\delta W \to C$, therefore it does not satisfy the condition (\ref{def1}) for it to constitute a small deformation. We conclude that the bounces analyzed in section \ref{sssec:bW} do not admit small deformations.


\section{Stability of fluctuations around holographic RG flows}
\label{app:reg}

In this appendix we study linear fluctuations around a bounce. We concentrate on scalar perturbations, but the arguments we present here can be repeated for tensor perturbations as well.  Specifically, we show that the  spectrum of linear scalar perturbations  is determined by a quantum mechanical problem  governed  by a Hamiltonian which is strictly positive if we assume standard quantization in the UV,  and for a regular (or singular but computable)  IR.
Positivity of the Hamiltonian in turn implies absence of tachyonic instabilities.

\subsection{Scalar linear perturbations}
\label{ssec:back}

Consider the following background configuration:
\eql{d0}{
	\phi(r,x^\m)=\phi(r),\quad ds^2=e^{2A(r)}\le(dr^2+\eta_\mn dx^\m dx^\n \ri)
}
The background Einstein equations are:
\begin{subequations}\label{EEr}
	\begin{align}
		&A_{,r}^2
		={1\over d(d-1)}
			\le[
				\ha { {\phi_{,r}}}^2- e^{2A(r)}V(\phi)
			\ri]
			\label{EEr1}\\
		&A_{,rr}-A_{,r}^2+{1 \over 2(d-1)} { \phi_{,r}}^2=0
			\label{EEr2}
	\end{align}
\end{subequations}

The background Klein-Gordon equation is:
\be
	\phi_{,rr}+(d-1)A_{,r}\phi_{,r}-e^{2A}V'(\phi)=0 \label{KGb}
\ee
and can actually be deduced from \eqref{EEr}.

Now we consider fluctuations around \eqref{d0}:
\begin{subequations} \label{d1}
\begin{align}
	&\phi(X)=\phi(r)+\chi (X),\label{d1a}\\
	&ds^2=a^2(r)\le[
			 dr^2(1+2D)
			+2 B_\m dx^\m dr
			+\le(\emn +h_\mn\ri)dx^\m dx^\n
		\ri].\label{d1b}
\end{align}
\end{subequations}
$B_\m$, $D$ and $h_\mn$ are functions of $(r,x^\m)$. Greek indexes are raised and lowered with $\emn$.
Under infinitesimal diffeomorphisms $\delta X^A=(\delta r, \delta x^\m)=(\xi^r,\xi^\m)$ these perturbations transform as:
\begin{subequations}\label{gauge}
	\begin{align}
		\delta h_\mn&=-2\p_{(\m} \xi_{\n)} -2{d A\over dr}\emn\xi^r ,\label{g1}\\
		\delta B_\m&=-2\p_r \xi_\m -\p_\m\xi^r,\label{g2}\\
		\delta D&=-\p_r \xi_r -{d A\over dr}\xi^r,\label{g3}\\
		\delta \chi&=-\phi'(r)\xi^r.\label{g4}
	\end{align}
\end{subequations}
The scalar fluctuations contained in $h_{\mu\nu}$ and $B_\mu$ are:
\begin{subequations}
\begin{align}
	&B_\m\equiv \p_\m B,\\
	&h_\mn\equiv\emn C +\p_\m \p_\n E
\end{align}
\end{subequations}

Under infinitesimal diffeomorphisms of the form $\delta X^A=(\xi^r,\p^\m\xi)$ the perturbations $B$, $C$ and $E$ have the following transformations:
\begin{subequations}\label{gauge2}
\begin{align}
&B\to B-\xi_{,r}\label{gauge2B}\\
&C\to C-{dA\over dr}\xi^r\label{gauge2C}\\
&E\to E-\xi\label{gauge2E}
\end{align}
\end{subequations}
Assume that the perturbations on constant $\phi(r)$ slices have an invariant d-dimensional mass, i.e.
\begin{subequations}\label{mass}
\begin{align}
	&\Box_d h_\mn=\p_\s \p^\s h_\mn=\o^2  h_\mn(r,x^\mu)\label{massh},\\
	&\Box_d B=\o^2  B,\label{massB}\\
	&\Box_d D=\o^2  D\label{massD}.
\end{align}
\end{subequations}
We can also rewrite the scalar perturbations $C$ and $E$ with well-defined, non-vanishing $d$-dimensional mass $\o^2 $ directly from $h_\mn(r,x^\mu)$ as:
\begin{subequations}
\begin{align}
&C(r,x^\mu)
	={1\over 2(d-1)}
		\le(
			\eta^\mn-{\p^\m \p^\n \over \o^2 }
		\ri)h_\mn
		\label{C0-a}\\
&E(r,x^\mu)
	=-{1\over 2(d-1)\o^2 }
		\le(
			\eta^\mn-(2d-1){\p^\m \p^\n \over \o^2 }
		\ri)h_\mn.
		\label{C0-b}
\end{align}
\end{subequations}

Using the transformations \eqref{gauge} and \eqref{gauge2} we can define the following gauge-invariant scalar perturbations:
\begin{subequations}\label{Bardeen}
	\begin{align}
		 &\Psi(r)\equiv C+A_{,r}(E_{,r}-B),\label{Bardeena}\\
		 &\Phi(r)\equiv D+A_{,r}(E_{,r}-B)+(E_{,r}-B)_{,r},\label{Bardeenb}\\
		 &\z(r,x^\mu)
		\equiv C(r,x^\mu)
				-{A_{,r}\over \phi_{,r}}\chi(r,x^\mu)\label{Bardeenc}.
	\end{align}
\end{subequations}
The variables $\Psi$ and $\Phi$ are called Bardeen variables.


\subsection{Equations of motion for the linear perturbations}

For concreteness, {\bf from now on we work in $d=4$ }but the results generalize to any $d>1$. The Einstein equations for the perturbations are:
\eql{EEp}{
	\delta G^\m_\n=\ha \delta T^\m_\n
}
with $8\pi G_N^{(5)}=1/2$, where $G_N^{(5)}$ is the five-dimensional Newton-constant.

In d=4 equations \eqref{EEp} take de form:
\begin{subequations}\label{EP0}
	\begin{align}
		{}^r_r: \quad&
		3\le[\Box_4\Psi
		+4A_{,r}\le(C_{,r}
		-A_{,r}D\ri)\ri]
		=-\ha\le(
							 {\phi_{,r}}^2D-{\phi_{,r}}{\chi_{,r}}+a^2V_{,\phi}\chi
							\ri),\label{rr}\\
		%
		{}^r_\m:\quad&
		-3\p_\m\le(
				 C_{,r}
				-A_{,r} D
				\ri)
		=\ha{\phi_{,r}}\p_\m\chi,\label{rm}\\
		%
		{}^{~~~\m}_{\n\neq\m}:\quad&
		-\p^\m\p_\n
		\le(2\Psi+\Phi\ri)=0,\label{mn}\\
		{}^{\m}_{\m}:\quad&
		\le(\Box_4-\p^{\hat \m}\p_{\hat \m}\ri)
				\le(2\Psi+\Phi\ri)
				+\le( 3A_{,r}+\p_r\ri)3C_{,r}
							-3A_{,r}D_{,r}
							-6D\le[(A_{,r})^2+A_{,rr}\ri]\nonumber\\
		&\qquad\qquad\qquad\qquad
		=\ha\le(
							{\phi_{,r}}^2D-{\phi_{,r}}{\chi_{,r}}
							- a^2V_{,\phi}\chi
							\ri),\label{mm}
	\end{align}
\end{subequations}
In equation \eqref{d1a} we defined the scalar perturbation $\chi$. It obeys the following perturbed Klein-Gordon equation in d=4:
\begin{align}
			&\Box_4 \chi +\chi_{,rr}+3A_{,r}\chi_{,r}
			+\phi_{,r}\Box_4 (E_{,r}-B)+\nonumber\\
			 &\quad+4\phi_{,r}C_{,r}-\phi_{,r}D_{,r}-2D(3A_{,r}\phi_{,r}+\phi_{,rr})=e^{2A(r)}V''(\phi_0)\chi \label{k0}
\end{align}
Equations \eqref{rm} and \eqref{mn} are constraints. Using these constraints we can show that the equation \eqref{mm} is trivially satisfied. In terms of $\Phi$, $\Psi$, and $\z$, equations \eqref{EP0} and \eqref{k0} can be reorganized and reduced to:
\begin{subequations} \label{EPF}
	\begin{align}
		0=&\Phi(r,x^\m)+2\Psi(r,x^\m),
		\label{ea4}\\
		%
		0=&
			\z-
				{
				\le(A_{,rr}-3A_{,r}^2\ri) \Psi
				-A_{,r}\Psi_{,r}
				\over
				A_{,rr}-A_{,r}^2}
		\label{eb4}\\
		%
		0=&
		\Box_4 \Psi
		+\Psi_{,rr}
		+\Psi_{,r}
			\le(
				3A_{,r}
				-2{\phi_{,rr}\over \phi_{,r}}
			\ri)
		+4\Psi
			\le(
				A_{,rr}
				-A_{,r}{\phi_{,rr}\over \phi_{,r}}
			\ri)
		\label{ec4}\\
		0=&
		\z_{,rr}
		+\le[
			3A_{,r}
			+2 \le(
				{\phi_{,rr}\over \phi_{,r}}
				-{A_{,rr}\over A_{,r}}
				\ri)
		\ri]\z_{,r}
		+\Box_4\z.
		\label{zeta1}
	\end{align}
\end{subequations}
in agreement with \cite{Kiritsis06}.
Equation \eqref{eb4} can be inverted to give $\Psi$ as an integral of $\z$:
\begin{subequations}\label{Psi0}
	\begin{align}
		\Psi(r,x^\m)
	&=e^{-3A(r)}A_{,r}
	\int^r
		 e^{3A(\hat r)}
		 \z(\hat r)
		 \le(
		 	1-{A_{,\hat r\hat r}\over A_{,\hat r}^2}
		\ri)
		 d\hat r
		 \label{Psi1}\\
	&={1\over 6}e^{-3A(r)}A_{,r}
	\int^r
		 e^{3A(\hat r)}
		 \z(\hat r)
		 \le(
		 	{\phi_{,\hat r}(\hat r)\over A_{,\hat r}(\hat r)}
		\ri)^2
		 d\hat r
		 \label{Psi2}
	\end{align}
\end{subequations}
where in the second line we used the Einstein equation \eqref{EEr2}.

The fluctuations of the lowest derivative curvature invariants can be written solely in terms of $\Psi$, from \eqref{Bardeena}, and the background fields $A$ and $\phi$:
\begin{subequations}\label{curv}
	\begin{align}
		&\delta R
			=-2e^{-2A(r)}\Big\{\Box_4 \Psi
				+4\le[\Psi_{,rr}+2A_{,r}\Psi_{,r}\ri]
				+8\le[3\le(A_{,r}\ri)^2+2A_{,rr} \ri]\Psi
			\Big\}
						\label{PhiRicci_s}\\
		\nonumber\\
		&\delta (R_\mn R^\mn)
			=4e^{-4A(r)}\Big\{
			\le[
				6(A_{,r})^2-2A_{,rr}
			\ri]\Box_4 \Psi
			+2\le[
				5A_{,rr}+3(A_{,r})^2
			\ri]\Psi_{,rr}
			+
			\nonumber\\
			&\quad\qquad
			+2\le[
			21A_{,rr}+27(A_{,r})^2
			\ri]A_{,r}\Psi_{,r}+
			\nonumber\\
			&\quad\qquad\qquad+8\le[
					9(A_{,r})^4
					+6(A_{,r})^2A_{,rr}
					+5\le(A_{,rr}\ri)^2
			\ri]\Psi
			\Big\}
			\label{PhiRicci_t2}\\
		\nonumber\\
		&\delta (R_{\r\s\mn} R^{\r\s\mn})
						=8e^{-4A(r)}\Big\{
				3(A_{,r})^2\le[
								\Box_4 \Psi+4A_{,r}\Psi_{,r}
							\ri]
				+4A_{,rr}\le(
								\Psi_{,rr}+\Psi_{,r}A_{,r}
							\ri)\nonumber\\
				&\qquad-2A_{,rr}\le(
								\Box_4 \Psi-4A_{,r}\Psi_{,r}
							\ri)
				+8\Psi\le[2\le(A_{,rr}\ri)^2+3(A_{,r})^4\ri]
			\Big\}
			\label{PhiRiemann_t2}
	\end{align}
\end{subequations}
Curvature invariants with higher derivatives can also be written in terms of $\Psi$ and its derivatives

To analyze the properties of linear fluctuations, it is convenient to recast   equations (\ref{ec4}) and (\ref{zeta1}) in the form of a Schr\"odinger equation.

We define the functions $L(r)$ and $G(r)$ by:
\eql{LG}{
\qquad L(r)\equiv \exp({-G(r)}):={\phi_{,r}\over A_{,r}}e^{3A(r)/2}}
Equations \eqref{ec4} and \eqref{zeta1} can be put in the form  of Schr\"odinger equations through the following change of variables:
\begin{subequations}\label{ch}
\begin{align}
	&{\cal U}
	=
	 {e^{3A/2}\over \phi_{,r}}\Psi\label{chPsi}\\
	&{\cal Z}=e^{-G(r)}\z \label{chz}
\end{align}
\end{subequations}
with $G(r)$ given by equation \eqref{LG}.

We define the operators $P$ and $\Pt$:
\eql{PPt}{P:=\p_r+G_{,r}
	\quad\text{and}\quad
	\Pt:=-\p_r+G_{,r},
}
We consider states with well-defined 4-dimensional mass:
\begin{subequations}\label{f4}
\bea
&\Box_4 \Psi=\m^2\Psi,\quad \Psi(r,x)=\Psi(r)e^{ik_\n x^\n},\label{f4a}\\
&\Box_4 \z=\m^2\Psi,\quad \z(r,x)=\z(r)e^{ik_\n x^\n},\quad k^2=\m^2.\label{f4b}
\eea
\end{subequations}

With the change of variables \eqref{ch} and the definitions \eqref{LG} and \eqref{PPt}, the equations of motion \eqref{ec4} and \eqref{zeta1} become:
\begin{subequations}\label{H1}
\begin{align}
	H_u{\cal U}\equiv P \Pt{\cal U} =\m^2 {\cal U} \label{Hu1}\\
	H_z{\cal Z}\equiv\Pt P{\cal Z} =\m^2 {\cal Z} \label{Hz1}
\end{align}
\end{subequations}
where
\begin{subequations}\label{H2}
\begin{align}
	H_u=-{d^2\over dr^2}+V_u(r),
	\quad&\text{with}\quad
	V_u(r)
		\equiv \le(G_{,r}\ri)^2+G_{,rr}
		\equiv \le(L^{-1}\ri)_{,rr}L,\label{Hu2}
	\\
	H_z=-{d^2\over dr^2}+V_z(r),
	\quad&\text{with}\quad
	V_z(r)
		\equiv \le(G_{,r}\ri)^2-G_{,rr}
		\equiv L_{,rr}L^{-1}.\label{Hz2}
\end{align}
\end{subequations}
and $L$ is given by \eqref{LG}. The constraint equation \eqref{eb4} assumes a simple form in these variables:
\eql{Sp2}{{{\cal Z}=6(\p_r-G_{,r}) {\cal U}=-6 \Pt {\cal U}}.}
Equation \eqref{Sp2} shows $\cal Z$ and $\cal U$ are not independent. Equations \eqref{H1} and \eqref{H2} show that the spectra of these fluctuations is given by different Schr\"odinger equations. The spectral problem is consistent if both spectra agree and in the next subsection we will show that this is the case.


\subsection{Consistency}
\label{consist}

Here we will show that the spectra of $H_u$ and $H_z$ coincide, and therefore the fluctuations $\cal Z$ and $\cal U$ describe the same physical set of modes. This is necessary for consistency as there is a single physical bulk scalar fluctuation.

The Hamiltonian operators $H_u$ and $H_z$ from \eqref{H1} are different in form and act on different sets of functions. We know however that $\Psi$ and $\zeta$ are related by the constraint \eqref{eb4} which in terms of $\cal U$ and $\cal Z$ is \eqref{Sp2}.
Clearly, for every $\cal U$ there is a corresponding $\cal Z$, which is true in particular for the eigenfunctions of $H_u$ of eigenvalue $\m^2 $, denoted by ${\cal U}_\m$ :
\eql{Sp3}{H_u {\cal U}_\m\equiv(P\Pt) {\cal U}_\m=\m^2 ~{\cal U}_\m.}
Therefore, there for every ${\cal U}_\m$ satisfying \eqref{Sp3}, via \eqref{Sp2} there is a unique function ${\cal Z}_\m$ such that:
\eql{Sp4}{
	H_z{\cal Z}_\m
	=-6H_z \Pt {\cal U}_\m
	=-6 \Pt (P~\Pt) {\cal U}_\m
	=-6 \Pt \m^2  {\cal U}_\m
	=\m^2 {\cal Z}_\m.
}
We stablished that every eigenfunction of $H_u$  has a corresponding eigenfunction of $H_z$ with the same eigenvalue.

For the spectra to agree we also need the converse: for every eigenfunction of $H_z$ there must be a corresponding eigenfunction of $H_u$ with the same eigenvalue.  This time the hypothesis is:
\eql{Sp5}{
	H_z{\cal Z}_\m
	=\Pt P{\cal Z}_\m
	=\m^2 {\cal Z}_\m.
}
Multiplying both sides of \eqref{Sp5} by $P$ we obtain:
\eql{Sp6}{
	(P\Pt) P{\cal Z}_\m
	=H_u  P{\cal Z}_\m
	=\m^2  P{\cal Z}_\m.
}
therefore showing that to every eigenstate ${\cal Z}_\m$ of $H_z$ there is a corresponding eigenstate of $H_u$, $P{\cal Z}_\m$, with the same eigenvalue.
From equations \eqref{Sp2} and \eqref{Sp3} we can fix the relation  between $U_\m$ and $P{\cal Z}_\m$:

\eql{Sp7}{
		{\cal U}_\m
		=-{1\over 6\m^2 }P{\cal Z}_\m
		=-{1\over 6\m^2 }\le(\p_r+G_{,r}\ri){\cal Z}_\m
}
Expression \eqref{Sp7} is valid only in the absence of zero modes of $P$, because they lead to a vanishing $\m^2$. We have reduced the problem of the uniqueness of the spectrum to the problem of finding normalizable zero modes of $P$. Below we show that in fact these modes are absent.

Let ${\cal Z}_0$ be a zero mode of $P$. Then,
\eql{Sp8}{{\cal Z}_0=Ce^{-G(r)}\equiv C_z L(r)}
using the definition \eqref{LG}, with $C_z$ an arbitrary integration constant. For a solution $(\phi(r), A(r))$ of an holographic RG flow that has a regular IR behavior, as described in subsection \ref{sssec:min}, the zero mode \eqref{Sp8} is not normalizable.

The zero mode of $H_u$ associated with $\Pt$ is:
\eql{Sp9}{{\cal U}_0=Ce^{G(r)}\equiv C_u \le(L(r)\ri)^{-1}}
For solutions that start are asymptotically AdS at the UV, as described in subsection \ref{sssec:max}, the zero mode \eqref{Sp9} is not normalizable.

Therefore, for RG flows that interpolate between AdS fixed points our operators $H_u$ and $H_z$ from \eqref{H1} will not contain zero modes, and their spectra coincide.


\subsection{Stability of linear perturbations}
\label{pos}

 Here we show that for RG flows with a regular UV fixed point with no multi-trace deformations, and a IR which is regular or has a good singularity, the spectrum of fluctuations is positive.

The eigenvalues $\m^2$ of the operators $H_u$ and $H_z$ from equation \eqref{H2} correspond to the $4-dimensional$ rest-mass of the fluctuations. Equation \eqref{f4}, where $\m^2$ is shown to correspond to the square of the momentum 4-vector, implies that a negative $\m^2$ leads to an exponentially growing mode in time and, therefore, to an instability. {\em Stable fluctuations are those without negative $\m^2$ modes.} In this subsection we present sufficient conditions for the stability of the fluctuations: computability at the IR and the absence of multi-trace deformations at the UV.

In the present subsection we denote derivatives with respect to $r$ by a prime.
\eql{'}{{df(r)\over dr}=f'(r)}
The spectra of $H_u$ and $H_z$ coincide, as shown in subsection \ref{consist}, therefore we study only the positivity of the operator $H_u$ defined in equation \eqref{Hu2}:
\eql{Hu}{
	H_u=-{d^2\over dr^2}+V_u(r)=(\p_r+G')(-\p_r+G',),
	\quad\text{with}\quad
	V_u(r)
		\equiv G'^2+G''.
	}
where $G(r)$ was defined in \eqref{LG}.

Consider one eigenfunction ${\cal U}_\m(r)$ of $H_u$ with eigenvalue $\m^2$. It satisfies:
\begin{align}
	\m^2\int_0^\infty{\cal U}_\m^*(r){\cal U}_\m(r) dr
	&=
	\int_0^\infty{\cal U}_\m^*(r)~H_u ~{\cal U}_\m(r) dr\nonumber\\
	&=
	\int_0^\infty{\cal U}_\m(r)~H_u ~{\cal U}^*_\m(r) dr\nonumber\\
	&=\int_0^\infty
		{\cal U}_\m^*(r)\le(
				-{d^2\over dr^2}+G'^2+G''
				\ri)
		{\cal U}_\m(r) dr\nonumber\\
	&=\int_0^\infty
		\le|
			{\cal U}_\m'(r)
			-G'(r){\cal U}_\m
		\ri|^2
		dr
		-\le[{\cal U}^*_\m(r)\le({\cal U}_\m'
		+G'(r){\cal U}_\m
		 \ri)
		 \ri]_0^\infty\nonumber\\
	&=\int_0^\infty
		\le|
			\tilde P{\cal U}_\m
		\ri|^2
		dr
		-\le[{\cal U}^*_\m(r)P{\cal U}_\m(r)
		 \ri]_0^\infty
	\label{pos1}
\end{align}
In the last line we used the definitions \eqref{PPt} of $P$ and $\Pt$.
The only potentially negative term in \eqref{pos1} is the boundary term, if it vanishes the spectrum is strictly positive.

To calculate the boundary term in \eqref{pos1} we consider the asymptotic behavior of ${\cal U}_\m$ as $r$ goes to $0$ corresponding to a solution leaving a UV fixed point which respects the BF bound. In this case, $\phi\approx\phi_0r^\Delta$ and $A\approx-\log(r/\ell)$ for small $r$ and:
\begin{subequations} \label{pos2}
\begin{align}
&G(r)=\le(d-1\over2\ri)\log\le(r\over \ell\ri)
-\log\le(\phi_0r^\Delta\Delta\ri)\label{pos2a}
+\cO(r)\\
&V_u(r)=
{\a_u \over r^2} +\cO(r^{-1}),
\quad
\text{with}
\quad
\a_u= {(d-2\Delta-1)(d-2\Delta-3)\over 4.}
\label{pos2b}
\end{align}
\end{subequations}
Because the potential is divergent at $r=0$ can approximate the equations of motion near the origin by:
\be
	{\cal U}_{\m}''\approx V_u(r){\cal U}_\m. \label{posa}
\ee
Notice that the coefficient $\a_u$ can change sign according to the dimension $\Delta$.
Close to $r=0$, assuming that the unitarity bound is respected, i.e. $\Delta \geqslant (d-2)/2$, we obtain:
\eql{pos3}{
\a_u<0 ~\text{ if }~{d-2\over 2}\leqslant \Delta<{d-1\over 2}
\quad \text{ and }\quad
\a_u\geqslant0 ~\text{ if }~\Delta\geqslant{d-1\over 2}
}
Equation \eqref{pos2} has two linearly independent solutions:
\eql{pos4}{
{\cal U}_{\m,1}\approx r^{(d-2\Delta-1)/2},
\qquad
{\cal U}_{\m,2}\approx r^{(3-d+2\Delta)/2}
}

From \eqref{PPt} and \eqref{pos4} we have:
\begin{subequations}\label{pos5}
\begin{align}
{\cal U}^*_{\m,1}~P~{\cal U}_{\m,1}
	&\approx(d-2\Delta-1)r^{d-2\Delta-2}
	\label{pos5a}\\
{\cal U}^*_{\m,2}~P~{\cal U}_{\m,2}
	&\approx r^{2-d+2\Delta}
	\label{pos5b}\\
{\cal U}^*_{\m,2}P{\cal U}_{\m,1}+{\cal U}^*_{\m,1}P{\cal U}_{\m,2}
	&\approx d-2\Delta
	\label{pos5c}
\end{align}
\end{subequations}
For dimensions $\Delta$ above the unitarity bound, \eqref{pos5a} diverges as $r$ goes to $0$ while \eqref{pos5b} vanishes. In this case only the mode with asymptotic behavior ${\cal U}_{\m,2}$, given by equation \eqref{pos3}, is acceptable. When the unitarity bound is saturated, the potential vanishes to leading order and sub-leading terms should be considered. Saturation of this bound implies that the operator $\cO$ at the dual field theory is a free scalar field \cite{saturation}. For concreteness we keep here $\Delta$ above the unitarity bound.
Therefore, the UV boundary condition is:
\begin{align}
&\cU_{\m}\approx r^{(3-d+2\Delta)/2}
	\quad\text{with }\quad
		\Delta>{d-2\over 2}
	\quad\text{ as }\quad
		r\to0
.\label{pos6}
\end{align}
A fluctuation is normalizable if $\cal Z$ is square-normalizable \cite{Kiritsis06}. From the constraint equation \eqref{Sp2} we deduce that the fluctuation $\cU$ is normalizable iff
\be
\int|\Pt \cU|^2dr<\infty.\label{pos7}
\ee

Now we show the assumption of existence of normalizable negative $\m^2$ modes for regular geometries leads to a contradiction.

When $r$ asymptotes to infinity, the potential $V_u(r)$ vanishes and we can approximate \eqref{Hu1} by:
\eql{pos8}{\cU''\approx\mu^2 \cU \implies
\begin{cases}
 \cU_{\m^2}^+ \approx e^{+\sqrt{|\m^2|}r}\\
  \cU_{\m^2}^- \approx e^{-\sqrt{|\m^2|}r}
\end{cases}
.}
We need the solution to be normalizable according to \eqref{pos7}, so we need to calculate the asymptotic form of $\Pt\cU_{\m^2}$.
For every $\cU(r)$, the definition \eqref{PPt} allows us to write:
\eql{pos9}{\Pt\cU\equiv-{d\over dr}\le(e^{-G(r)}\cU\ri)}
If the solution ends at an IR fixed point, $G(r)$ will be given by \eqref{pos2a} with $\Delta\equiv\Delta_-<0$ and the $\cO(r)$ replaced by terms that vanish as $r$ asymptotes to $\infty$.
\eql{pos10}{\Pt\cU_{\m^2}^\pm\propto r^{\Delta_--(d-1)/2}e^{\pm2\sqrt{|\m^2|} r}\le[\pm\sqrt{|\m^2|}+{1\over r}\le(\Delta_--{d-1\over 2}\ri)\ri]+...}
and only the $\cU_{\m^2}^-$ solution is normalizable. For this solution,
\eql{pos11}{\cU_{\m^2}^{-*}P \cU_{\m^2}^-\equiv \cU_{\m^2}^{-*}{d\over dr}\le(e^G\cU_{\m^2}^-\ri)\to
-\sqrt{|\m^2|} e^{-2\sqrt{|\m^2|} r}r^{-\Delta_-+(d-1)/2}+...
}
which vanishes as $r$ asymptotes to infinity. If the $UV$ fixed point also satisfies the boundary condition \eqref{pos6}, then the boundary term in \eqref{pos1} vanishes, leaving us with:
\be
\m^2=
		{\int_0^\infty
		\le|
			\tilde P{\cal U}_\m(r)
		\ri|^2
		dr
		\over
		\int_0^\infty\le|{\cal U}_\m(r) \ri|^2dr
		}
		>0
	\label{pos12}
\ee
in contradiction with the hypothesis that $\m^2<0$.
In this case any two differentiable functions $\cU_a(r)$ and $\cU_b(r)$ with boundary conditions \eqref{pos6} and normalizable according to \eqref{pos7} satisfy:
\eql{pos13}{
\int_0^\infty\cU_a P \cU_b dr =\int_0^\infty\cU_b\Pt\cU_a dr
}
therefore showing that, in this space of functions,
\eql{pos14}{\Pt=P^\dagger\quad\text{and}\quad P=\Pt^\dagger.}
Therefore we can write equations \eqref{H1} in the following way:
\eql{pos15}{H_u=PP^\dagger \quad\text{and}\quad H_z=P^\dagger P}
showing explicitly the positivity of these operators. This generalizes to the case where the IR is singular but computable in the sense explained in subsection \ref{sssec:specc}.

Consider now a solution with a singular but computable IR. For concreteness, consider the case of the asymptotically exponential potential \eqref{good1} satisfying the computability bound \eqref{spec8}. In that case the scalar field profile and the scale factor in domain-wall coordinates are given in equation \eqref{good3}. Only the scale factor gives a divergent contribution to the potential $V_u(r)$ defined in equation \eqref{Hu2} and it is therefore the leading term. Using the scale factor in conformal coordinates, \eqref{spec2} we obtain (here $\g_c^2=1/3$  from equation \eqref{spec8}):
\begin{equation}
	G'(r)\approx
	\begin{cases}
		{q(r_*-r)^{-1}},\quad r\to r_*,\quad{1\over 6}<\g^2<{1\over 3}\\
		{-q r^{-1}},\qquad\qquad r\to \infty,\quad\g^2<{1\over 6}
	\end{cases}\label{pos16a}
\end{equation}
and the following form of the potential:
\begin{equation}
	V_u(r)\approx
	\begin{cases}
		{q(q-1)(r_*-r)^{-2}},\quad r\to r_*,\quad{1\over 6}<\g^2<{1\over 3}\\
		{q(q+1) r^{-2}},\qquad\qquad r\to \infty,\quad\g^2<{1\over 6}
	\end{cases}\label{pos16b}
\end{equation}
 and $q$ was defined in equation \eqref{spec6}. For $d=4$:
\be
q={3\over 2}{1\over 6\g^2-1}.\label{pos17}
\ee
We must distinguish the situation  $0<r<r_*$, in which the potential $V_u(r)$ is divergent at $r=r_*$,  from the one  in which $0<r<\infty$ and the potential $V_u(r)$ vanishes at infinity. In the second case   positivity of the spectrum is guaranteed  by applying same analysis from equations \eqref{pos7} to \eqref{pos12}. In the first case, however, the divergent potential allows us to use the approximation \eqref{posa} so that:
\eql{pos18}{\cU(r)\approx A(r_*-r)^{1-q}+B(r_*-r)^{q},\quad r\to r_*,\quad{3\over 2}<q<\infty.}
From equations \eqref{PPt}, \eqref{pos16a} and \eqref{pos18} we deduce:
\bea
\label{pos19}
\Pt\cU(r)
	\approx
	 2A(r_*-r)^{-q}+2qB(r_*-r)^{q-1},\quad q>{3\over2}.
\eea
Therefore, $\cU$ is normalizable in the IR, according to \eqref{pos7}, iff $A=0$. This condition leads to a vanishing $\cU P\cU$ at the IR:
\eql{pos20}{
\cU P\cU\approx(2q-1)A\le[A(r_*-r)^{1-2q}+B+...\ri] \to 0, \quad  r\to r_*}
therefore ensuring that the spectrum is positive, as follows from \eqref{pos1}.
Thus, the spectrum is  positive  if the computability bound is respected and if the boundary conditions \eqref{pos6} are satisfied at the UV.

\subsection{Tensor perturbations}

For completeness we briefly discuss tensor modes.
Besides scalar perturbations, the Einstein-dilaton theories allow for tensor perturbations of the transverse traceless part of $h_{\mu\nu}$ in equation (\ref{d1b}),
\be
\de^\mu h_{\mu\nu} = h^\mu_{\, \mu} = 0.
\ee
These are gauge invariant modes and contain $d(d-1)/2 -1$ components. As it is well known, Einstein's equation for these modes reduces to  the scalar massless wave equation  in the background metric (\ref{d0}),
\be
\de_r e^{(d-1)A}\de_r h_{\mu\nu} +   e^{(d-1)A} \de^\rho\de_\rho  h_{\mu\nu} = 0.
\ee
As in the case of the scalar mode, this equation can be turned into a Schr\"odinger-like  equation by considering $d$-momentum eigenmodes with $p^\mu p_\mu = -\mu^2$ and by defining:
\be
h_{\mu\nu} = e^{-(d-1)A/2}\psi_t(r) \exp(i p_\rho x^\rho) \epsilon_{\mu\nu}
\ee
where $\epsilon_{\mu\nu}$ is a  constant polarization tensor. The resulting equation for the radial wave-function $\psi(r)$ is:
\be
\left[-\de_r^2  + V_t(r) \right]\psi_t = \mu^2 \psi_t \, , \quad V_t(r) = {(d-1)^2\over 4}\left({d A \over dr}\right)^2 - {(d-1)\over 2}{d^2 A \over dr^2}.
\ee
The same analysis we have applied to the Hamiltonian for scalar modes in section (\ref{pos}) also applies here. In particular, in the UV the modes behave in the same way as scalar modes but with $\Delta=4$, and the IR asymptotics are also  of the same type. Therefore, stability in the tensor sector does not add extra conditions.



\section{Regularity of bouncing solutions}
\label{app:stab}

In this section we  show that $\Psi$ has a regular power series expansion around a bounce. This implies, via equations (\ref{curv}),   that no curvature  singularity can arise from the linear fluctuations.

The fluctuation $\Psi(r)$ obeys equation \eqref{ec4}.
After a change of variables to $\cal U$ via equation \eqref{chPsi} we obtain the Schr\"odinger equation \eqref{Hu1} with Hamiltonian \eqref{Hu2}. We expand $L(r)$ from \eqref{LG} near a bounce at $r=r_B$:
\eql{f12}{
	L(r)=\sum_{n=1}^3{L_n \over n!}(r-r_B)^n+\cO (r-r_B)^4
	}
	We also expand the background scalar field and scale factor as power series around a bounce:
\eql{phibr}{
\phi(r)=\phi_B+\sum_{n=2}^{\infty}{\phi^{(n)}_B\over n!}(r-r_B)^n,
}
\eql{Abr}{A(r)=A_B+\sum_{n=1}^{\infty}{A^{(n)}_B\over n!}(r-r_B)^n.}
Substituting \eqref{phibr} into \eqref{f12} we obtain:
\eql{L1}{
	L_1
	=
	e^{ 3 A_B /2}{\phi_B^{(2)}\over A_B^{(1)}}
	}
\eql{L2}{
	L_2
	=
	e^{3A_B /2}
	\le[
		 \phi_B^{(2)}
		 \le(
		 	3
			-2 {A_B^{(2)}\over \le(A_B^{(1)}\ri)^2}
		\ri)
		+{\phi_B^{(3)}\over A_B^{(1)} }
	\ri]
	}
\bea \label{L3}
	L_3
	=
	&e^{ 3 A_B /2}
	\Big\{
	{\phi_B^{(2)}}
	\le[
		\frac{27}{4} A_B^{(1)}
		-\frac{9}{2} {A_B^{(2)}\over A_B^{(1)}}
		-3\frac{ A_B^{(3)}}{\le(A_B^{(1)}\ri)^2}
		+\frac{6 \le(A_B^{(1)}\ri)^2}{\le(A_B^{(1)}\ri)^3}
	\ri]\nonumber\\
	&+{3\over2}{\phi_B^{(3)}}
	\le[
	 3 - 2 {A_B^{(2)}\over \le(A_B^{(1)}\ri)^2}
	\ri]
	\Big\}
\eea

Solving the Einstein equations \eqref{EEr} order by order in $(r-r_B)$, the first coefficients on \eqref{phibr} and on \eqref{Abr} are:
\begin{subequations}\label{Coeff}
	\begin{align}
		&\phi_B^{(2)}=V'(\phi_B)e^{2A_B},
		\\&\phi_B^{(3)}=-e^{3 A_B} V'(\phi_B) \sqrt{-\frac{ V(\phi_B)}{12}}
		\\&A_B^{(1)}=-\sqrt{-V(\phi_B) \over 12}e^{A(r)},
		\\&A_B^{(2)}=-{V(\phi_B) \over 12}e^{2A(r)}
	\end{align}
\end{subequations}
Replacing the coefficients \eqref{Coeff} into the expression \eqref{L2} one obtains \eql{f13}{L_2=0.}
The potential $V_{\cal U}$ from \eqref{Hu2} has a simple expression in terms of $L_n$:
\eql{f14}{
	V_{\cal U}(r)
		={2\over (r-r_B)^2}+{L_3\over 3 L_1}+\cO(r-r_B)^2.
}
The two linearly independent solutions of \eqref{Hu1} with the potential \eqref{f14} are:
\begin{subequations}\label{f15}
	\begin{align}
		{\cal U}_1(r)
			=
			&\frac{1}{r-r_B}
			+\frac{
				3 L_1 m^2-L_3
				}{
				6 L_1
				}
				(r-r_B)
				+\cO(r-r_B)^3
			\label{f15a}\\
			{\cal U}_2(r)
			=
			&(r-r_B)^2+\cO(r-r_B)^3.
			\label{f15b}
	\end{align}
\end{subequations}
The only singularity is the simple pole of ${\cal U}_1$ at $r_B$. However, the quantity appearing in the curvature invariants is $\Psi(r)$ which is proportional to $\phi_{,r}{\cal U}_1$, from its definition \eqref{chPsi}. As $\phi_{,r}$ vanishes linearly at the bounce, the simple pole of ${\cal U}_1$ is cancelled leading to a $\Psi(r)$ with a regular series expansion around $r_B$:
\begin{align}
	\Psi(r)
	=
	&C_1
	\le[
		1
		-
		{5\over 4\sqrt{3}}
		\sqrt{-V_ B}e^{A_B}(r-r_B)
	+\left(
		{m^2\over 2}
		-{5\over 32}e^{2A_B}V_ B
	\right)
	(r-r_B)^2
	+\cO(r-r_B)^3\ri]
	\nonumber\\
	&+C_2\le[(r-r_B)^3+\cO(r-r_B)^4\ri]
	\label{f18}
\end{align}
with $V_B\equiv V(\phi_ B)$. Therefore, all the geometrical invariants built from the Riemann tensor are finite to linear order in the perturbations, showing that the geometries generated by perturbations around a bounce are regular. The explicit form of the linear perturbation of some of the invariants \eqref{curv} is:
\begin{subequations}
	\begin{align}
		\delta R
		=&
		C_1\left(
		\frac{35}{6}V(\phi _B)
		-10e^{-2 A_B}m^2\right)
		+\cO(r-r_B),\\
		\delta (R_\mn R^\mn)
		=&
		-C_1V(\phi _B)
			\le(
				{5\over9} V(\phi _B)
				+{23\over 3} e^{-2 A_B} m^2
			\ri)
		+\cO(r-r_B),\\
		\delta (R_{\r\s\mn} R^{\r\s\mn})
		=&
		- C_1V\left(\phi _B\right)
			\le(
				{5\over18}V(\phi _B)
				+{10\over3} e^{-2 A_B}m^2
			\ri)
		+\cO(r-r_B),
	\end{align}
\end{subequations}
all of them regular.

\section{Spectral properties of  scalar modes near AdS extrema}
\label{BF}
In this section we (re)derive the spectrum of scalar  fluctuations around an AdS solution  and around a solution flowing out (or cascading out, if the BF bound is violated) of an AdS extremum located at $\phi=0$,  where the potential takes the approximate form:
\be\label{s1}
V(\phi) = V_0 + {1\over 2}m^2 \phi^2 + O(\phi^4)   \qquad V_0 = -{d(d-1) \over \ell^2} \, .
\ee

We start by considering  scalar fluctuation $\phi(r,x_\mu)$ around an exact AdS solution. We work in conformal coordinates:
\be\label{s2}
ds^2 = {\ell^2 \over r^2} \left( dr^2 - dt^2 + dx_i^2\right)
\ee
with $i=1,...,d-1$.
The field equation for the scalar fluctuation is
\be\label{s3}
r^{d-1}\de_r {1\over r^{d-1}}  \de_r \phi - {\ell^2 m^2 \over r^2} \phi + \Box_d \phi = 0\, , \quad \Box_d \equiv \de_\m \de^\m.
\ee
We concentrate  on  a mode with frequency $\omega$ and no spatial dependence (or equivalently on a mode with $d$-dimensional mass $\mu$, satisfying $\Box_d \phi= \mu^2 \phi$,  in the rest frame where $\omega^2 = \mu^2$). Then the problem reduces to an ordinary differential equation:
\be \label{s4}
\phi'' - {(d-1)\over r} \phi' - {\ell^2 m^2 \over r^2}\phi + \omega^2 \phi =0  \,.
\ee

One has the well-know solutions around $r=0$ (where the $\omega^2$ term is negligible)
\be \label{s5}
\phi(r) \sim a_- r^{\Delta_-} + a_+ r^{\Delta_+}, \qquad \Delta_{\pm} = {d\over 2} \pm {d\over 2} \sqrt{1 + {4 m^2 \ell^2 \over d^2}}.
\ee

One way to solve equation \eqref{s4} is to eliminate the linear term and transform it in a Schr\"odinger equation through the change of variables:
\be \label{sch1}
\psi (r)  = r^{-(d-1)/2}\phi(r).
\ee
Then equation (\ref{s4}) turns into the wave equation:
\be \label{sch2}
-\psi'' + {\a_{UV}\over r^2} \psi = \omega^2 \psi \qquad \a_{UV} \equiv \left(m^2\ell^2 + {d^2 - 1 \over 4}\right).
\ee
This is a $1d$ Schr\"odinger problem with potential $\a_{UV}/r^2$, which is well-known to be unstable for $\a_{UV}< -1/4$, \cite{Landau}, i.e. for $m^2\ell^2 < -d^2/4$ (the BF bound). We will re-derive this result in the following subsections.


\subsection{Perturbations around  RG-flows and  cascading solution} \label{cascade pert}

Here we show that solutions that cascade out of a BF bound violating maximum are unstable.

To show this we study linear perturbations around a solution which has the following asymptotic form in the UV:
\be \label{pert1}
A(r) \simeq -\log {r\over \ell}, \quad \phi(r) = \phi_0 r^{d/2} \cos \left[{|\nu |\over 2}  \log r + \varphi\right] \quad r\to 0.
\ee

The perturbations around this solutions admit an usual scalar-vector-tensor (SVT) decomposition and at linear order only the scalar perturbations are sensitive to the non-trivial $\phi(r)$ background.

Through the procedure of appendix \ref{ssec:back}, now in $d$ dimensions, we use the gauge invariant scalar fluctuation $\z$ from equation \eqref{Bardeenc}. We will use the formalism of appendix \ref{app:reg} to study the fluctuations and therefore restrict to $d=4$. We impose $\z$ has a well defined 4-dimensional mass $\m^2 $, i.e.,
\be
\Box_4\z(r,x^\m)=\m^2 \z(r,x^\m)
\ee
eliminating the $x^\m$ dependence, so we can write $\z=\z(r)$. The equation of motion is then:
\eql{eomz}{
	\z_{,rr}-2G_{,r}\z_{,r}+\o^2 \z(r)=0.
}
with
\eql{Gd}{
G(r) \equiv -{3 \over 2} A(r) -\log \left|{\phi_{,r} \over A_{,r}}\right|,
}
a generalization of \eqref{LG} to $d$ dimensions. We define a new variable $\psi_s(r)$ through:
\be
	\psi_s(r)=e^{-G(r)}\z(r) \label{inc3}.
\ee
The equation of motion \eqref{eomz} becomes:
\be
-\psi_{,rr}+V_s\psi=\o^2 \psi(r)
\quad
\text{with}
\quad
V_s(r) = (G_{,r})^2 - G_{,rr}
\ee

Therefore, the problem of finding the spectrum of scalar perturbations is analog to a Schr\"odinger problem with energy $\o^2 $ in a generalization of equations \eqref{Hz1} and \eqref{Hz2} for arbitrary dimensions.

Now we use \eqref{pert1} and write $\phi_{,r}/A_{,r}$ as follows:
\be
 \left({\phi_{,r} \over A_{,r}}\right)^2= r^4 f^2(r) , \qquad f(r) \equiv  2\left(1 + \le|\nu\over 2\ri|^2\right)^{1/2} \sin \left[{|\nu| \over 2} \log r + \tilde{\varphi}\right],
\ee
where $\tilde{\varphi}$ is another phase whose precise value is irrelevant.

Then, we have:
\be
G(r) = -{1\over 2} \log r f^2
\ee
and the Schr\"odinger potential:
\be
V_s(r) = -{1\over 4 r^2} + {f_{,rr} \over f} + {f_{,r} \over r f}.
\ee
Using the explicit expression for $f$ we find that the oscillating part cancels out, and we are left with:
\be
V_s(r) = - \left({1\over 4} + {|\nu|^2 \over 4} \right) {1\over r^2 }  = {\a_{UV} \over r^2}
\ee
This is again a $1/r^2$ potential with  coefficients less than $-1/4$, and has infinitely many negative energy normalizable states \cite{Landau}. In the last line,  we used the relation $\nu = \sqrt{1+4\a_{UV}}$ (where $\a_{UV} = m^2 \ell^2 + 15/4$ is the coefficient of the $1/r^2$ potential in the AdS solution for $d=4$, see equation (\ref{sch2})). Therefore,  the potential close to $r=0$ is the same as for scalar fluctuations around the $\phi=0$ AdS solution.


\section{The inverse formalism}
\label{app:inv}

In this section we provide a method to generate solutions of the superpotential equation \eqref{SuperP} by reverse-engineering: first one postulates a scalar field profile, then the corresponding geometry is derived together with the superpotential and the corresponding scalar field potential.

In our previous setup, presented in subsection \ref{ssec:setup}, we started with an Einstein-dilaton action in $d+1$ dimensions with a minimally coupled, self-interacting dilaton. We restricted to solutions satisfying $d$-dimensional Poincar\'e invariance, with the extra dimension being space-like. In subsection \ref{ss:superp}, we observed that the corresponding Einstein equations are automatically solved if there exists a (potentially multi-branched) function of the scalar field, called superpotential, satisfying a non-linear, first order differential equation involving the scalar field potential.

The idea here is to first postulate a function $\phi(u)$, with certain properties that we identify, and then use it to build the corresponding superpotential. From the superpotential we will construct the potentials that allow for such a solution and determine what are the possible scale factors $A(u)$ compatible with the ansatz for $\phi(u)$. This method will provide a fully back-reacted solution to the Einstein equations by simply postulating a scalar field profile.

This method will allow us to show that a generic solution containing minima or maxima of $\phi(u)$ will lead to a multi-branched potential $V(\phi)$.

Consider a function $f(u)$ as a solution for the dilaton that is a function of the domain wall coordinate $u\in (-\infty,+\infty)$.
\be
\phi(u)=f(u)
\label{i1}\ee
We have, by definition
\be
\phi_{UV}=f(-\infty)\sp \phi_{IR}=f(\infty)
\label{i2}\ee
and we assume the limits are finite, i.e. the flow is regular. Our conventions for derivatives are the same as before:
\be
f=f(u),~~{df\over du} \equiv \dot{f}(u) ~~~~and~~~~g=g(\phi),~~{dg\over d\phi} \equiv g'(\phi).
\ee

We will also assume that $f(u)$ remains finite during the flow.
From (\ref{phiW}) we have
\be
\dot\phi=\dot f= {dW\over d\phi}= {dW\over du}{du\over d\phi}
\label{i3}\ee
If for a given $u=u_B$ the derivative of $f(u)$ vanishes, the last step in \eqref{i3} holds in the limit $u\to  u_B$. From \eqref{i3} we obtain:
\be
{dW\over du}= (\dot f)^2.
\label{i4}\ee
We solve to obtain $W$ as a function of $u$,
\be
W(u)=\int_{-\infty}^{u} dv~(\dot f(v))^2+W_{UV} \label{i5}
\ee
where $W_{UV}$ is a constant. We also define
\be
W_{IR}\equiv W(+\infty)=W_{UV}\pm\int_{-\infty}^{+\infty} dv~(\dot f(v))^2.\label{i6}
\ee

The scale factor satisfies
\be
\dot A(u)=- {W(f)\over 2(d-1)}=-{1\over 2(d-1)}\left(\int_{-\infty}^u dv~\dot f^2(v)+ W_{UV}\right)\label{i7}
\ee
For flows that interpolate between AdS fixed points it follows that:
\be
	\dot{A}(-\infty)=-{1\over \ell_{UV}}=-{W_{UV}\over 2(d-1)},
	\qquad
	\dot{A}(\infty)=-{1\over \ell_{IR}}=-{W_{IR}\over 2(d-1)} \label{i8}
\ee
so the scale factor can be written as:
\be
A(u)=A(0)-{u\over \ell_{UV}}-{1\over 2(d-1)}\int_0^u dv\int_{-\infty}^v dw~\le(\dot f(w)\ri)^2.\label{i9}
\ee
where $A(0)$ is an integration constant coming from equation \eqref{i7} and $\ell_{UV}$ was defined in \eqref{i8}.
We use \eqref{i8} to rewrite \eqref{i5}:
\be
\boxed{W(u)=\frac{2(d-1)}{\ell_{UV}}+\int_{-\infty}^{u} dv~(\dot f(v))^2}\label{i10}
\ee
From \eqref{i8} and  \eqref{i10} we deduce the relation between the UV and IR $AdS$ lengths:
\eql{l_IR}{
{1\over \ell_{IR}}= \frac{1}{2(d-1)}\int_{-\infty}^{\infty} dv~(\dot f(v))^2+\frac{1}{\ell_{UV}}
}

To obtain the true superpotential $W(f)$ we need the inverse of $f(u)$. Assume that $f(u)$ has $n$ points\footnote{The case where $f(u)$ goes to a local maximum of $V$ that violates the BF bound has $n=\infty$.} such that
\be
\dot f(u)\text{ changes sign at }u=\hat u_1,...,\hat u_n \label{i11}
\ee
and define
\be
\hat u_0\equiv-\infty,\quad \hat u_{n+1}\equiv+\infty. \label{i12}
\ee
Then there will be $n+1$ functions $u_i(f)$ such that:
\be
u_i(f):=f^{-1}(u)\text{ for } u\in (\hat u_{i-1},\hat u_{i}). \label{i13}
\ee

The multi-branched super-potential that results from \eqref{i5} and \eqref{i13} will have branches labeled $W_1,...,W_n$ defined by:
\be
W_i(f):= W(u_i(f))=\int_{-\infty}^{u_{i}(f)} dv~(\dot f(v))^2+W_{UV}\equiv \int_{\hat u_{i-1}}^{u_{i}(f)} dv~(\dot f(v))^2+W(\hat u_{i-1}) \label{i14}
\ee


Each branch $W_i(f)$ must satisfy the superpotential equation:
\be
V(f)={1\over 2} W_i'^2(f)-\frac{d}{4(d-1)}W_i^2(f), \quad i=1,...,n+1. \label{i15}
\ee
which can be alternatively expressed in terms of $u$ as:
\begin{align}
V(f(u))=:V(u)&={1\over 2} \le({{\dot W(u)}\over \dot f(u)}\ri)^2-\frac{d}{4(d-1)}(W(u))^2
	\\&={1\over 2}\dot f^2-\frac{d}{4(d-1)}W^2
	\\&={1\over 2}\dot f^2
		-\frac{d}{4(d-1)}
			\le(
			\int_{-\infty}^{u} dv~(\dot f(v))^2+W_{UV}
			\ri)^2
	\label{i16}
\end{align}

While the right hand side of equation \eqref{i16} is single valued in $u$, this is not obvious that the right-hand side of \eqref{i15} is single-valued as a function of $f$. For this to be true, we must impose:
\be
\ha W_i'^2(f)-\frac{d}{4(d-1)}W_i^2(f)={1\over 2} W_j'^2(f)-\frac{d}{4(d-1)}W_j^2(f), \quad i,j=1,...,n+1 \label{i17}
\ee
Equation \eqref{i17} places restrictions on the possible functions $f(u)$ in a highly non-local way. In terms of $f(u)$ equation \eqref{i17} becomes:
\begin{align}\label{i18}
	V_i(f)\equiv
	&-\frac{d}{4(d-1)}\left(\pm\int_{-\infty}^{u_i(f)} dv~(\dot f(v))^2+W_{UV} \right)^2
		+{1\over 2}\dot f^2(u_i(f))\nonumber\\
	=&-\frac{d}{4(d-1)}\left(\pm\int_{-\infty}^{u_j(f)} dv~(\dot f(v))^2+W_{UV} \right)^2
		+{1\over 2}\dot f^2(u_j(f))\equiv V_j(f).
\end{align}
for $ i,j=1,...,n+1$. In general $V_i(f)\neq V_j(f)$.

\subsection{Regularity}

A holographic RG flow is regular if $W(\phi)$ and $V(\phi)$ are finite along the flow $\phi(u)$. The approach in this appendix is to build $V(\phi)$ from an ansatz $f(u)=\phi(u)$, with $f(u)$ of class $C^2(\mathbb{R})$. A finite $W(u)\equiv W(f(u))$ along the flow is then equivalent to a finite $W_{IR}$ and a finite $W_{UV}$ in equation \eqref{i6}. In other words, {\em the geometry is regular if the integral of $(\dot f(u))^2$ on the real line is convergent}. This implies that $\dot f$ should vanish faster than $|u|^{-1/2}$ as $|u|\to \infty$.

\subsection{Fixed points}

A regular RG flow interpolates between two fixed points. For this to be true certain conditions must be imposed on $f(u)$. We know from regularity that $\dot f (\pm\infty)=0$ in these cases, therefore:
\begin{subequations}\label{i19}
\begin{align}
	V_{UV}
		\equiv &-\frac{d}{4(d-1)}W^2_{UV}+{1\over 2}\dot f^2(-\infty)
		=-\frac{d(d-1)}{\ell_{UV}}\label{i19a}\\
	V_{IR}
		\equiv&-\frac{d}{4(d-1)}W^2_{IR}+{1\over 2}\dot f^2(+\infty)
		=-\frac{d(d-1)}{\ell_{IR}}\label{i19b}
\end{align}
\end{subequations}
where we used \eqref{i8}.
We also need to impose appropriate asymptotic behavior to ensure that the flow is from one maximum of the potential to a minimum or that it interpolates between two minima.

We now compute
\be
\dot V=\dot f\ddot f -\frac{d}{2(d-1)}W\dot W=\dot f\ddot f- \frac{d}{2(d-1)}W\dot f^2=\dot f(\ddot f- \frac{d}{2(d-1)}W\dot f)
\label{i20}
\ee
On the other hand
\be
\dot V={dV\over d\phi}\dot \phi=V'\dot f \label{V.}
\ee
so that
\be
V'\equiv {dV\over d\phi}
	=\ddot f -\frac{d}{2(d-1)}W\dot f
	=\dot f \le(
		{\ddot f\over \dot f} -\frac{d}{2(d-1)}W
		\ri)
\label{V'f}
\ee
Using W$''$=$\ddot f/\dot f$ we can put \eqref{V'f} in the same form as equation \eqref{V'} from subsection \ref{ssec:crit}:
\be
	V'=W'\left(W''-\frac{d}{2(d-1)}W\right)\label{i21}
\ee
There are two possibilities for $V'(f_*)$ to vanish:
\begin{enumerate}
\item $W'(f_*)$=0 and $W''(f_*)$ is finite : the solution $W(f)$ reaches a fixed point at $f_*$,
\item $W'(f_*)\neq$0 and $W''(f_*)=dW(f_*)/[2(d-1)]$ : the solution W misses a fixed point but there is an extremum of $V(\phi)$ at $\phi=f_*$.
\end{enumerate}
These possibilities are respectively equivalent to:
\begin{enumerate}
\item $\dot f(u_*) = 0, ~\ddot f(u_*)= 0$ and  ${\ddot f/ \dot f}\big|_{u=u_*}\left(=\pm W''(\phi_*)\right)~\text{finite} $

\item $\dot f(u_*)\neq0$ and ${\ddot f/ \dot f}\big|_{u=u_*}= \pm\frac{d}{2(d-1)}W\big|_{\phi=\phi_*}$ and W$'$($\phi_*$)

\end{enumerate}
where $\phi_*\equiv\phi(u_*)$ and $|u_*|=\infty$.

This implies that we can have a non-vanishing W$'$ at a point where V$'$ is zero, so the flow of $\phi(u)$ does not stop at the extremum of V.

We now compute the second derivative
\be
	V''
		={d^2V\over d\phi^2}
		={d\over d\phi}{\dot V\over \dot\phi}
		={1\over \dot \phi}{d\over du}\le[
								\ddot f
								- \frac{d}{2(d-1)}W\dot f
								\ri]
		={\dddot f-\frac{d}{2(d-1)}\le(
					 \dot f^3+ W\ddot f
					\ri)
					\over \dot f}
	\label{V''}
\ee

If we want a potential with a non-zero mass term at the fixed points, we impose:
\be
V''(f_*)=-\left({d\over 2(d-1)}\right)^2W^2(f_*)+{\dddot f\over \dot f}\Big|_{f_*}=m^2
\label{i22}
\ee
where $f_*$ may be a UV or an IR fixed point and we used the fact that $\dot f$ vanishes at a fixed point.
We define $\Delta_\pm$ as in subsection \ref{ssec:crit}:
\be
\Delta_{\pm} = {d\over 2} \pm {d\over 2} \sqrt{1 + {4 m^2 \ell^2 \over d^2}}
\label{i23}
\ee
and we denote $\Delta_+$ by $\Delta$.
With the conditions \eqref{i19} and \eqref{i22}, the results from subsection \ref{ssec:crit} tell us that the flow given by $f(u)$ reach a UV fixed point as $u$ asymptotes to $-\infty$ if:
\begin{align}
f(u) =
\begin{cases}
C_+ e^{u \Delta/\ell_{UV}}+C_- e^{u (d-\Delta)/\ell_{UV}}+...,\text{if }{d\over 2}<\Delta<d.\\
C_+ e^{u \Delta/\ell_{UV}}+...,\text{if }\Delta>d.
\end{cases}
\end{align}
For an IR fixed point, the condition is that, when $u$ asymptotes to $+\infty$,
\begin{align}
f(u) = \tilde C_- e^{u (d-\Delta)/\ell_{IR}}+...,\text{and }\Delta>d.
\end{align}


\section{Cyclic RG flows}
\label{app:cycles}
Here we show that in an Einstein-scalar gravity with a minimally coupled, self-interacting scalar there can be cycles only if the scalar field potential is multi-valued in a very specific way.

The approach of this appendix proceeds along similar lines with appendix \ref{app:inv}:  we will use reverse-engineering. We start with a superpotential $W(\phi)$ and deduce for which potential $V(\phi)$ the ansatz for $W$ will solve the superpotential equation \eqref{SuperP}:
\eql{cy0}{V(\phi)=\ha W'^2(\phi)-\frac{d}{4(d-1)}W^2(\phi)}
We first study an example from the literature and then derive a property that can be used to put constraints on multi-branched potentials coming from monodromy \cite{mono}.

\subsection{A case study of cyclic flow}

One case in which the potential is ill-defined is the case of cyclic RG flows of the form given in \cite{cycles}. In this case we have a coupling $\phi$ with a multi-branched $\b$-function:
\eql{cy1}{
	{d\phi\over dA}=(-1)^n\sqrt{1-\phi^2}=-2(d-1){d\over d\phi}\log W_n, \qquad n\in\mathbb{Z}
}
with solutions which are RG cycles for a single coupling:
\eql{cy2}{
\phi(A)=\sin(A)
}
In our holographic setup, RG flows are gradient flows described by a super-potential $W(\phi)$ and in the presence of turning, or bounces, the superpotential becomes multi-valued. For cyclic flows there will be infinitely many branches. The multi-branched super-potential leading to \eqref{cy1}, represented by the functions $W_n(\phi)$ in \eqref{cy1}, is given by:
\eql{cy3}{
W_n(\phi)=\exp
		\le\{
			{1\over 4(d-1)}
			\le[
				{\pi\over 2}(1+2n)
				+(-1)^n\le(
						\arcsin(\phi)+\phi\sqrt{1-\phi^2}
						\ri)
			\ri]
		\ri\}
}
where $\arcsin$ is calculated in the principal branch. The resulting potentials are:
\begin{align}
V_n(\phi)=&{1\over 4(d-1)}
		\le(
		{1-\phi^2\over 2(d-1)}-d
		\ri)\times
		\nonumber\\
		&\times\exp
		\le\{
			{1\over 4(d-1)}
			\le[
				{\pi\over 2}(1+2n)
				+(-1)^n\le(
						\arcsin(\phi)+\phi\sqrt{1-\phi^2}
						\ri)
			\ri]
		\ri\}.\label{V_n}
\end{align}
All the potentials \eqref{V_n} are different, negative and none of them is well defined for $|\phi|>1$. After each cycle of $\phi(A)$ there is a rescaling of the potential:
\be
V_{n+2}(\phi)=e^{\pi\over 2(d-1)}V_n(\phi).
\ee
Hence their impossibility in a holographic setup of the form given in subsection \ref{ssec:setup}. The fact that $V(\phi)$ does not exist for $|\phi|>1$ is specific to this case, but the multi-branched nature of the potential and the periodic scaling are general features that we prove below.

\subsection{Generalization}

We now generalize the above results to show that any cyclic RG flow can only arise from multi-branched potentials that increases amplitude by a constant factor after each cycle.

Assume that $\phi(A)$ is a solution to the RG flow equation:
\eql{gen1}{{d\phi\over dA}=\b(\phi).}

Suppose also that $\phi(A)$ is periodic, with period $\t$. If $\phi(A)$ is continuous, periodicity implies the existence of turning points of the RG flow. These turning points correspond to bounces, points where the superpotential switches branches. We have:
\eql{gen2}{
	{d\phi(A)\over dA}={d\phi(A+\t)\over dA}
	\implies
	\b(\phi(A))=\b(\phi(A+\t))
	}
We denote again each branch of the superpotential by $W_n(\phi)$, $n\in\mathbb{Z}$. If the periodicity of the scalar field appears after $N$ branches of the superpotential, with $N\geqslant2$, we have:
\eql{gen3}{
	{d\over d\phi}\log W_n(\phi)={d\over d\phi}\log W_{n+N}(\phi)
	\implies
	W_{n+N}(\phi)=C W_n(\phi)
	}
with a real constant $C$. The superpotential equation \eqref{cy0} then implies:
\eql{gen4}{
V_{n+N}(\phi)=C^2V_n(\phi), \quad C\in\mathbb{R}.
}
Notice that if $V_n(\phi)$, $n=0,1,...,N-1$ are strictly negative, then all the branches of $V$ will be strictly negative. In this case the superpotential cannot change sign, as follows from equations \eqref{BoundW} and \eqref{Wpos}, and we conclude that when $V_n(\phi)$ is strictly negative the constant $C$ in equations \eqref{gen3} and \eqref{gen4} is strictly positive.

Furthermore, as $W(\phi(u))$ increases monotonically as a function of $u$, without loss of generality we can label the branches $W_n$ with $n$ increasing as $W$ increases. If a given branch $W_n(\phi)$ changes sign, it is impossible to make $W(\phi(u))$ continuous in the presence of the rescaling \eqref{gen3}. If $W_n$ does not change sign, cyclic solutions which are continuous and monotonic in $u$ must have a positive $C$. The labeling of the branches in order of increasing $u$ implies $C>1$, ruling out single-valued potentials.

If instead we impose $C=1$, we obtain $W_{n+N}(\phi)=W_n(\phi)$, for all $n\in\mathbb{Z}$ and all $\phi$ in the domain of $W_n$. Because $W(u)$ cannot decrease and $W(\phi)$ must be continuous in order to satisfy \eqref{cy0} with a regular potential, the condition $C=1$ implies that $W(\phi)$ is single-branched and the labeling $n$ is a redundancy. Therefore only single-branched potentials are allowed and cycles are ruled out.

The behavior \eqref{gen4} is in contrast with the multi-valued potential found in \cite{mono}, where the different branches correspond solely to discrete translations in field space and not an overall rescaling of the potential.

\section{RG projection on single trace operators}
\label{single}

In this appendix we will explore a simple set of flows generated by a QFT single trace operator $O(x)$ of scaling dimension $\Delta$ and its double trace cousin $O^2$ of dimension $2\Delta$ at large $N$.
We consider a perturbation of the CFT by
\be
S_{pert}=N\l_1\int d^dx~ O(x)+  \l_2\int d^dx~ O^2(x)
\label{a1}\ee
The associated flow equations are
\be
\dot \l_1=\beta_1(\l_1,\l_2)\sp \dot \l_2=\beta_2(\l_1,\l_2)
\label{a2}\ee
with the associated (one-loop and large-N) $\beta$-functions given as
\be
\beta_1=(d-\Delta)\l_1-C_{111}\l_1^2-C_{112}\l_1\l_2\sp \beta_2=(d-2\Delta)\l_2-C_{112}\l_1^2-C_{111}\l_1\l_2-C_{222}\l_2^2
\label{a3}\ee
At large $N$, these $\beta$-functions are one-loop exact\footnote{For a discussion also of the subleading large N corrections see \cite{kn}.} and with appropriate normalizations $C_{112}=C_{222}=1$, but we will keep the OPE coefficients general for demonstrative purposes.

There are four fixed points of the system (\ref{a2}).
\begin{itemize}

\item  The trivial fixed point that we denote by $F_0$.

\be
\l_1=\l_2=0
\label{a4}\ee
The flows in the vicinity of the fixed point $\l_i=\l_i^*+\delta\l_i$ are of the form
\be
\left(\begin{matrix} \dot {\delta\l_1} \\ \dot {\delta\l_2}\end{matrix}\right)=M_0 ~\left(\begin{matrix} \delta\l_1 \\ \delta\l_2\end{matrix}\right)
\label{a6}\ee
with the stability matrix being 
\be
M_0=\left(\begin{matrix} d-\Delta & 0 \\ 0&d-2\Delta\end{matrix}\right).
\label{a7}\ee
For ${d\over 2}<\Delta <d$,  $O$ is relevant and $O^2$ is irrelevant, making the $\l_1$ direction unstable while the $\l_2$ is stable; on the other hand, for  ${d\over 2}-1<\Delta < {d\over 2}$, both $O$ and $O^2$ are relevant and the origin is unstable in both $\l_1$ and $\l_2$ directions.

\item A non-trivial fixed point in the double trace direction that we denote by $F_2$.  
\be
\l_1=0\sp \l_2={d-2\Delta\over C_{222}} \, .
\label{a5}\ee

The stability matrix here is 
\be
M_2=\left(\begin{matrix} -{(d-2\Delta)C_{112}-(d-\Delta)C_{222}\over C_{222}} & 0 \\ -{C_{111}\over C_{222}}(d-2\Delta)&2\Delta-d\end{matrix}\right)\, .
\label{a8}\ee
The diagonal elements coincide with the eigenvalues.

\item A pair of non-trivial fixed points that we denote by $F_{\pm}$
\be
\l^{\pm}_1=-{(d-\Delta)C_{111}\over 2C_{112}^2}\pm{\sqrt{(d-\Delta)^2C_{111}^2+4C_{112}(d-\Delta)(C_{112}(d-2\Delta)-C_{222}(d-\Delta)}\over 2C_{112}^2}  
\label{a6}\ee
\be
\l^{\pm}_2={d-\Delta\over C_{112}}-C_{111}\l_1^{\pm}
\label{a12} \ee
with stability matrix
\be
M_{\pm}=\left(\begin{matrix} d-\Delta-2C_{111}\l_1^{\pm}-C_{112}\l_2^{\pm}& -C_{112}\l_1^{\pm}\\ -2C_{112}\l_1^{\pm}-C_{111}\l_2^{\pm} &
d-2\Delta-C_{111}\l_1^{\pm}-2C_{222}\l_2^{\pm}
\end{matrix}\right)\, .
\label{a9}\ee
These exist only if 
\be
(d-\Delta)^2C_{111}^2+4C_{112}(d-\Delta)(C_{112}(d-2\Delta)-C_{222}(d-\Delta))\geq 0\;.
\ee

\end{itemize}

We can ``integrate out" the double trace coupling at the expense of making the flow equation for the single trace coupling second order in the flow parameter.
For this, we differentiate the first equation in (\ref{a2}) once, substitute $\dot \l_2$ from the second equation and substitute $\l_2$ by resolving the first equation in terms of $\dot \l_1$ and $\l_1$. We obtain,

\be
\l_1\ddot\l_1-{C_{112}+C_{222}\over C_{112}}{\dot\l_1^2}+3C_{111}\l_1\dot \l_1-{(d-2\Delta)C_{112}-2(d-\Delta)C_{222}\over C_{112}}\l_1\dot \l_1-C_{112}^2\l_1^4-
\label{a10}\ee
$$
-(d-\Delta)C_{111}\l_1^3+(d-\Delta){(C_{112}(d-2\Delta)-C_{222}(d-\Delta)\over C_{112}}\l_1^2=0\, .
$$
The fixed points of this projected RG equation can be obtained from the relation 
\be
\left[C_{112}^2\l_1^2+ (d-\Delta)C_{111}\l_1 -(d-\Delta){(C_{112}(d-2\Delta)-C_{222}(d-\Delta)\over C_{112}}\right]\l_1^2=0\, ,
\label{a11}\ee
and coincide with the associated fixed point values of the four original fixed points $F_0,F_2, F_{\pm}$.
Note however that the double zero fixed point corresponds to two distinct fixed points of the original system $F_0$ and $F_2$ that are only distinguished by the value of the double trace coupling $\l_2$.

It is clear that the original system needs two integration constants: the initial values $\l_1^{(0)}, \l_2^{(0)}$ of $\l_1,\l_2$.
The second order equation for the single trace coupling needs as initial conditions $\l_1^{(0)}, \dot\l_1^{(0)}$. These are however equivalent 
via the relation
\be
\l_2=  {-\dot \l_1+ (d-\Delta)\l_1-C_{111}\l_1^2\over C_{112}\l_1}\, .
\ee
\begin{figure}[h!]
\centering
\includegraphics[width=0.6\textwidth]{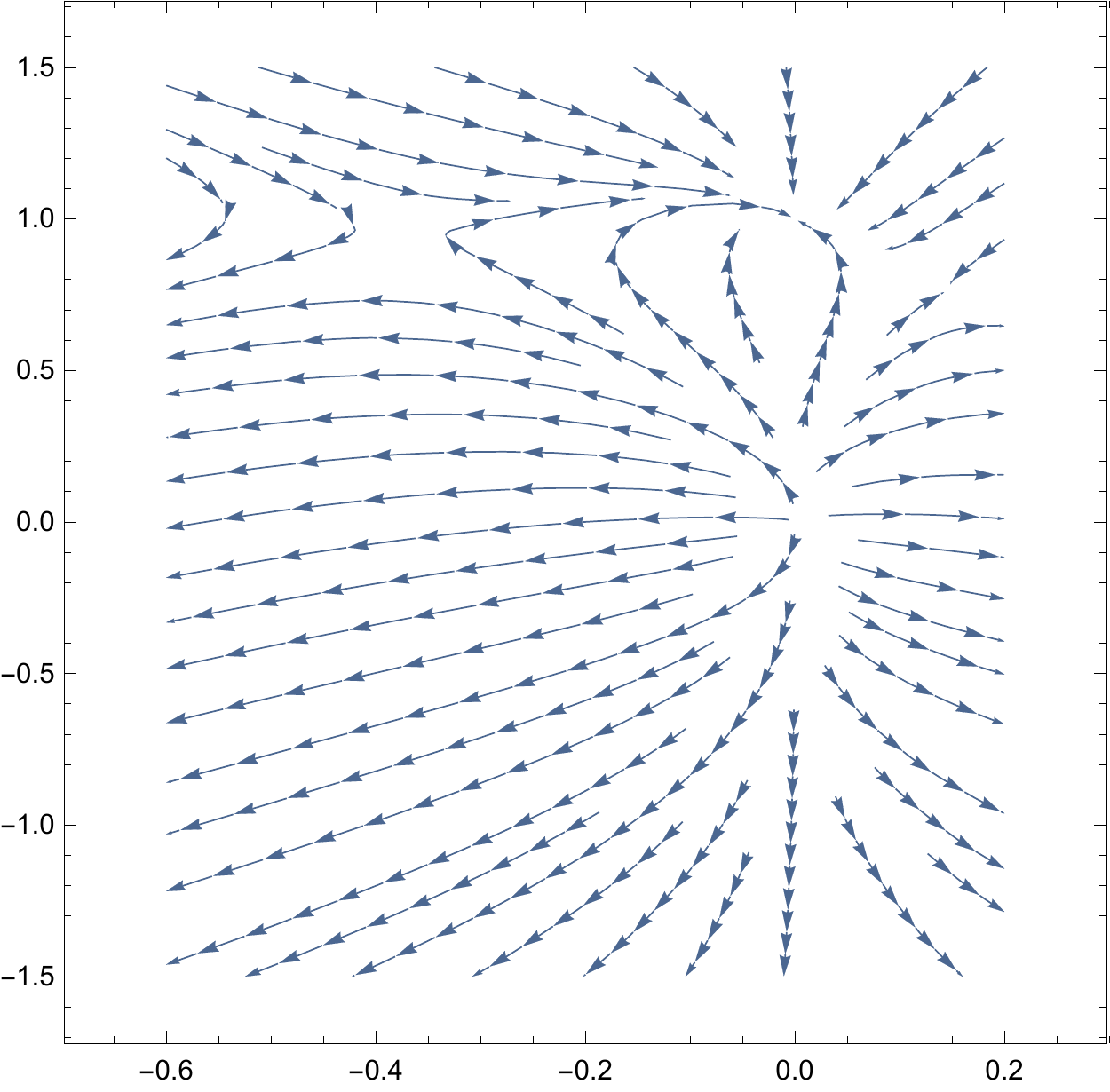}
\caption{
The two-coupling  flow pattern}
\label{flow}.
\end{figure}

For a specific choice of numbers $d=4,\Delta=3/2,C_{111}=C_{222}=1,C_{112}=3$,  the system has four fixed points at (0,0) (fully unstable), (0,1) (fully stable) and (0.14,0.78) and (-0.36,0.95) (both saddles). The two dimensional flow pattern is shown in figure \ref{flow}. It is in the same topological class as figure 7 of \cite{gukov}.

It is clear now that the projected flows on the $\l_1$ axis described by (\ref{a9}) show some similarities to some of the exotic flows we have found on the holographic side.In particular, depending on the initial conditions the flow of $\l_1$ may change direction without stopping.
This will happen for example if $\l^{(0)}_1$ is slightly negative and $\dot \l^{(0)}_1$ is sufficiently positive.
In that case $\l_1$ initially decreases, then turns around and ends up at the stable fixed point at 0. Moreover, as we have seen, this is really  a collection of two distinct fixed points in the two coupling diagram, distinguished by the value of the double trace coupling.

\end{document}